\documentclass[12pt]{article}
\usepackage{a4wide,amsmath,amssymb,bm,graphicx}
\newcommand{\D}{\rlap{\hspace{0.2em}/}D}
\newcommand{\MS}{\ensuremath{\overline{\text{MS}}}}
\DeclareMathOperator{\Tr}{Tr}
\numberwithin{equation}{section}
\newcommand{\1}[1]{\mathbf{1}^{#1}}
\newcommand{\2}[1]{\mathbf{2}^{#1}}
\newcommand{\3}[1]{\mathbf{3}^{#1}}
\newcommand{\4}[1]{\mathbf{4}^{#1}}
\newcommand{\5}[1]{\mathbf{5}^{#1}}
\newcommand{\Li}[1]{\mathop{\mathrm{Li}}\nolimits_{#1}}

\begin{document}

\renewcommand{\thefootnote}{\fnsymbol{footnote}}

\begin{center}
\LARGE Introduction to effective field theories%
\footnote{Extended version of lectures
given at Karlsruher Institut f\"ur Technologie, 2010,
and Novosibirsk State University, 2011.}\\[0.5em]
\Large 3. Bloch--Nordsieck effective theory, HQET\\[1.5em]
\large Andrey Grozin\\[0.5em]
\large Budker Institute of Nuclear Physics SB RAS\\
\large and Novosibirsk State University
\end{center}
\vspace{1.5em}

\begin{abstract}
This is a continuation of the lectures~\cite{eft1,eft2}.
In this part we discuss interaction of electrons with soft photons
in QED~\cite{BN:37}
and Heavy Quark Effective Theory (HQET, see~\cite{N:94,MW:00,G:04}).
\end{abstract}

\tableofcontents
\renewcommand{\thefootnote}{\arabic{footnote}}
\setcounter{footnote}{0}

\section{Bloch--Nordsieck effective theory}
\label{S:BN}

\subsection{Heavy electron effective theory (HEET)}
\label{S:Photonia}

Photonia has imported a single electron from Qedland,
and physicists are studying its interaction with soft photons
(both real and virtual) which they can produce and detect so well.
The aim is to construct a theory describing states
with a single electron plus soft photon fields.

The ground state (``vacuum'') of the theory is the electron at rest
(and no photons).
It is natural to define its energy to be $0$.
When the electron has momentum $\vec{p}$, its energy is
\begin{equation}
\varepsilon(\vec{p}\,) = \frac{\vec{p}\,^2}{2M}\,,
\label{Photonia:kinetic}
\end{equation}
where $M$ is the electron mass (in the on-shell renormalization scheme),
our large mass scale.
The electron velocity is
\begin{equation}
\vec{v} = \frac{\partial\varepsilon(\vec{p}\,)}{\partial\vec{p}}
= \frac{\vec{p}}{M}\,.
\label{Photonia:velocity}
\end{equation}

At the leading ($0$-th) order in $1/M$,
the mass shell of the free electron is
\begin{equation}
\varepsilon(\vec{p}\,) = 0\,.
\label{Photonia:mshell}
\end{equation}
At this order, the electron velocity is
\begin{equation}
\vec{v} = \frac{\partial\varepsilon(\vec{p}\,)}{\partial\vec{p}}
= \vec{0}\,.
\label{Photonia:zero}
\end{equation}
The electron does not move; it always stays in the point
where it has been put initially.
The Lagrangian
\begin{equation}
L = h^+ i \partial_0 h\,,
\label{Photonia:L0}
\end{equation}
where $h$ is the 2-component spinor electron field,
leads to the equation of motion
\begin{equation}
i \partial_0 h = 0\,.
\label{Photonia:EOM0}
\end{equation}
This means that the energy of an on-shell electron
is $\varepsilon=0$.
Thus the Lagrangian~(\ref{Photonia:L0})
reproduces the mass shell~(\ref{Photonia:mshell}),
and can be used to describe the free electron
at the leading order in $1/M$.

The electron has charge $-e$.
Therefore, when placed in an external electromagnetic field,
it has energy
\begin{equation}
\varepsilon = - e A_0
\label{Photonia:E}
\end{equation}
instead of~(\ref{Photonia:mshell}).
Therefore, the equation of motion is
\begin{equation}
i D_0 h = 0
\label{Photonia:EOM}
\end{equation}
instead of~(\ref{Photonia:EOM0}), where
\begin{equation}
D_\mu = \partial_\mu - i e A_\mu
\label{Photonia:covariant}
\end{equation}
is the covariant derivative.
It can be obtained from the HEET Lagrangian~\cite{EH:90}
\begin{equation}
L = h^+ i D_0 h\,.
\label{Photonia:L}
\end{equation}
This Lagrangian is not Lorentz-invariant.
It is invariant with respect to the gauge transformation
\begin{equation}
A_\mu \to A_\mu + \partial_\mu \alpha(x)\,,\quad
h \to e^{i e \alpha(x)} h\,.
\label{Photonia:gauge}
\end{equation}

Of course, the full Lagrangian is the sum of~(\ref{Photonia:L})
and the Lagrangian of the photon field.
This gives the equation of motion for the electromagnetic field
\begin{equation}
\partial_\mu F^{\mu\nu} = j^\nu\,,
\label{Photonia:Maxwell}
\end{equation}
where the current $j^\mu$ has only $0$-th component
\begin{equation}
j^0 = - e h^+ h
\label{Photonia:j}
\end{equation}
(the interaction term in the Lagrangian~(\ref{Photonia:L}) is $-j^\mu A_\mu$).
The electron produces the Coulomb field.

At the leading order in $1/M$,
the electron spin does not interact with electromagnetic field.
We can rotate it without affecting physics.
Speaking more formally, the Lagrangian~(\ref{Photonia:L}) has,
in addition to the $U(1)$ symmetry $h\to e^{i\alpha}h$,
also the $SU(2)$ spin symmetry~\cite{IW:90}:
it is invariant with respect to transformations
\begin{equation}
h \to U h\,,
\label{Photonia:SU2}
\end{equation}
where $U$ is a $SU(2)$ matrix ($U^+ U=1$).

In fact, the electron has magnetic moment $\vec{\mu}=\mu\vec{\sigma}$
proportional to its spin $\vec{s}=\vec{\sigma}/2$,
and this magnetic moment interacts with magnetic field:
the interaction Hamiltonian is $-\vec{\mu}\cdot\vec{B}$.
But by dimensionality the magnetic moment $\mu\sim e/M$,
and this interaction only appears at the level of $1/M$ corrections.
Namely, $\mu=-\mu_B$ (up to small radiative corrections), where
\begin{equation}
\mu_B = \frac{e}{2M}
\label{Photonia:mu}
\end{equation}
is the Bohr magneton.
The Lagrangian thus has an additional term
describing this magnetic interaction,
\begin{equation}
L_m = - \frac{e}{2M} h^+ \vec{B}\cdot\vec{\sigma} h\,.
\label{Photonia:Lm}
\end{equation}
This term violates the $SU(2)$ spin symmetry at the $1/M$ level.

If we assume that there are $n_f$ flavours of heavy fermions,
\begin{equation}
L = \sum_{i=1}^{n_f} h_i^+ i D_0 h_i\,,
\label{Photonia:nf}
\end{equation}
then the Lagrangian has $U(1)\times SU(2n_f)$ symmetry
(even when the masses $M_i$ are different).
The spin-flavour symmetry is broken at the $1/M_i$ level
by both the kinetic-energy term and the magnetic-interaction term.

At the leading order in $1/M$,
not only the spin direction but also its magnitude is irrelevant.
We can, for example, switch the electron spin off:
\begin{equation}
L = \varphi^* i D_0 \varphi\,,
\label{Photonia:spin0}
\end{equation}
where $\varphi$ is a scalar field (with charge $-e$).
This is the most convenient form of the Lagrangian in all cases
when we are not interested in $1/M$ corrections.
If we consider the scalar and the spinor fields together,
\begin{equation}
L = \varphi^* i D_0 \varphi + h^+ i D_0 h\,,
\label{Photonia:spins}
\end{equation}
then this Lagrangian has $U(1)\times SU(3)$ symmetry~\cite{GW:90}.
The superflavour $SU(3)$ symmetry contains, in addition to $SU(2)$
spin transformations~(\ref{Photonia:SU2})
and phase rotations $\varphi\to e^{2i\alpha}\varphi$, $h\to e^{-i\alpha}h$,
also transformations which mix spin-0 and spin-$\frac{1}{2}$ fields.
In the infinitesimal form,
\begin{equation}
\delta \left(\begin{array}{c}\varphi\\h\end{array}\right) =
i \left(\begin{array}{cc}0&\varepsilon^+\\\varepsilon&0\end{array}\right)
\left(\begin{array}{c}\varphi\\h\end{array}\right)\,,
\label{Photonia:super}
\end{equation}
where $\varepsilon$ is an infinitesimal spinor parameter.
So, this $SU(3)$ is a supersymmetry group.
If we want, we can consider, e.\,g., spins $\frac{1}{2}$ and 1;
the corresponding Lagrangian has $SU(5)$ superflavour symmetry.
The superflavour symmetry is broken at the $1/M$ level
by the magnetic-interaction term in the Lagrangian~(\ref{Photonia:Lm}).

\subsection{Feynman rules}
\label{S:Feyn}

For now, we are working at the leading order in $1/M$.
The HEET Lagrangian expressed via the bare fields and parameters is
\begin{equation}
L = \varphi_0^* i D_0 \varphi_0
- \frac{1}{4} F_{0\mu\nu} F_0^{\mu\nu}
- \frac{1}{2 a_0} (\partial_\mu A_0^\mu)^2\,,\quad
D_\mu = \partial_\mu - i e_0 A_{0\mu}\,.
\label{Feyn:L}
\end{equation}
It gives the usual photon propagator.
From the free electron part $\varphi_0^* i \partial_0 \varphi_0$
we obtain the momentum-space free electron propagator
\begin{equation}
\raisebox{-3.5mm}{\begin{picture}(22,6.5)
\put(11,4.5){\makebox(0,0){\includegraphics{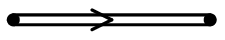}}}
\put(11,0){\makebox(0,0)[b]{$p$}}
\end{picture}}
= i S_0(p)\,,\qquad
S_0(p) = \frac{1}{p_0+i0}\,.
\label{Feyn:Sp}
\end{equation}
It depends only on $p_0$, not on $\vec{p}$.
If we use the spin-$\frac{1}{2}$ field $h_0$
instead of the spin-0 field $\varphi_0$,
then the unit $2\times2$ spin matrix is assumed here.
The coordinate-space propagator is its Fourier transform:
\begin{equation}
\raisebox{-3.5mm}{\begin{picture}(22,6.5)
\put(11,4.5){\makebox(0,0){\includegraphics{fr2.eps}}}
\put(1,0){\makebox(0,0)[b]{$0$}}
\put(21,0){\makebox(0,0)[b]{$x$}}
\end{picture}}
= i S_0(x)\,,\qquad
S_0(x) = S_0(x_0) \delta(\vec{x}\,)\,,\qquad
S_0(t) = - i \theta(t)\,.
\label{Feyn:Sx}
\end{equation}
The infinitely heavy (static) electron does not move:
it always stays at the point where it has been placed initially.
Alternatively, instead of Fourier-transforming~(\ref{Feyn:Sp}),
we can obtain~(\ref{Feyn:Sx}) by direct solving the equation
\begin{equation}
i \partial_0 S_0(x) = \delta(x)
\label{Feyn:propeq0}
\end{equation}
for the free $x$-space propagator.
Finally, the interaction term $e_0 \varphi_0^* \varphi_0 A_0^0$
in~(\ref{Feyn:L}) produces the vertex
\begin{equation}
\raisebox{-1mm}{\begin{picture}(22,15)
\put(11,6.5){\makebox(0,0){\includegraphics{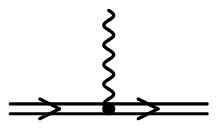}}}
\put(11,15){\makebox(0,0)[t]{$\mu$}}
\end{picture}}
= i e_0 v^\mu\,,
\label{Feyn:vert}
\end{equation}
where
\begin{equation}
v^\mu = (1,\vec{0}\,)
\label{Feyn:v}
\end{equation}
is the 4-velocity of our laboratory frame
(in which the electron is nearly at rest all the time).

The static field $\varphi_0$ (or $h_0$)
describes only particles, there are no antiparticles.
Therefore, there are no pair creation and annihilation
(even virtual).
In other words, there are no loops formed by propagators
of the static electron.
The electron propagates only forward in time~(\ref{Feyn:Sx});
the product of $\theta$ functions along a loop vanishes.
We can also see this in momentum space:
all poles of the propagators~(\ref{Feyn:Sp}) in such a loop
are in the lower $p_0$ half-plane,
and closing the integration contour upwards, we get 0.

It is easy to find the propagator of the static electron
an an arbitrary external electromagnetic field $A^\mu(x)$.
It satisfies the equation
\begin{equation}
i D_0 S(x,x') = (i \partial_0 + e_0 A^0(x)) S(x,x') = \delta(x-x')
\label{Feyn:propeq}
\end{equation}
instead of~(\ref{Feyn:propeq0})
(the derivative $\partial_0$ acts on $x$).
Its solution is
\begin{equation}
S(x,x') = S(x_0,x_0') \delta(\vec{x}-\vec{x}\,')\,,\qquad
S(x_0,x_0') = S_0(x_0-x_0') W(x_0,x_0')\,,
\label{Feyn:SA}
\end{equation}
where
\begin{equation}
W(x_0,x_0') = \exp i e_0 \int\limits_{x_0'}^{x_0} A^\mu(t,\vec{x}\,) v_\mu dt
\label{Feyn:W}
\end{equation}
is the straight Wilson line from $x'$ to $x$ (along $v$).
The same formula can be used when the electromagnetic field
is quantum (operator $A_0^\mu(x)$),
but the exponent~(\ref{Feyn:W}) has to be path-ordered:
operators referring to earlier points (along the path)
are placed to the right from those for later points.
This is usually denoted by $P\exp$;
when the path is directed to the future,
$P$-ordering coincides with $T$-ordering.
The Wilson line has a useful property
\begin{equation}
D_0 W(x,x') \varphi_0(x) = W(x,x') \partial_0 \varphi_0(x)\,.
\label{Feyn:D0W}
\end{equation}
Properties of Wilson lines were investigated in many papers,
see, e.\,g., \cite{P:79,A:80,DV:80}.
Many results now considered classics of HQET
were derived in the course of these studies
before HQET was invented in 1990.
In particular, the HQET Lagrangian~(\ref{Photonia:spin0})
has been introduced~\cite{A:80} as a technical device
for investigation of Wilson lines.

\begin{figure}[ht]
\begin{center}
\begin{picture}(33,33)
\put(16,16){\makebox(0,0){\includegraphics{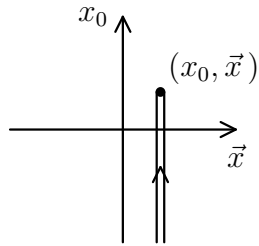}}}
\put(31,14){\makebox(0,0)[t]{$\vec{x}$}}
\put(14,31){\makebox(0,0)[r]{$x_0$}}
\put(22,22){\makebox(0,0)[bl]{$(x_0,\vec{x}\,)$}}
\end{picture}
\end{center}
\caption{A straight Wilson line from the infinite past to $x$ along time}
\label{F:Winf}
\end{figure}

If we choose the gauge $A^0(x)=0$,
then the field $\varphi_0(x)$ in~(\ref{Feyn:L})
does not interact with the electromagnetic field
(and thus becomes free).
However, this gauge is rather pathological.
The static electron creates the Coulomb electric field $\vec{E}$
(because is has the charge density~(\ref{Photonia:j})).
In the $A^0=0$ gauge, $\vec{A}$ has to depend on $t$ linearly
in order to reproduce this electric field.
Imagine trying to solve the hydrogen atom problem in this gauge:
the Hamiltonian is time-dependent,
there are no stationary states, etc.
We can formally express the field $\varphi_0(x)$ in any gauge
via a free field $\varphi^{(0)}(x)$:
\begin{equation}
\varphi_0(x) = W(x) \varphi^{(0)}(x)\,,
\label{Feyn:viafree}
\end{equation}
where
\begin{equation}
W(x_0,\vec{x}\,) = P \exp i \int\limits_{-\infty}^{x_0}
A_0^\mu(t,\vec{x}\,) v_\mu dt
\label{Feyn:Winf}
\end{equation}
is the straight Wilson line from $-\infty$
to the point $x$ along $v$ (Fig.~\ref{F:Winf}).
Then from~(\ref{Feyn:D0W}) we have
$W^{-1}(x) D_0 W(x) = \partial_0$,
and the leading-order Lagrangian becomes free:
\begin{equation*}
L = \varphi^{(0)*} i \partial_0 \varphi^{(0)}\,.
\end{equation*}

As discussed in Sect.~\ref{S:Photonia},
the lowest-energy state (``vacuum'') in HEET is a single electron at rest,
and it is convenient to use its energy as the zero level.
In the full theory, its energy is $M$, and
\begin{equation}
E = M + \varepsilon\,,
\label{Feyn:E}
\end{equation}
where $E$ is the energy of some state
(containing a single electron) in the full theory,
and $\varepsilon$ is its energy in HEET
(it is called the residual energy).
We can re-write this relation in a relativistic form:
\begin{equation}
P^\mu = M v^\mu + p^\mu\,,
\label{Feyn:P}
\end{equation}
where $P^\mu$ is the 4-momentum of some state
(containing a single electron) in the full theory,
$p^\mu$ is its momentum in HEET (the residual momentum),
and $v^\mu$ is 4-velocity of a reference frame
in which the electron always stays approximately at rest.
In other words, HEET is applicable if there exists such a 4-velocity $v$
that, after decomposition~(\ref{Feyn:P}),
the components of the electron residual momentum $p$ are always small,
and components of all photon momenta $p_i$ are also small:
\begin{equation}
p^\mu \ll M\,,\quad
p_i^\mu \ll M\,.
\label{Feyn:appl}
\end{equation}
This condition does not fix $v$ uniquely;
it can be varied by $\delta v \sim p/M$.
Effective theories corresponding to different choices of $v$
must produce identical physical predictions.
This requirement is called reparametrization invariance~\cite{LM:92}.
It produces relations between some quantities
of different orders in $1/M$, as we'll see later.

We can re-write the Lagrangian~(\ref{Feyn:L})
in a relativistic form~\cite{G:90}:
\begin{equation}
L = \varphi_0^* i v \cdot D \varphi_0
+ (\text{light fields})\,.
\label{Feyn:Lv}
\end{equation}
This Lagrangian is not Lorentz-invariant,
because it contains a fixed vector $v$.
It gives the free propagator
\begin{equation}
S_0(p) = \frac{1}{p\cdot v + i0}\,.
\label{Feyn:S}
\end{equation}
The mass shell of the static electron is
\begin{equation}
p\cdot v = 0\,.
\label{Feyn:mshell}
\end{equation}
If we want to consider the spin-$\frac{1}{2}$ electron,
it is described by the 4-component (Dirac) spinor field $h_v$
which satisfies the condition
\begin{equation}
\rlap/v h_v = h_v
\label{Feyn:hv}
\end{equation}
(so that in the $v$ rest frame the field has only 2 upper components
non-vanishing).
The Lagrangian~\cite{G:90}
\begin{equation}
L = \bar{h}_{v0} i v\cdot D h_{v0}
+ (\text{light fields})
\label{Feyn:Lhv}
\end{equation}
gives the propagator
\begin{equation}
S_0(p) = \frac{1 + \rlap/v}{2}\,\frac{1}{p\cdot v + i0}
\label{Feyn:Sh}
\end{equation}
and the vertex $i e_0 v^\mu$~(\ref{Feyn:vert}).

And what can our friends from Qedland say about this theory?
They are not surprised.
The finite-mass free electron propagator $S_0(P)$
with $P=Mv+p$~(\ref{Feyn:P}), $M\to\infty$ can be approximated as
\begin{equation}
S_0(Mv+p) =
\frac{M + M\rlap/v + \rlap/p}{(Mv+p)^2-M^2+i0} =
\frac{1+\rlap/v}{2}\,\frac{1}{p\cdot v+i0}
+ \mathcal{O}\left(\frac{p}{M}\right)\,.
\label{Feyn:SP}
\end{equation}
Diagrammatically, it is related to the HEET propagator:
\begin{equation}
\raisebox{-3.5mm}{\begin{picture}(22,6.5)
\put(11,4.5){\makebox(0,0){\includegraphics{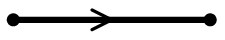}}}
\put(11,0){\makebox(0,0)[b]{$Mv+p$}}
\end{picture}} =
\raisebox{-3.5mm}{\begin{picture}(22,6.5)
\put(11,4.5){\makebox(0,0){\includegraphics{fr2.eps}}}
\put(11,0){\makebox(0,0)[b]{$p$}}
\end{picture}}
+ \mathcal{O}\left(\frac{p}{M}\right)\,.
\label{Feyn:QED}
\end{equation}
When a QED vertex $i e_0 \gamma^\mu$ is sandwiched
between two propagators~(\ref{Feyn:Sh}),
it can be replaced by the HEET vertex $i e_0 v^\mu$:
\begin{equation}
\frac{1+\rlap/v}{2} \gamma^\mu \frac{1+\rlap/v}{2} =
\frac{1+\rlap/v}{2} v^\mu \frac{1+\rlap/v}{2}\,.
\label{Feyn:gamma}
\end{equation}
But what if there is an external spinor $u(P)$ after the vertex
(or $\bar{u}(P)$ before it)?
From the Dirac equation we have
\begin{equation*}
\rlap/v u(Mv+p) = u(Mv+p) + \mathcal{O}\left(\frac{p}{M}\right)\,,
\end{equation*}
so that we may insert the projectors $(1+\rlap/v)/2$
before $u(P_i)$ and after $\bar{u}(P_i)$, too,
and the replacement~(\ref{Feyn:gamma}) is applicable.
We have derived the HEET Feynman rules from the QED ones
in the limit $M\to\infty$.
Therefore, we again arrive at the HEET Lagrangian~(\ref{Feyn:Lhv})
which corresponds to these Feynman rules.

We have thus proved that at the tree level any QED diagram
is equal to the corresponding HEET diagram up to
$\mathcal{O}(p/m)$ corrections.
This is not true at loops,
because loop momenta can be arbitrarily large.
Renormalization properties of HEET
(anomalous dimensions, etc.) differ from those in QED.
In Sect.~\ref{S:El} we shall see that QED loop diagrams
can be decomposed into integration regions,
with some loops hard (momenta $\sim M$)
and some soft (momenta $\sim p$).
Then hard loops produce local interactions
(in the effective theory language,
they follow from local operators in the HEET Lagrangian);
soft loops can be calculated as in HEET.

\subsection{One-loop diagrams}
\label{S:L1}

\begin{figure}[ht]
\begin{center}
\begin{picture}(54,24)
\put(27,12.5){\makebox(0,0){\includegraphics{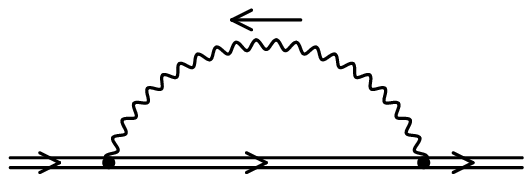}}}
\put(27,0){\makebox(0,0)[b]{$k+p$}}
\put(27,24){\makebox(0,0)[t]{$k$}}
\put(6,0){\makebox(0,0)[b]{$p$}}
\put(48,0){\makebox(0,0)[b]{$p$}}
\put(27,9){\makebox(0,0)[t]{$n_1$}}
\put(27,13){\makebox(0,0)[b]{$n_2$}}
\end{picture}
\end{center}
\caption{One-loop propagator diagram}
\label{F:h1}
\end{figure}

Let's calculate the simplest one-loop diagram (Fig.~\ref{F:h1})
\begin{equation}
\begin{split}
&\frac{1}{i\pi^{d/2}} \int \frac{d^d k}{D_1^{n_1} D_2^{n_2}}
= (-2\omega)^{d-n_1-2n_2} I(n_1,n_2)\,,\\
&D_1 = -2(k+p)_0-i0\,,\quad
D_2 = -k^2-i0\,.
\end{split}
\label{L1:def}
\end{equation}
It depends only on the residual energy $\omega=p_0$, not $\vec{p}$;
the power of $-2\omega$ is clear from dimensional counting.
If $\omega>0$, real pair production is possible, and we are on a cut.
We shall consider the case $\omega<0$,
when the integral is an analytic function of $\omega$.
If $n_1$ is integer and $n_1\le0$,
$I(n_1,n_2)=0$ because this is a massless vacuum diagram.
If $n_2$ is integer and $n_2\le0$,
$I(n_1,n_2)=0$ because the diagram contains an HQET loop.

\begin{figure}[b]
\begin{center}
\begin{picture}(32,15.5)
\put(16,7.75){\makebox(0,0){\includegraphics{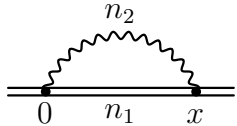}}}
\put(16,0){\makebox(0,0)[b]{$n_1$}}
\put(16,15.75){\makebox(0,0)[t]{$n_2$}}
\put(6,0){\makebox(0,0)[b]{$0$}}
\put(26,0){\makebox(0,0)[b]{$x$}}
\end{picture}
\end{center}
\caption{One-loop propagator diagram in coordinate space}
\label{F:Coord}
\end{figure}

It is easy to calculate this diagram in coordinate space.
Static propagators in $p$ and $x$ space are related by
\begin{align}
&\int_{-\infty}^{+\infty} \frac{e^{-i\omega t}}{(-2\omega-i0)^n}
\frac{d\omega}{2\pi} =
\frac{i}{2\Gamma(n)} \left(\frac{it}{2}\right)^{n-1}
e^{-0t} \theta(t)\,,
\label{L1:Fh1}\\
&\int_0^\infty e^{(i\omega-0)t} \left(\frac{it}{2}\right)^{n-1} dt =
- \frac{2 i \Gamma(n)}{(-2\omega-i0)^n}\,;
\label{L1:Fh2}
\end{align}
massless propagators --- by
\begin{align}
&\int \frac{e^{-ip\cdot x}}{(-p^2-i0)^n} \frac{d^d p}{(2\pi)^d} =
\frac{i}{(4\pi)^{d/2}} \frac{\Gamma(d/2-n)}{\Gamma(n)}
\left(\frac{4}{-x^2+i0}\right)^{d/2-n}\,,
\label{L1:Fl1}\\
&\int \left(\frac{4}{-x^2+i0}\right)^n e^{ip\cdot x} d^d x =
- i (4\pi)^{d/2} \frac{\Gamma(d/2-n)}{\Gamma(n)}
\frac{1}{(-p^2-i0)^{d/2-n}}\,.
\label{L1:Fl2}
\end{align}
Our diagram in $x$ space (Fig.~\ref{F:Coord}, $x=vt$)
is just the product of the heavy propagator~(\ref{L1:Fh1})
and the light one~(\ref{L1:Fl1}) (where $-x^2/4=-t^2/4=(it/2)^2$):
\begin{equation*}
- \frac{1}{2} \frac{1}{(4\pi)^{d/2}}
\frac{\Gamma(d/2-n_2)}{\Gamma(n_1) \Gamma(n_2)}
\left(\frac{it}{2}\right)^{n_1+2n_2-d-1} \theta(t)\,.
\end{equation*}
The inverse Fourier transform~(\ref{L1:Fh2})
gives our diagram~(\ref{L1:def}) in $p$ space
\begin{equation*}
\frac{i}{(4\pi)^{d/2}} I(n_1,n_2) (-2\omega)^{d-n_1-2n_2}\,,
\end{equation*}
where
\begin{equation}
I(n_1,n_2) =
\frac{\Gamma(n_1+2n_2-d) \Gamma\bigl(\frac{d}{2}-n_2\bigr)}%
{\Gamma(n_1) \Gamma(n_2)}\,.
\label{L1:I}
\end{equation}

Now we shall re-calculate the one-loop diagram~(\ref{L1:def})
(Fig.~\ref{F:h1}) using $\alpha$ parametrization
\begin{equation}
\frac{1}{a^n} = \frac{1}{\Gamma(n)}
\int_0^\infty d\alpha\,\alpha^{n-1}\,e^{-a\alpha}\,.
\label{alpha:alpha}
\end{equation}
We get
\begin{equation*}
\frac{1}{\Gamma(n_1) \Gamma(n_2)}
\int d\alpha\,\alpha^{n_2-1}\,d\beta\,\beta^{n_1-1}\,d^d k\,e^X\,,\quad
X = \alpha k^2 + 2 \beta (k+p)\cdot v
\end{equation*}
($\alpha$ has dimensionality $1/m^2$, and $\beta$ --- $1/m$).
We shift the integration momentum $k = k' - \frac{\beta}{\alpha} v$
to eliminate the linear term in the exponent:
\begin{equation*}
X = \alpha k^{\prime2} - \frac{\beta^2}{\alpha} + 2 \beta \omega\,.
\end{equation*}
The Wick rotation $k_0 = i k_{E0}$
brings us into Euclidean momentum space ($k^2=-k_E^2$).
Now it is easy to calculate the momentum integral:
\begin{equation}
\int d^d k\,e^{\alpha k^2}
= i \int d^d k_E\,e^{-\alpha k_E^2}
= i \left(\frac{\pi}{\alpha}\right)^{d/2}\,.
\end{equation}
Therefore,
\begin{equation*}
(-2\omega)^{d-n_1-2n_2} I(n_1,n_2) =
\frac{1}{\Gamma(n_1) \Gamma(n_2)}
\int d\alpha\,\alpha^{n_2-1}\,d\beta\,\beta^{n_1-1}\,\alpha^{-d/2}
\exp\left(-\frac{\beta^2}{\alpha}+2\beta\omega\right)\,.
\end{equation*}
Now we make the substitution $\beta=\alpha y$ and integrate in $\alpha$:
\begin{equation}
\frac{\Gamma\bigl(n_1+n_2-\frac{d}{2}\bigr)}{\Gamma(n_1) \Gamma(n_2)}
\int_0^\infty d y\,y^{n_1-1}
\bigl[y(y-2\omega)\bigr]^{d/2-n_1-n_2}\,.
\label{L1:y}
\end{equation}
The HQET Feynman parameter $y$ has the dimensionality of energy
and varies from 0 to $\infty$.
The $y$ integral can be easily calculated in $\Gamma$ functions,
and we again obtain~(\ref{L1:I}).

Equivalently, we can use the HQET Feynman parametrization from the beginning.
Multiplying two copies of~(\ref{alpha:alpha}),
\begin{equation*}
\frac{1}{a_1^{n_1} a_2^{n_2}} = \frac{1}{\Gamma(n_1) \Gamma(n_2)}
\int d\beta\,\beta^{n_1-1}\,d\alpha\,\alpha^{n_2-1}\,e^{-a_1\beta-a_2\alpha}\,,
\end{equation*}
substituting $\beta=\alpha y$ and integrating in $\alpha$,
we obtain the HQET Feynman parametrization
\begin{equation}
\frac{1}{a_1^{n_1} a_2^{n_2}} = \frac{\Gamma(n_1+n_2)}{\Gamma(n_1) \Gamma(n_2)}
\int_0^\infty \frac{y^{n_1-1} dy}{(a_1 y + a_2)^{n_1+n_2}}\,.
\label{L1:Feyn}
\end{equation}
For~(\ref{L1:def}) this gives
\begin{equation*}
\frac{\Gamma(n_1+n_2)}{\Gamma(n_1) \Gamma(n_2)}
\int\frac{y^{n_1-1}\,dy\,d^d k}{(-k^2-2y(k+p)\cdot v)^{n_1+n_2}}\,.
\end{equation*}
Shifting the integration momentum $k=k'-yv$ and using~(2.10) from~\cite{eft1},
we again get~(\ref{L1:y}).
One more method of calculating this diagram is discussed in Appendix~\ref{S:App1}.

Tensor integrals similar to~(\ref{L1:def})
but with $k^{\mu_1}\cdots k^{\mu_n}$ in the numerator
can be expressed via $g^{\mu\nu}$ and $v^\mu$.
Writing down a general form of the result with unknown coefficients
and solving the linear system, we can find any such integral.
However, it may be easier to use the explicit finite sum~\cite{G:00}
\begin{equation}
\begin{split}
&\frac{1}{i\pi^{d/2}} \int \frac{P_n(k) d^d k}{D_1^{n_1} D_2^{n_2}} = {}\\
& (-2\omega)^{d-n_1-2n_2} \sum_m I(n_1,n_2;n,m) \frac{(-2\omega)^{2m}}{m!}
\left.
\left( - \frac{1}{4}
\frac{\partial}{\partial k_\mu}
\frac{\partial}{\partial k^\mu}
\right)^m P(k) \right|_{k\to2\omega v}\,,
\end{split}
\label{L1:tensor}
\end{equation}
where
\begin{equation*}
P_n(\lambda k) = \lambda^n P(k)
\end{equation*}
is a homogeneous polynomial (it may contain tensor indices), and
\begin{equation}
I(n_1,n_2;n,m) = \frac{\Gamma(n_1+2n_2-n-d)
\Gamma\left(\frac{d}{2}-n_2+n-m\right)}%
{\Gamma(n_1) \Gamma(n_2)}\,.
\label{L1:tensorI}
\end{equation}
Some other kinds of one-loop diagrams are considered in Appendix~\ref{S:App1},
see also~\cite{G:07,G:08}.

\subsection{Renormalization}
\label{S:Prop}

The full propagator of the static electron $S(p)$
depends only on the residual energy $\omega=p_0$, not on $\vec{p}$.
It has the structure
\begin{equation}
\raisebox{-1.5mm}{\includegraphics{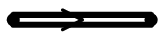}}
= \raisebox{-1.5mm}{\includegraphics{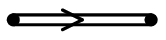}}
+ \raisebox{-4mm}{\includegraphics{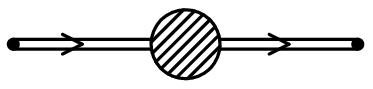}}
+ \raisebox{-4mm}{\includegraphics{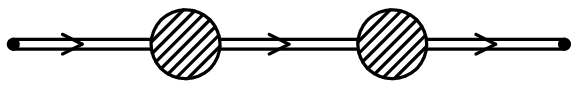}}
+ \cdots
\label{Prop:Dyson0}
\end{equation}
where the electron self-energy
\begin{equation}
\raisebox{-4mm}{\includegraphics{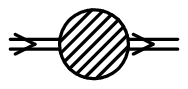}}
= - i \Sigma(\omega)
\label{Prop:Sigma}
\end{equation}
is the sum of one-particle-irreducible self-energy diagrams
(which cannot be separated into two disconnected parts
by cutting a single electron line).
We have
\begin{equation}
i S(\omega) = i S_0(\omega)
+ i S_0(\omega) (-i) \Sigma(\omega) i S_0(\omega)
+ i S_0(\omega) (-i) \Sigma(\omega) i S_0(\omega)
  (-i) \Sigma(\omega) i S_0(\omega)
+ \cdots
\label{Prop:Dyson1}
\end{equation}
where $S_0(\omega)=1/\omega$ is the free propagator~(\ref{Feyn:Sp}).
This series implies
$S(\omega) = S_0(\omega) + S_0(\omega) \Sigma(\omega) S(\omega)$, or
$S^{-1}(\omega) = S_0^{-1}(\omega) - \Sigma(\omega)$:
\begin{equation}
S(\omega) = \frac{1}{\omega - \Sigma(\omega)}\,.
\label{Prop:Dyson2}
\end{equation}

\begin{figure}[ht]
\begin{center}
\begin{picture}(54,24)
\put(27,12.5){\makebox(0,0){\includegraphics{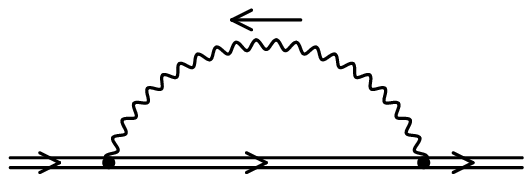}}}
\put(27,0){\makebox(0,0)[b]{$k+p$}}
\put(27,24){\makebox(0,0)[t]{$k$}}
\put(6,0){\makebox(0,0)[b]{$p$}}
\put(48,0){\makebox(0,0)[b]{$p$}}
\end{picture}
\end{center}
\caption{One-loop static-electron self-energy}
\label{F:se1}
\end{figure}

At one loop (Fig.~\ref{F:se1})
\begin{equation}
\Sigma(\omega) = i \int \frac{d^d k}{(2\pi)^d}
i e_0 v^\mu \frac{i}{k_0+\omega} i e_0 v^\nu
\frac{-i}{k^2} \left(g_{\mu\nu} - \xi \frac{k_\mu k_\nu}{k^2}\right)\,,
\label{Ren:se1}
\end{equation}
where $\xi=1-a_0$.
In the numerator, we may replace
$(k\cdot v)^2=(k_0+\omega-\omega)^2\to\omega^2$,
because if we cancel $k_0+\omega$ in the denominator
the integral vanishes.
Using~(\ref{L1:I}), we obtain
\begin{equation}
\begin{split}
\Sigma(\omega) &= \frac{e_0^2 (-2\omega)^{1-2\varepsilon}}{(4\pi)^{d/2}}
\left[2 I(1,1) + \frac{\xi}{2} I(1,2)\right]\\
&= \frac{e_0^2 (-2\omega)^{1-2\varepsilon}}{(4\pi)^{d/2}}
\frac{\Gamma(1+2\varepsilon) \Gamma(1-\varepsilon)}{d-4}
\left(\xi + \frac{2}{d-3}\right)\,.
\end{split}
\label{Ren:Sigma1}
\end{equation}
This correction vanishes in the $d$-dimensional Yennie~\cite{FY:58} gauge
\begin{equation}
a_0 = \frac{2}{d-3} + 1\,.
\label{Prop:Yennie}
\end{equation}

Let us also re-derive this result in $x$ space.
Using the heavy-quark propagator~(\ref{Feyn:Sx}) and the gluon propagator
\begin{equation}
D^0_{\mu\nu}(x) = \frac{i\Gamma(d/2-1)}{8\pi^{d/2}}
\frac{(1+a_0) x^2 g_{\mu\nu} + (d-2) (1-a_0) x_\mu x_\nu}{(-x^2+i0)^{d/2}}\,,
\label{Prop:Dx}
\end{equation}
we obtain
\begin{equation}
\Sigma(x) = - e_0^2 D^0_{\mu\nu}(vt) v^\mu v^\nu \theta(t)
= i e_0^2 \frac{\Gamma(d/2-1)}{8\pi^{d/2}} (d-3)
\left(\xi + \frac{2}{d-3}\right) (i t)^{2-d} \theta(t)\,.
\label{Ren:Six}
\end{equation}
Transforming this to $p$ space~(\ref{L1:Fh2}), we recover~(\ref{Ren:Sigma1}).

The static quark propagator up to one loop is
\begin{equation}
S(\omega) = S_0(\omega) \left[ 1
- \frac{e_0^2 (-2\omega)^{-2\varepsilon}}{(4\pi)^{d/2}}
\frac{2 \Gamma(1+2\varepsilon) \Gamma(1-\varepsilon)}{d-4}
\left( \xi + \frac{2}{d-3} \right)
+ \mathcal{O}(e_0^4) \right]\,.
\label{Prop:Sw1}
\end{equation}
In $x$ space~(\ref{L1:Fh1})
\begin{equation}
S(t) = S_0(t) \left[ 1
- \frac{e_0^2}{(4\pi)^{d/2}} \left(\frac{it}{2}\right)^{2\varepsilon}
\Gamma(-\varepsilon) \left( \xi + \frac{2}{d-3} \right)
+ \mathcal{O}(e_0^4) \right]\,,
\label{Prop:St1}
\end{equation}
where $S_0(t)=-i\theta(t)$~(\ref{Feyn:Sx}).
It is real in the Euclidean space $t=-i\tau$.
Re-expressing~(\ref{Prop:Sw1}) via renormalized quantities we obtain
\begin{equation*}
S(\omega) = S_0(\omega) \left[ 1
+ \frac{\alpha}{4\pi\varepsilon} e^{-2L\varepsilon}
\left(3 - a + 4\varepsilon + \mathcal{O}(\varepsilon^2)\right)
+ \mathcal{O}(\alpha^2) \right]\,,
\end{equation*}
where
\begin{equation*}
L = \log\frac{-2\omega}{\mu}\,.
\end{equation*}
This should be equal to $Z_h(\alpha(\mu),a(\mu)) S_r(\omega;\mu)$
where the renormalization constant $Z_h$ has the minimal form,
and the renormalized propagator $S_r(\omega;\mu)$ is finite at $\varepsilon\to0$.
We obtain
\begin{equation}
Z_h(\alpha,a) = 1 - (a-3) \frac{\alpha}{4\pi\varepsilon}
+ \mathcal{O}(\alpha^2)\,,
\label{Prop:Zh}
\end{equation}
and the anomalous dimension of the static electron field is
\begin{equation}
\gamma_h(\alpha,a) = 2 (a-3) \frac{\alpha}{4\pi}
+ \mathcal{O}(\alpha^2)\,.
\label{Prop:gamma}
\end{equation}
It vanishes in the Yennie gauge~\cite{FY:58} $a=3$.
We can analyze~(\ref{Prop:St1}) in a similar way.
Re-expressing it via renormalized quantities we obtain
\begin{equation*}
S(t) = S_0(t) \left[ 1
+ \frac{\alpha}{4\pi\varepsilon} e^{2L_t\varepsilon}
\left(3 - a + 4\varepsilon + \mathcal{O}(\varepsilon^2)\right)
+ \mathcal{O}(\alpha^2) \right]\,,
\end{equation*}
where
\begin{equation*}
L_t = \log\frac{i\mu t}{2} + \gamma_E
\end{equation*}
($\gamma_E$ is the Euler constant).
This should be equal to $Z_h(\alpha(\mu),a(\mu)) S_r(t;\mu)$,
and we again arrive at~(\ref{Prop:Zh}).

In fact, the static electron propagator can be calculated exactly~\cite{YFS:61}!
Suppose we calculate the one-loop correction to the static electron propagator
in coordinate space.
Let us multiply this correction by itself.
We obtain an integral in $t_1$, $t_2$, $t_1'$, $t_2'$
with $0<t_1<t_2<t$, $0<t_1'<t_2'<t$.
The ordering of primed and non-primed integration times can be arbitrary.
The integration area is subdivided into six regions,
corresponding to the six diagrams:
\begin{equation}
\begin{split}
&\raisebox{-6.25mm}{\begin{picture}(23,14.5)
\put(11.5,9.875){\makebox(0,0){\includegraphics{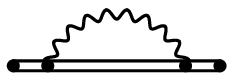}}}
\put(1,4.5){\makebox(0,0){$0$}}
\put(22,4.5){\makebox(0,0){$t$}}
\put(4.5,4.1){\makebox(0,0){$t_1$}}
\put(18.5,4.1){\makebox(0,0){$t_2$}}
\end{picture}}
\times
\raisebox{-6.25mm}{\begin{picture}(23,14.5)
\put(11.5,4.625){\makebox(0,0){\includegraphics{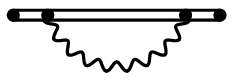}}}
\put(1,10.3){\makebox(0,0){$0$}}
\put(22,10.3){\makebox(0,0){$t$}}
\put(4.5,9.9){\makebox(0,0){$t_1'$}}
\put(18.5,9.9){\makebox(0,0){$t_2'$}}
\end{picture}}\\
&{}= \raisebox{-6.75mm}{\includegraphics{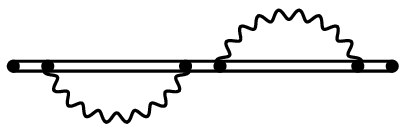}}
+ \raisebox{-6.75mm}{\includegraphics{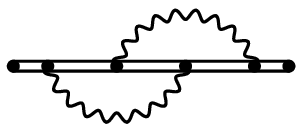}}
+ \raisebox{-6.75mm}{\includegraphics{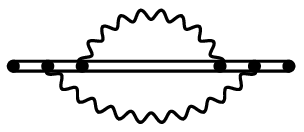}}\\
&{}+ \raisebox{-6.75mm}{\includegraphics{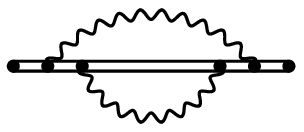}}
+ \raisebox{-6.75mm}{\includegraphics{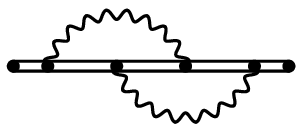}}
+ \raisebox{-6.75mm}{\includegraphics{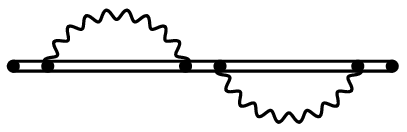}}
\end{split}
\label{Prop:exp}
\end{equation}
This is twice the 2-loop correction to the propagator.
Continuing this drawing exercise,
we see that the one-loop correction cubed
is $3!$ times the 3-loop correction, and so on.
Therefore, the exact all-order propagator
is the exponential of the one-loop correction:
\begin{equation}
S(t) = S_0(t) \exp \left[ - \frac{e_0^2}{(4\pi)^{d/2}}
\left(\frac{it}{2}\right)^{2\varepsilon} \Gamma(-\varepsilon)
\left(\xi + \frac{2}{d-3}\right) \right]\,.
\label{HEET:S}
\end{equation}
In particular, in the $d$-dimensional Yennie gauge~(\ref{Prop:Yennie})
the exact propagator~(\ref{HEET:S}) is free.

There are no corrections to the photon propagator in HEET~(\ref{Feyn:L})
because static-electron loops don't exist%
\footnote{This argument works up to the order $1/M^3$.
At $1/M^4$ a 4-photon interaction appears, see~\cite{eft1}.
However, the only correction to the photon propagator
at this order vanishes (eq.~(2.9) in~\cite{eft1}).
The first non-vanishing correction involves two 4-photon vertices,
and appears at $1/M^8$.}.
Therefore, the photon field is not renormalized: $Z_A=1$
(this also means that the gauge-fixing parameter is not renormalized,
$a=a_0$).

Now let's discuss the operator $J_0=\varphi^*\varphi$.
The integral
\begin{equation}
Q_0 = \int J_0(x_0,\vec{x}) d^3 \vec{x}
\label{HEET:N}
\end{equation}
is the operator of the full number of static electrons
(we are considering the space of eigenstates of this operator
having the eigenvalue 1).
If we write $Q_0 = Z_J(\alpha(\mu)) Q(\mu)$ then $Z_J=1$
because $Q_0$ needs no renormalization.
The same is true for the current: $J_0=J(\mu)$.

We can also prove this using the Ward identity.
The Green function
\begin{equation}
{<}0| \varphi_0^*(x) J_0(0) \varphi(x') |0{>} =
\delta(\vec{x}) \delta(\vec{x}\,') G(x_0,x'_0) =
\raisebox{-3.25mm}{\begin{picture}(29,9)
\put(14.5,4.5){\makebox(0,0){\includegraphics{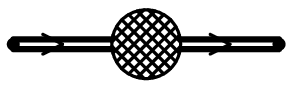}}}
\put(1,1){\makebox(0,0){$\vphantom{y_0'}x_0$}}
\put(29,1){\makebox(0,0){$\vphantom{y_0'}x'_0$}}
\put(10,1){\makebox(0,0){$\vphantom{t'}t$}}
\put(20,1){\makebox(0,0){$\vphantom{t'}t'$}}
\end{picture}}
\label{HEET:G}
\end{equation}
consists of the vertex function $\Gamma(t,t')=\delta(t'-t)+\Lambda(t,t')$
(the sum of one-particle-irreducible diagrams not including the external line)
and two full propagators.
Starting from each diagram for $\Sigma$,
we can obtain a set of diagrams for $\Lambda$
by inserting the $J_0$ vertex into each electron propagator.
For example,
\begin{equation}
\begin{split}
&\raisebox{-8.75mm}{\begin{picture}(38,14)
\put(19,5.5){\makebox(0,0){\includegraphics{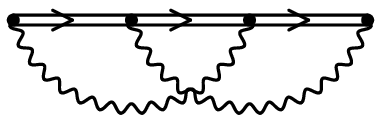}}}
\put(1,12.5){\makebox(0,0){$\vphantom{t_1'}t$}}
\put(37,12.5){\makebox(0,0){$\vphantom{t_1'}t'$}}
\put(13,12.5){\makebox(0,0){$\vphantom{t_1'}t_1$}}
\put(25,12.5){\makebox(0,0){$\vphantom{t_1'}t_2$}}
\end{picture}} \Rightarrow{}\\
&\raisebox{-8.75mm}{\begin{picture}(38,14)
\put(19,5.5){\makebox(0,0){\includegraphics{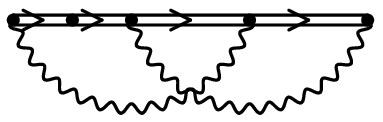}}}
\put(1,12.5){\makebox(0,0){$\vphantom{t_1'}t$}}
\put(37,12.5){\makebox(0,0){$\vphantom{t_1'}t'$}}
\put(13,12.5){\makebox(0,0){$\vphantom{t_1'}t_1$}}
\put(25,12.5){\makebox(0,0){$\vphantom{t_1'}t_2$}}
\put(7,12.5){\makebox(0,0){$\vphantom{t_1'}0$}}
\end{picture}} +
\raisebox{-8.75mm}{\begin{picture}(38,14)
\put(19,5.5){\makebox(0,0){\includegraphics{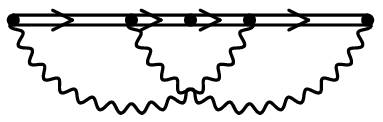}}}
\put(1,12.5){\makebox(0,0){$\vphantom{t_1'}t$}}
\put(37,12.5){\makebox(0,0){$\vphantom{t_1'}t'$}}
\put(13,12.5){\makebox(0,0){$\vphantom{t_1'}t_1$}}
\put(25,12.5){\makebox(0,0){$\vphantom{t_1'}t_2$}}
\put(19,12.5){\makebox(0,0){$\vphantom{t_1'}0$}}
\end{picture}} +
\raisebox{-8.75mm}{\begin{picture}(38,14)
\put(19,5.5){\makebox(0,0){\includegraphics{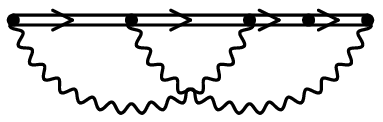}}}
\put(1,12.5){\makebox(0,0){$\vphantom{t_1'}t$}}
\put(37,12.5){\makebox(0,0){$\vphantom{t_1'}t'$}}
\put(13,12.5){\makebox(0,0){$\vphantom{t_1'}t_1$}}
\put(25,12.5){\makebox(0,0){$\vphantom{t_1'}t_2$}}
\put(31,12.5){\makebox(0,0){$\vphantom{t_1'}0$}}
\end{picture}}\\
&{} = \theta(-t) \theta(t')
\raisebox{-8.75mm}{\begin{picture}(38,14)
\put(19,5.5){\makebox(0,0){\includegraphics{wc0.eps}}}
\put(1,12.5){\makebox(0,0){$\vphantom{t_1'}t$}}
\put(37,12.5){\makebox(0,0){$\vphantom{t_1'}t'$}}
\put(13,12.5){\makebox(0,0){$\vphantom{t_1'}t_1$}}
\put(25,12.5){\makebox(0,0){$\vphantom{t_1'}t_2$}}
\end{picture}}
\end{split}
\label{HEET:Wardcoord}
\end{equation}
(the 3 vertex diagrams have the integration regions
$t \le 0 \le t_1 \le t_2 \le t'$,
$t \le t_1 \le 0 \le t_2 \le t'$,
$t \le t_1 \le t_2 \le 0 \le t'$;
their union the the integration region of the self-energy diagram
$t \le t_1 \le t_2 \le t'$).
Therefore,
\begin{equation}
\Lambda(t,t') = - i \theta(-t) \theta(t') \Sigma(t'-t)\,.
\label{HEET:Ward1}
\end{equation}
Alternatively, we can start from diagrams for $S(t,t')$
(including one-particle-reducible ones), and obtain
\begin{equation}
G(t,t') = i \theta(-t) \theta(t') S(t'-t)\,.
\label{HEET:Ward2}
\end{equation}
Here the left-hand side should be equal to $Z_h Z_J G_r$
and the right-hand side is $Z_h S_r$,
where the renormalized Green functions $G_r$, $S_r$ are finite at $\varepsilon\to0$.
Then $Z_J$ is finite;
but the only minimal renormalization constant finite at $\varepsilon\to0$
is $Z_J=1$.

We can also consider this Green function in momentum space:
\begin{equation}
G(\omega,\omega') =
\raisebox{-3.25mm}{\begin{picture}(29,15)
\put(14.5,7.5){\makebox(0,0){\includegraphics{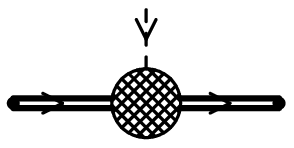}}}
\put(6,2){\makebox(0,0){$\vphantom{\omega'}\omega$}}
\put(23,2){\makebox(0,0){$\vphantom{\omega'}\omega'$}}
\put(16.5,11){\makebox(0,0){$q$}}
\end{picture}}
= i S(\omega)\,\Gamma(\omega,\omega')\,i S(\omega')\,,
\label{HEET:Gmom}
\end{equation}
where $q$ is the momentum entering the $J_0$ vertex ($q_0=\omega'-\omega$)
and $\Gamma(\omega,\omega') = 1 + \Lambda(\omega,\omega')$.
Starting from each diagram for $\Sigma(\omega)$,
we can obtain a set of diagrams for $\Lambda(\omega,\omega')$
by inserting the $J_0$ vertex into each electron propagator.
Due to the elementary identity
\begin{equation}
\raisebox{-2.75mm}{\begin{picture}(22,11)
\put(11,7){\makebox(0,0){\includegraphics{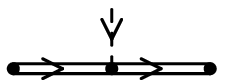}}}
\put(6,1.5){\makebox(0,0){$\vphantom{\omega'}\omega$}}
\put(16,1.5){\makebox(0,0){$\omega'$}}
\put(13,7){\makebox(0,0){$q$}}
\end{picture}}
= - \frac{i}{\omega' - \omega} \Biggl[
\raisebox{-2.75mm}{\begin{picture}(12,11)
\put(6,7){\makebox(0,0){\includegraphics{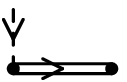}}}
\put(6,1.5){\makebox(0,0){$\omega'$}}
\end{picture}} -
\raisebox{-2.75mm}{\begin{picture}(12,11)
\put(6,7){\makebox(0,0){\includegraphics{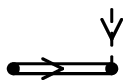}}}
\put(6,1.5){\makebox(0,0){$\vphantom{\omega'}\omega$}}
\end{picture}}
\Biggr]\,,
\label{HEET:Warde}
\end{equation}
each diagram in this set becomes a difference.
For example,
\newlength{\lea}\settowidth{\lea}{$\displaystyle{}=-\frac{i}{\omega'-\omega}\Biggl[\Biggr.$}
\newlength{\leb}\settowidth{\leb}{$\displaystyle{}+{}$}
\addtolength{\lea}{-\leb}
\begin{align}
&\raisebox{-8.75mm}{\begin{picture}(46,14)
\put(23,5.5){\makebox(0,0){\includegraphics{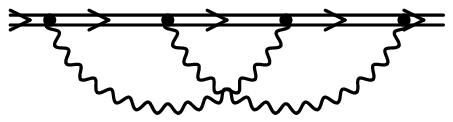}}}
\put(3,12.5){\makebox(0,0){$\vphantom{\omega'}\omega$}}
\put(43,12.5){\makebox(0,0){$\vphantom{\omega'}\omega$}}
\end{picture}} \Rightarrow {}
\nonumber\\
&\raisebox{-8.75mm}{\begin{picture}(46,17)
\put(23,8.5){\makebox(0,0){\includegraphics{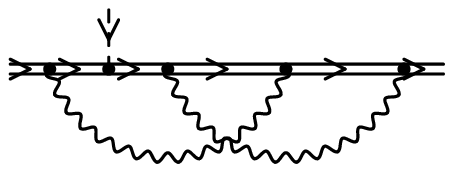}}}
\put(3,12.5){\makebox(0,0){$\vphantom{\omega'}\omega$}}
\put(43,12.5){\makebox(0,0){$\omega'$}}
\put(13,13){\makebox(0,0){$q$}}
\end{picture}} +
\raisebox{-8.75mm}{\begin{picture}(46,17)
\put(23,8.5){\makebox(0,0){\includegraphics{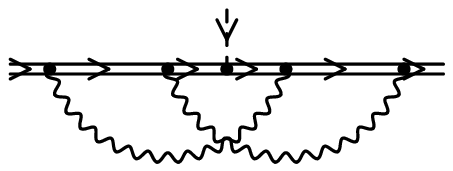}}}
\put(3,12.5){\makebox(0,0){$\vphantom{\omega'}\omega$}}
\put(43,12.5){\makebox(0,0){$\omega'$}}
\put(25,13){\makebox(0,0){$q$}}
\end{picture}} +
\raisebox{-8.75mm}{\begin{picture}(46,17)
\put(23,8.5){\makebox(0,0){\includegraphics{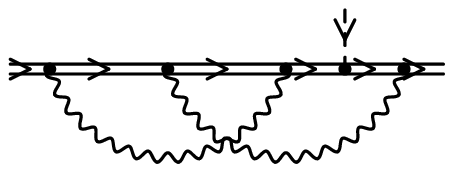}}}
\put(3,12.5){\makebox(0,0){$\vphantom{\omega'}\omega$}}
\put(43,12.5){\makebox(0,0){$\omega'$}}
\put(37,13){\makebox(0,0){$q$}}
\end{picture}}
\nonumber\\
&{} = - \frac{i}{\omega' - \omega} \Biggl[
\raisebox{-8.75mm}{\begin{picture}(46,17)
\put(23,8.5){\makebox(0,0){\includegraphics{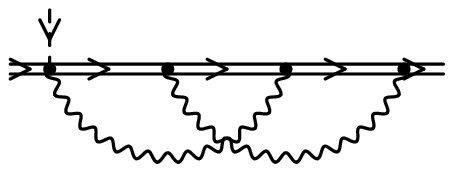}}}
\end{picture}} -
\raisebox{-8.75mm}{\begin{picture}(46,17)
\put(23,8.5){\makebox(0,0){\includegraphics{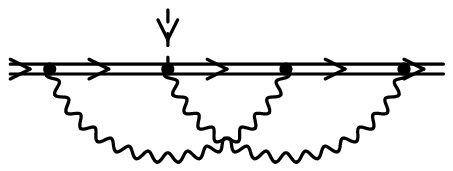}}}
\end{picture}}
\nonumber\\
&\strut\hspace{\lea} +
\raisebox{-8.75mm}{\begin{picture}(46,17)
\put(23,8.5){\makebox(0,0){\includegraphics{wm5.eps}}}
\end{picture}} -
\raisebox{-8.75mm}{\begin{picture}(46,17)
\put(23,8.5){\makebox(0,0){\includegraphics{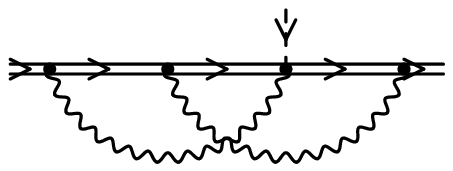}}}
\end{picture}}
\nonumber\\
&\strut\hspace{\lea} +
\raisebox{-8.75mm}{\begin{picture}(46,17)
\put(23,8.5){\makebox(0,0){\includegraphics{wm6.eps}}}
\end{picture}} -
\raisebox{-8.75mm}{\begin{picture}(46,17)
\put(23,8.5){\makebox(0,0){\includegraphics{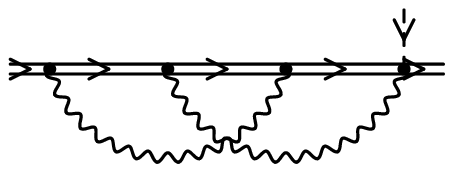}}}
\end{picture}}
\Biggr]
\nonumber\\
&{} = - \frac{i}{\omega' - \omega} \Biggl[
\raisebox{-8.75mm}{\begin{picture}(46,14)
\put(23,5.5){\makebox(0,0){\includegraphics{wm0.eps}}}
\put(3,12.5){\makebox(0,0){$\omega'$}}
\put(43,12.5){\makebox(0,0){$\omega'$}}
\end{picture}} -
\raisebox{-8.75mm}{\begin{picture}(46,14)
\put(23,5.5){\makebox(0,0){\includegraphics{wm0.eps}}}
\put(3,12.5){\makebox(0,0){$\vphantom{\omega'}\omega$}}
\put(43,12.5){\makebox(0,0){$\vphantom{\omega'}\omega$}}
\end{picture}}
\Biggr]\,.
\label{HEET:Wardmom}
\end{align}
All terms cancel each other, except the extreme ones,
and we obtain the Ward identity
\begin{equation}
\Lambda(\omega,\omega') =
- \frac{\Sigma(\omega') - \Sigma(\omega)}{\omega' - \omega}
\quad\text{or}\quad
\Gamma(\omega,\omega') =
\frac{S^{-1}(\omega') - S^{-1}(\omega)}{\omega' - \omega}
\label{HEET:Ward3}
\end{equation}
(this equality can also be derived by Fourier transforming~(\ref{HEET:Ward1})).
Therefore, the Green function~(\ref{HEET:Gmom}) is
\begin{equation}
G(\omega,\omega') = \frac{S(\omega') - S(\omega)}{\omega' - \omega}
\label{HEET:Ward4}
\end{equation}
(this equality can also be derived by considering all diagrams for $G(\omega,\omega')$,
including one-particle-reducible ones,
and using the identity~(\ref{HEET:Warde}),
or by Fourier transforming~(\ref{HEET:Ward2})).

The electron--photon vertex function in HEET is
\begin{equation}
\raisebox{-3.75mm}{\begin{picture}(25,15)
\put(12.5,8.75){\makebox(0,0){\includegraphics{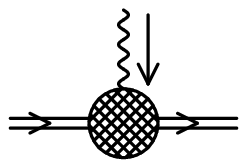}}}
\put(5,0){\makebox(0,0)[b]{$\omega$}}
\put(20,0){\makebox(0,0)[b]{$\omega'$}}
\put(17,12){\makebox(0,0)[l]{$q$}}
\end{picture}} = i e_0 v^\mu \Gamma(\omega,\omega')
\label{Ren:Vdef}
\end{equation}
(the external photon is always attached to the static electron line).
The Ward identity~(\ref{HEET:Ward3}) implies $Z_\Gamma Z_h=1$
(see Sect.~3.6 in~\cite{eft1}).
Therefore, the charge renormalization constant
$Z_\alpha = (Z_\Gamma Z_h)^{-2} Z_A^{-1} =  Z_A^{-1} = 1$ ---
the electron charge is not renormalized in HEET
(see Sect.~2.3, 2.4 in~\cite{eft1}).
Owing to the absence of charge and photon-field renormalization,
we may replace $e_0\to e$, $a_0\to a$ in the bare propagator~(\ref{HEET:S}).
This propagator is made finite by the minimal renormalization constant $Z_h$,
which is just the exponential of the one-loop term
\begin{equation}
Z_h = \exp \left[ -(a-3)\frac{\alpha}{4\pi\varepsilon} \right]\,,
\label{HEET:ZQ}
\end{equation}
and the anomalous dimension is exactly equal to the one-loop contribution
\begin{equation}
\gamma_h = 2 (a-3) \frac{\alpha}{4\pi}\,.
\label{HEET:gammaQ}
\end{equation}
Note that it vanishes in the Yennie gauge
where divergences in the propagator are absent.

\subsection{Electron field in QED and HEET}
\label{S:El}

Operators of full QED can be written as series in $1/M$ via HEET operators:
\begin{equation}
O(\mu) = C(\mu) \tilde{O}(\mu)
+ \frac{1}{2M} \sum_i B_i(\mu) \tilde{O}_i(\mu)
+ \cdots
\label{El:O}
\end{equation}
The coefficients $C(\mu)$, $B_i(\mu)$,\,\dots{}
are obtained by matching on-shell matrix elements:
the effective theory must reproduce matrix elements calculated in the full theory
and expanded in $1/M$ up to some finite order.

In particular, the bare electron field in QED can be written as
\begin{equation}
\psi_0(x) = e^{-i M v\cdot x}
\left[ z_0^{1/2} h_{v0}(x) + \cdots \right]\,,
\label{El:psi0}
\end{equation}
where the factor $e^{-i M v\cdot x}$ reflects the difference
in the momentum definitions~(\ref{Feyn:P}) of the fields in the two theories.
The on-shell matrix elements of the QED and HEET fields are
\begin{equation}
{<}0|\psi_0|e(P){>} = \left(Z_\psi^{\text{os}}(e_0)\right)^{1/2} u(P)\,,\qquad
{<}0|h_{v0}|e(p){>} = \left(Z_h^{\text{os}}(e_0')\right)^{1/2} u_v(p)\,,
\label{El:os}
\end{equation}
where the Dirac spinor $u(P)$ ($P=Mv+p$) can be expressed via the HEET spinor $u_v(p)$
(satisfying $\rlap/v u_v(p)=u_v(p)$) by the Foldy--Wouthuysen transformation (see Sect.~\ref{S:L2});
$e_0$ is the bare charge in QED and $e_0'$ in HEET (it is the same as in QPD~\cite{eft1}).
At the moment we don't consider $1/M$ corrections;
the leading bare matching coefficient is
\begin{equation}
z_0 = \frac{Z_\psi^{\text{os}}(e_0)}{Z_h^{\text{os}}(e_0')}\,.
\label{El:z0}
\end{equation}
The on-shell wave function renormalization constant in the effective theory is $Z_h^{\text{os}}=1$
(all loop corrections contain no scale);
in QED it is gauge invariant to all orders~\cite{JZ:59,MR:00}.
The renormalized fields $\psi(\mu)$ and $h_v(\mu)$ are related by the formula
similar to~(\ref{El:psi0}) but with the renormalized matching coefficient
\begin{equation}
z(\mu) =
\frac{Z_h(\alpha'(\mu),a'(\mu))}%
{Z_\psi(\alpha(\mu),a(\mu))} z_0\,,
\label{El:zmu}
\end{equation}
where $\alpha(\mu)$, $a(\mu)$ are the \MS{} renormalized QED quantities,
and $\alpha(\mu)$, $a'(\mu)$ are those in HEET
(in fact they don't depend on $\mu$ and are equal to the bare ones,
see~\cite{eft1});
$Z_h(\alpha',a')$ is given by~(\ref{HEET:ZQ}).

Now we shall prove that $z(\mu)$ is gauge invariant~\cite{G:10}.
The bare matching coefficient $z_0 = Z_\psi^{\text{os}}$ is gauge invariant;
$\log Z_h=(3-a') \alpha'/(4\pi\varepsilon)$~(\ref{HEET:ZQ}),
where $\alpha'=\alpha_{\text{os}}$;
in Appendix~\ref{S:Appe} we demonstrate that
$\log Z_\psi = - a(\mu) \alpha(\mu)/(4\pi\varepsilon) + (\text{gauge invariant})$;
finally, decoupling relations~\cite{eft1} state that
$a(\mu)\alpha(\mu)=a' \alpha'$,
and the gauge dependence cancels in $\log(Z_h/Z_\psi)$.

Collecting together 2-loop results for $Z_\psi^{\text{os}}$ (see~\cite{G:07}),
$Z_\psi$ (Appendix~\ref{S:Appe}), and $Z_h$ (\ref{HEET:ZQ}),
we obtain
\begin{equation}
z(M) = 1 - \frac{\alpha}{\pi}
+ \biggl(\pi^2 \log2 - \frac{3}{2} \zeta_3 - \frac{55}{48} \pi^2 + \frac{5957}{1152}\biggr)
\left(\frac{\alpha}{\pi}\right)^2
+ \cdots
\end{equation}
The 3-loop correction has been obtained in~\cite{G:10}.

We can look at the relation between the electron fields in the two theories
from a slightly different point of view.
Let's consider the QED electron propagator near the mass shell,
$P=(M + \omega) v$ where the on-shell mass is $M = M_0 + \delta M$
and $\omega\ll M$.
The electron self-energy has 2 Dirac structures
\begin{equation}
\Sigma(P) = \Sigma_0(\omega) + \Sigma_1(\omega) (\rlap/v - 1)\,.
\label{El:Sigma}
\end{equation}
The propagator is
\begin{equation*}
S(P) = \frac{1}{\rlap{\,/}P - M_0 - \Sigma(P)}
= \frac{1}{\left[ M + \omega - \Sigma_1(\omega) \right] \rlap/v
- M + \delta M - \Sigma_0(\omega) + \Sigma_1(\omega)}\,.
\end{equation*}
Its denominator
\begin{equation*}
\left[ M + \omega - \Sigma_1(\omega) \right]^2
- \left[ M - \delta M + \Sigma_0(\omega) - \Sigma_1(\omega) \right]^2
\end{equation*}
should vanish at $\omega=0$,
therefore the mass counterterm is given by
\begin{equation}
\delta M = \Sigma_0(0)\,.
\label{El:dM}
\end{equation}
The propagator is
\begin{equation*}
\begin{split}
S(P) &{}= \frac{1}{\left[ M + \omega - \Sigma_1(\omega) \right] \rlap/v
- M - \Sigma_0(\omega) + \Sigma_0(0) + \Sigma_1(\omega)}\\
&{}= \frac{\left[ M + \omega - \Sigma_1(\omega) \right] \rlap/v
+ M + \Sigma_0(\omega) - \Sigma_0(0) - \Sigma_1(\omega)}%
{\left[ M + \omega - \Sigma_1(\omega) \right]^2
- \left[ M + \Sigma_0(\omega) - \Sigma_0(0) - \Sigma_1(\omega) \right]^2}\,;
\end{split}
\end{equation*}
its denominator at $\omega\to0$ is
\begin{equation*}
\begin{split}
&\left[ M - \Sigma_1(0) + \omega - \Sigma_1(\omega) + \Sigma_1(0) \right]^2
- \left[ M - \Sigma_1(0)
+ \Sigma_0(\omega) - \Sigma_0(0)
- \Sigma_1(\omega) + \Sigma_1(0) \right]^2\\
&{}\approx 2 \left(M - \Sigma_1(0)\right)
\left[ \omega - \Sigma_0(\omega) + \Sigma_0(0) \right]\,,
\end{split}
\end{equation*}
and its numerator at $\omega\to0$ is $\left(M - \Sigma_1(0)\right) \left(1+\rlap/v\right)$.
Finally, the electron propagator at $\omega\to0$ can be written as
\begin{equation}
S(P) \approx \frac{1+\rlap/v}{2}
\frac{1}{\omega - \Sigma_0(\omega) + \Sigma_0(0)}\,.
\label{El:S}
\end{equation}

\begin{figure}[b]
\begin{center}
\begin{picture}(54,24)
\put(27,12.5){\makebox(0,0){\includegraphics{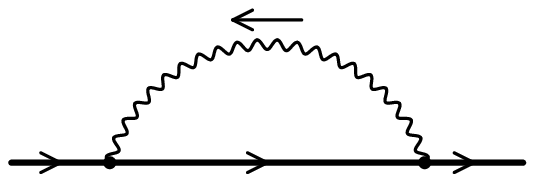}}}
\put(27,0){\makebox(0,0)[b]{$k+P$}}
\put(27,24){\makebox(0,0)[t]{$k$}}
\put(6,0){\makebox(0,0)[b]{$P$}}
\put(48,0){\makebox(0,0)[b]{$P$}}
\end{picture}
\end{center}
\caption{One-loop electron self-energy}
\label{F:el1}
\end{figure}

At one loop (Fig.~\ref{F:el1})
\begin{equation*}
\begin{split}
&\Sigma_0(\omega) = \frac{1}{4} \Tr (1+\rlap/v) \Sigma(P)
= - i e_0^2 \int \frac{d^d k}{(2\pi)^d} \frac{1}{2 D_1 D_2}\\
&\biggl[ (d+2) M - (d-2) \omega - (d-2) \frac{D_2 + M^2}{M+\omega}
+ \frac{\xi \omega^2}{D_2} \frac{D_2 + 4 M \omega + \omega^2}{M+\omega} \biggl]\,,
\end{split}
\end{equation*}
where $P = (M + \omega) v$,
\begin{equation*}
D_1 = M^2 - (k+P)^2\,,\qquad
D_2 = - k^2\,.
\end{equation*}
According to the method of regions~\cite{S:02,J:11},
it is the sum of two contributions, hard and soft.

In the hard region $k\sim M$,
\begin{equation*}
D_1 = D_h - (D_2 - D_h + 2 M^2) \frac{\omega}{M} - \omega^2\,,
\end{equation*}
where $D_h = M^2 - (k + Mv)^2$;
$D_h\sim M^2$, $D_2\sim M^2$, and we can expand the integrand
in Taylor series in $\omega$.
Each term is a loop integral with a single scale $M$:
\begin{equation}
\Sigma_h(\omega) = \frac{e_0^2 M^{1-2\varepsilon}}{(4\pi)^{d/2}}
\Gamma(\varepsilon) \frac{d-1}{d-3} \left(1 - \frac{\omega}{M} + \cdots\right)\,.
\label{El:Sh}
\end{equation}
Thus we obtain the on-shell mass renormalization
\begin{equation}
\delta M = M \left[
\frac{e_0^2 M^{-2\varepsilon}}{(4\pi)^{d/2}}
\Gamma(\varepsilon) \frac{d-1}{d-3}
+ \cdots \right]
\label{El:dM1}
\end{equation}
(it is gauge invariant to all orders)
and the on-shell wave-function renormalization
\begin{equation}
Z_\psi^{\text{os}} = \frac{1}{1 - \Sigma_0'(0)}
= 1 - \frac{e_0^2 M^{-2\varepsilon}}{(4\pi)^{d/2}} \Gamma(\varepsilon) \frac{d-1}{d-3}
+ \cdots
\label{El:Zos}
\end{equation}
(in QED it is also gauge invariant to all orders).

In the soft region $k\sim\omega$,
\begin{equation*}
D_1 = M D_s - (k + \omega v)^2\,,
\end{equation*}
where $D_s = - 2 (k\cdot v + \omega)$;
$D_s\sim\omega$, $D_2\sim\omega^2$, and we can expand the integrand
in Taylor series in $1/M$.
Each term is a loop integral with a single scale $\omega$:
\begin{equation}
\Sigma_s(\omega) = \Sigma(\omega)
\left[1 + \mathcal{O}\left(\frac{\omega}{M}\right) \right]\,,
\label{El:Ss}
\end{equation}
where $\Sigma(\omega)$ is the HEET self-energy~(\ref{Ren:Sigma1}).

We arrive at the following conclusion.
The full QED propagator near the mass shell
\begin{equation}
S(p) = \frac{1 + \rlap/v}{2}
\frac{1}{\omega - \Sigma_h'(0) \omega - \Sigma_s(\omega)}
= z_0 S(\omega)\,,
\label{El:SP}
\end{equation}
where
\begin{equation}
S(\omega) = \frac{1 + \rlap/v}{2}
\frac{1}{\omega - \Sigma(\omega)}
\label{El:Sw}
\end{equation}
is the HEET propagator, and
\begin{equation}
z_0 = Z_\psi^{\text{os}} = \frac{1}{1 - \Sigma_h'(0)}\,.
\label{El:zSh}
\end{equation}
Higher terms in $\Sigma_h$ lead to $1/M^n$ corrections
to the expression for $\psi_0$ via $h_{v0}$;
higher terms in $\Sigma_s$ lead to corrections to $S(\omega)$
due to $1/M^n$ terms in the HEET Lagrangian.

Now we shall discuss power counting.
We are considering QED processes with small characteristic residual momenta $p$,
and the small parameter is $\lambda\sim p/M$.
When acting on soft fields ($h$, $A$), $\partial_\mu\sim\lambda$;
also $A\sim\lambda$ (see~\cite{eft1}),
so that the covariant derivative is homogeneous: $D\sim\lambda$.
The static electron propagator is
\begin{equation*}
{<}T\{h(x) h^+(0)\}{>} \sim
\int \frac{d^4 p}{(2\pi)^4} e^{-ip\cdot x} \frac{1}{p\cdot v+i0}\,,
\end{equation*}
and from $p\sim\lambda$ we obtain $h\sim\lambda^{3/2}$.
The leading-order Lagrangian scales as $h^+ iD_0 h \sim \lambda^4$,
this means that the characteristic action is of order $1$.
The first power corrections to the Lagrangian
$h^+ \vec{D}\,^2 h \sim \lambda^5$, $h^+ \vec{B}\cdot\vec{\sigma} h \sim \lambda^5$,
and their contributions to the action are $\sim\lambda$.

\subsection{Heavy--heavy current}
\label{S:HH}

\begin{figure}[ht]
\begin{center}
\begin{picture}(42,22)
\put(21,11){\makebox(0,0){\includegraphics{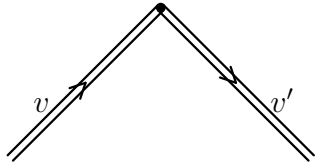}}}
\put(5,7){\makebox(0,0)[b]{$v$}}
\put(37,7){\makebox(0,0)[b]{$v'$}}
\end{picture}
\end{center}
\caption{Heavy--heavy current}
\label{F:HH0}
\end{figure}

Suppose the electron substantially changes its 4-velocity
(due to some hard-photon interaction).
In the HEET framework this can be described by the current (Fig.~\ref{F:HH0})
\begin{equation}
J_0 = \varphi_{v'0}^* \varphi_{\vphantom{v'}v0} = Z_J(\vartheta) J(\mu)\,,
\label{HH:J0}
\end{equation}
where $\cosh\vartheta = v\cdot v'$.
If $v'=v$ then $Z_J(0)=1$ (Sect.~\ref{S:Prop});
but for $\vartheta\ne0$ non-trivial renormalization appears.
This anomalous dimension of an angle (cusp) on a Wilson line
has been studied in a number of papers;
we shall see that in QED it is very simple.

\begin{figure}[b]
\begin{center}
\begin{picture}(90,22)
\put(21,11){\makebox(0,0){\includegraphics{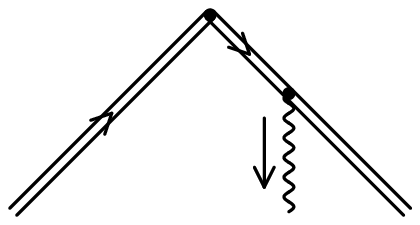}}}
\put(69,11){\makebox(0,0){\includegraphics{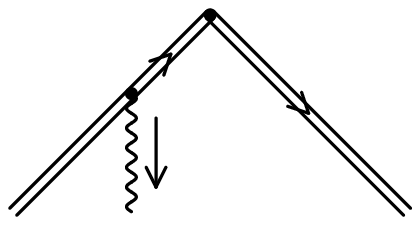}}}
\put(67,7){\makebox(0,0)[r]{$k$}}
\put(25,7){\makebox(0,0)[r]{$k$}}
\put(26,18){\makebox(0,0)[bl]{$k$}}
\put(64,18){\makebox(0,0)[br]{$-k$}}
\end{picture}
\end{center}
\caption{Real photon radiation}
\label{F:Real}
\end{figure}

We shall calculate the one-loop anomalous dimension of the current~(\ref{HH:J0})
by several methods.
The first one is based on considering real photon radiation (Fig.~\ref{F:Real}).
Its matrix element is
\begin{equation}
M^\mu = e \left(\frac{v^\mu}{k\cdot v} - \frac{v^{\prime\mu}}{k\cdot v'}\right)\,.
\label{HH:M}
\end{equation}
The probability to emit a photon with energy $\omega$ (in the $v$ rest frame)
integrated over directions is
\begin{align}
F(\omega) ={}& - e^2 \int \frac{d^d k}{(2\pi)^d}\,
2 \pi \delta(k^2)\,\delta(k\cdot v-\omega)
\left(\frac{v}{k\cdot v} - \frac{v'}{k\cdot v'}\right)^2
\nonumber\\
={}& - \frac{2}{\Gamma(1-\varepsilon)}
\frac{e^2}{(4\pi)^{d/2}} \frac{1}{\omega^{1+2\varepsilon}}
\int\limits_{-1}^{+1} dc\,
\left[ 1 + \frac{2\coth\vartheta}{c-\coth\vartheta}
+ \frac{1}{\sinh^2\vartheta} \frac{1}{(c-\coth\vartheta)^2}
\right]
\nonumber\\
{}={}& \frac{8}{\Gamma(1-\varepsilon)} \frac{e^2}{(4\pi)^{d/2}}
\frac{\vartheta\coth\vartheta - 1}{\omega^{1+2\varepsilon}}\,.
\label{HH:F}
\end{align}
This is the soft radiation function in classical electrodynamics~\cite{LL}.

Now we shall use Bjorken sum rule~\cite{Bj}.
Let $\xi$ be the amplitude \emph{not} to emit a photon.
The full probability is
\begin{equation*}
\xi^2 + \int\limits_0^\infty F(\omega)\,d\omega = 1\,.
\end{equation*}
Therefore,
\begin{equation*}
\xi = 1 - \frac{1}{2} \int\limits_\lambda^\infty F(\omega)\,d\omega
= 1 - 2 \frac{\alpha}{4\pi\varepsilon}
(\vartheta\coth\vartheta - 1)\,,
\end{equation*}
where $\lambda$ is an IR regulator,
and only the UV $1/\varepsilon$ pole is retained in the result.
Hence the renormalization constant is
\begin{equation}
Z_J = 1 - 2 \frac{\alpha}{4\pi\varepsilon}
(\vartheta\coth\vartheta - 1)\,,
\label{HH:Z1}
\end{equation}
and the one-loop anomalous dimension
\begin{equation}
\Gamma(\vartheta) = (\vartheta\coth\vartheta - 1)
\frac{\alpha}{\pi}
\label{HH:Gamma1}
\end{equation}
is given by the classical soft radiation function.
It should be included in The Guinness Book of Records
as the anomalous dimension being known for a longest time
(probably, $>100$ years).

\begin{figure}[t]
\begin{center}
\begin{picture}(94,28)
\put(21,14){\makebox(0,0){\includegraphics{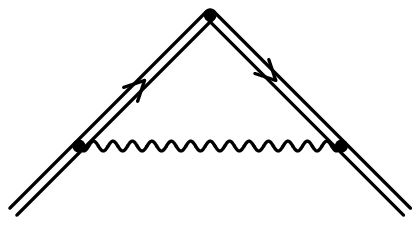}}}
\put(21,26){\makebox(0,0)[b]{0}}
\put(6,10){\makebox(0,0)[br]{$-vt$}}
\put(36,10){\makebox(0,0)[bl]{$v't'$}}
\put(21,0){\makebox(0,0)[b]{a}}
\put(73,14){\makebox(0,0){\includegraphics{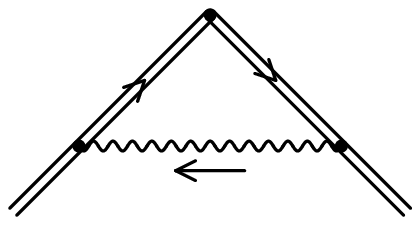}}}
\put(73,7.5){\makebox(0,0)[t]{$k$}}
\put(65,17){\makebox(0,0)[br]{$k+p$}}
\put(81,17){\makebox(0,0)[bl]{$k+p'$}}
\put(73,0){\makebox(0,0)[b]{b}}
\end{picture}
\end{center}
\caption{One-loop vertex of the heavy--heavy current
in coordinate and momentum space}
\label{F:HH1}
\end{figure}

Next we shall calculate it again in coordinate space~\cite{P:79}.
The one-loop contribution to the vertex function is (Fig.~\ref{F:HH1}a; see~(\ref{Prop:Dx}))
\begin{align}
&\Lambda(t,t';\vartheta) = i e^2 D^0_{\mu\nu}(x) v^\mu v^{\prime\nu}
\theta(t) \theta(t')
\label{HH:coord}\\
&{}= - \frac{e^2}{8\pi^{d/2}} \Gamma(1-\varepsilon) \theta(t) \theta(t')
\frac{(1+a) x^2 \cosh\vartheta + (d-2) (1-a) (t+t'\cosh\vartheta) (t'+t\cosh\vartheta)}%
{(-x^2+i0)^{d/2}}\,,
\nonumber
\end{align}
where $x = vt + v't'$.
In momentum space it is
\begin{equation*}
\Lambda(\omega,\omega';\vartheta) =
\int dt\,dt'\,e^{i\omega t+i\omega't'} \Lambda(t,t';\vartheta)\,.
\end{equation*}
Substituting $t = \tau (1+\xi)/2$, $t' = \tau (1-\xi)/2$, we get
\begin{equation*}
\begin{split}
\Lambda(0,0;\vartheta) ={}& - \frac{e^2}{16 \pi^{d/2}} \Gamma(1-\varepsilon)
\int\limits_0^T \frac{d\tau}{\tau^{1-2\varepsilon}}
\int\limits_{-1}^{+1} d\xi\\
&{}\times\frac{(1+a) \cosh\vartheta (c^2-s^2\xi^2)
+ (d-2) (1-a) (c^4-s^4\xi^2)}%
{(-c^2+s^2\xi^2)^{d/2}}\,,
\end{split}
\end{equation*}
where $T$ is an IR regulator, and
$c = \cosh(\vartheta/2)$, $s = \sinh(\vartheta/2)$.
Retaining only the UV $1/\varepsilon$ pole, we have
\begin{equation*}
Z_\Gamma(\vartheta) = 1 - \frac{\alpha}{4\pi\varepsilon}
\int\limits_{-1}^{+1} d\xi\,\Biggl[ (1+a)
\frac{\cosh\vartheta}{2\cosh^2(\vartheta/2)}
\frac{1}{1 - \xi^2 \tanh^2(\vartheta/2)}
+ (1-a)
\frac{1 - \xi^2 \tanh^4(\vartheta/2)}{[1 - \xi^2 \tanh^2(\vartheta/2)]^2}
\Biggr]\,.
\end{equation*}
Substituting $\xi = \tanh\psi/\tanh(\vartheta/2)$ we obtain
\begin{equation*}
Z_\Gamma(\vartheta) = 1 - \frac{\alpha}{4\pi\varepsilon}
\int\limits_{-\vartheta/2}^{+\vartheta/2} d\psi\,
\left[ 2 \coth\vartheta + \frac{1-a}{\sinh\vartheta} \cosh2\psi \right]
= 1 - \frac{\alpha}{4\pi\varepsilon}
(2 \vartheta \coth\vartheta + 1 - a)\,.
\end{equation*}
We see that $Z_J(\vartheta) = Z_\Gamma(\vartheta) Z_h$
is gauge invariant and coincides with~(\ref{HH:Z1}).

Finally, we shall do the calculation in momentum space.
The vertex function $\Lambda(\omega,\omega';\vartheta)$ (Fig.~\ref{F:HH1}b)
depends on $\omega=p\cdot v$ and $\omega'=p'\cdot v'$.
We only need the UV divergence; it is sufficient to consider the case $\omega'=\omega$.
In the Feynman gauge
\begin{equation*}
\Lambda(\omega,\omega;\vartheta)
= - i e_0^2 v\cdot v' \int \frac{d^d k}{(2\pi)^d}
\frac{1}{(k\cdot v+\omega) (k\cdot v'+\omega) k^2}
= 4 I(1,1,1) \cosh\vartheta \frac{e_0^2 (-2\omega)^{-2\varepsilon}}{(4\pi)^{d/2}}\,,
\end{equation*}
where
\begin{equation}
\begin{split}
&\frac{1}{i \pi^{d/2}} \int \frac{d^d k}{D_1^{n_1} D_2^{n_2} D_3^{n_3}} =
(-2\omega)^{d-n_1-n_2-2n_3} I(n_1,n_2,n_3)\,,\\
&D_1 = - 2 (k\cdot v + \omega)\,,\quad
D_2 = - 2 (k\cdot v' + \omega)\,,\quad
D_3 = - k^2\,.
\end{split}
\label{HH:Idef}
\end{equation}
This integral is considered in~\cite{GK:11} in detail.
Using HQET Feynman parametrization~(\ref{L1:Feyn}) we have
\begin{equation*}
\begin{split}
(-2\omega)^{-2\varepsilon} I(1,1,1)
&= 2 \int \frac{d^d k}{i \pi^{d/2}}
\frac{dy\,dy'}{\left[ - k^2 - 2 y (k\cdot v+\omega) - 2 (k\cdot v'+\omega)\right]^3}\\
&= \Gamma(1+\varepsilon) \int
\frac{dy\,dy'}{\left[(yv+y'v')^2 - 2\omega(y+y')\right]^{1+\varepsilon}}\,.
\end{split}
\end{equation*}
Substituting $y = zx$, $y' = z(1-x)$  and integrating in $z$, we get
\begin{equation*}
\begin{split}
&I(1,1,1) = \Gamma(2\varepsilon) \Gamma(1-\varepsilon)
\int\limits_0^1 \frac{dx}{A^{1-\varepsilon}}\,,\\
&A = x^2 + (1-x)^2 + 2 x (1-x) \cosh\vartheta
= \left[1 - (1-e^\vartheta) x\right] \left[1 - (1-e^{-\vartheta}) x\right]\,.
\end{split}
\end{equation*}
The result for this integral is~\cite{GK:11}
\begin{equation*}
\frac{I(1,1,1)}{\Gamma(2\varepsilon) \Gamma(1-\varepsilon)} = \,
{}_2 F_1\left(\left.
\begin{array}{c}
1,1-\varepsilon\\3/2
\end{array}
\right| \frac{1-\cosh\vartheta}{2} \right)
= \frac{\vartheta}{\sinh\vartheta} + \mathcal{O}(\varepsilon)
\end{equation*}
(it is easy to calculate the integral in $x$ at $\varepsilon=0$).
Finally,
\begin{equation*}
\Lambda(\omega,\omega;\vartheta)
= 2 \frac{\alpha}{4\pi\varepsilon} \vartheta \coth\vartheta + \mathcal{O}(1)\,,
\end{equation*}
and we arrive at the same result.

\begin{figure}[b]
\begin{center}
\begin{picture}(48,25)
\put(27,11){\makebox(0,0){\includegraphics{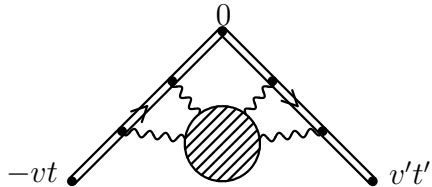}}}
\put(27,23){\makebox(0,0){0}}
\put(49,0.5){\makebox(0,0)[bl]{$v't'$}}
\put(5,0.5){\makebox(0,0)[br]{$-vt$}}
\end{picture}
\end{center}
\caption{Green function with the insertion of the heavy--heavy current}
\label{F:HHGreen}
\end{figure}

Now we shall prove that the exact anomalous dimension of the heavy--heavy current in HEET
is given by the one-loop term~(\ref{HH:Gamma1}),
just like the anomalous dimension of the static electron field~(\ref{HEET:gammaQ}).
To this end, let's consider the full Green function of $\varphi_{v0}^*$, $\varphi_{v'0}$, and $J_0$~(\ref{HH:J0}).
After singling out the obvious $\delta$-functions,
it can be written as $G(t,t';\vartheta)$ (Fig.~\ref{F:HHGreen}).
The exponentiation argument (see~(\ref{Prop:exp})) holds
for this heavy-quark world line with an angle, too.
Therefore,
\begin{equation*}
G(t,t';\vartheta) = \theta(t) \theta(t')
\exp \left[ \frac{e^2}{(4\pi)^{d/2}} F(t,t';\vartheta) \right]\,,
\end{equation*}
where $F(t,t';\vartheta)$ is just the one-loop correction.
Let us divide this by $G(t,t';0) = i S(t+t')\allowbreak\theta(t) \theta(t')$
at $t>0$, $t'>0$:
\begin{equation}
\mathcal{G}(t,t';\vartheta)
= \frac{G(t,t';\vartheta)}{i S(t+t')}
= \exp \left[ \frac{e_0^2}{(4\pi)^{d/2}}
\left( \mathcal{F}(t,t';\vartheta) - \mathcal{F}(t,t';0) \right)
\right]\,,
\label{HH:Ratio}
\end{equation}
where $\mathcal{F}(t,t';\vartheta)$ is the one-loop correction
which has the $J_0$ vertex inside (shown in Fig.~\ref{F:HH1});
corrections to the external legs cancel here.
If this ratio is re-expressed via the renormalized quantities
(this is trivial, because in this theory $e=e_0$ and $a=a_0$),
it should be equal to $Z_J(\vartheta) \mathcal{G}_r(t,t';\vartheta)$,
where $\mathcal{G}_{\text{r}}(t,t';\vartheta)$ is finite at $\varepsilon\to0$.
Therefore,
\begin{equation*}
Z_J(\vartheta) = \exp \left[ \frac{\alpha}{4\pi\varepsilon}
\left( f(\vartheta) - f(0) \right)
\right]\,,
\end{equation*}
where
\begin{equation*}
\varepsilon e^{\gamma\varepsilon} \mathcal{F}(t,t';\vartheta) =
f(\vartheta) + \mathcal{O}(\varepsilon)\,.
\end{equation*}
The anomalous dimension is exactly equal to the one-loop contribution:
\begin{equation}
\Gamma(\vartheta) = (\vartheta\coth\vartheta-1) \frac{\alpha}{\pi}\,.
\label{HH:Gamma}
\end{equation}

The anomalous dimension is obviously even: $\Gamma(-\vartheta)=\Gamma(\vartheta)$;
therefore, at $\vartheta\to0$
\begin{align}
&\Gamma(\vartheta) = \Gamma_0(\alpha) \vartheta^2 + \mathcal{O}(\vartheta^4)\,,
\label{HH:G0}\\
&\Gamma_0(\alpha) = \frac{\alpha}{3\pi}\,.
\label{HH:G01}
\end{align}
At $\vartheta\to\infty$
\begin{align}
&\Gamma(\vartheta) = \Gamma_\infty(\alpha) \vartheta + \mathcal{O}(\vartheta^0)\,,
\label{HH:Ginf}\\
&\Gamma_\infty(\alpha) = \frac{\alpha}{\pi}\,.
\label{HH:Ginf1}
\end{align}

\begin{figure}[ht]
\begin{center}
\begin{picture}(42,26)
\put(21,13){\makebox(0,0){\includegraphics{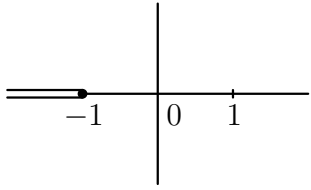}}}
\put(11,11.5){\makebox(0,0)[t]{$-1$}}
\put(31,11.5){\makebox(0,0)[t]{$1$}}
\put(22,11.5){\makebox(0,0)[tl]{$0$}}
\end{picture}
\end{center}
\caption{Complex plane of $\cos\vartheta_E$}
\label{F:cos}
\end{figure}

\begin{figure}[t]
\begin{center}
\begin{picture}(42,14)
\put(21,7){\makebox(0,0){\includegraphics{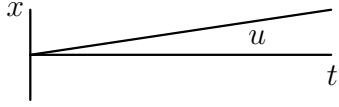}}}
\put(41,5.5){\makebox(0,0)[t]{$t$}}
\put(0,13){\makebox(0,0)[r]{$x$}}
\put(31,9.25){\makebox(0,0){$u$}}
\end{picture}
\end{center}
\caption{Wilson line for heavy particle--antiparticle production}
\label{F:delta}
\end{figure}

We could start from calculating the anomalous dimension of an angle $\vartheta_E$
on a Wilson line in Euclidean space.
The result is
\begin{equation}
\Gamma_E(\vartheta_E) = (\vartheta_E\cot\vartheta_E-1) \frac{\alpha}{\pi}\,.
\label{HH:GammaE}
\end{equation}
Here $\cos\vartheta_E\in[-1,1]$;
$\Gamma(\vartheta_E)$ can be analytically continued to the whole complex plane
with a cut from $-1$ to $-\infty$ (Fig.~\ref{F:cos}).
The region $\vartheta_E=i\vartheta$, $\cos\vartheta_E=\cosh\vartheta\ge1$
corresponds to the Minkowski result~(\ref{HH:Gamma}).
The branch point $\cos\vartheta_E=-1$ corresponds to a degenerate Wilson line:
it goes straight from infinity to some point,
then returns to infinity along the same ray.
When $\vartheta_E=\pi-\delta$, $\delta\ll1$ (Fig.~\ref{F:delta}),
the anomalous dimension behaves as
\begin{equation}
\Gamma_E(\pi-\delta) = - \frac{\alpha}{\delta} + \mathcal{O}(\delta^0)\,.
\label{HH:delta}
\end{equation}
We'll see in a moment that this behaviour is determined by the Coulomb potential.

The cut $\cos\vartheta_E\le-1$ corresponds to production (or annihilation)
of a heavy particle--antiparticle pair;
$v$ points to the future and $v'$ to the past (or vice versa),
and $v\cdot v'\le-1$.
The physical side of the cut is the lower one $v\cdot v'-i0$ (Fig.~\ref{F:cos}),
because this prescription~\cite{KR:87} reproduces the correct sign of $i0$ in~(\ref{HH:coord}).
This means $\vartheta=\vartheta_0-i\pi$ (or $\vartheta_E=\pi+i\vartheta_0$)
where $\vartheta_0$ is the angle between the particle velocity $v$ and the antiparticle one $-v'$.
The anomalous dimension
\begin{equation}
\Gamma(\vartheta_0-i\pi) = \left[ (\vartheta_0-i\pi) \coth\vartheta_0 - 1 \right] \frac{\alpha}{\pi}
\label{HH:Im}
\end{equation}
gets an imaginary part~\cite{KR:87}
\begin{equation*}
\mathop{\mathrm{Im}} \Gamma(\vartheta_0-i\pi) = - \alpha \coth\vartheta_0\,.
\end{equation*}
Matrix elements of processes where a heavy particle--antiparticle pair is produced (or annihilates)
satisfy renormalization group equations with this anomalous dimension;
its imaginary part produces Coulomb phase factors in such matrix elements~\cite{KMO:93}.

When $\vartheta_0=u\ll1$, the heavy particle--antiparticle pair is nonrelativistic,
and $u$ is the relative velocity:
\begin{equation*}
\mathop{\mathrm{Im}} \Gamma(u-i\pi) = - \frac{\alpha}{u} + \mathcal{O}(u^0)\,.
\end{equation*}
This result can be easily checked in the Coulomb gauge.
The heavy particle and the antiparticle interact via the instantaneous Coulomb potential
\begin{equation*}
V(r) = - \frac{e^2}{4\pi r^{1-2\varepsilon}}
\end{equation*}
(the power of $r$ is obvious from dimension counting;
there are no corrections to this formula,
because no loop can be inserted into the Coulomb photon propagator).
The Wilson line (Fig.~\ref{F:delta}) is
\begin{equation*}
W = \exp\left[-i \int_0^T V(ut)\,d t\right]\,,
\end{equation*}
where $T$ is an IR regulator.
Keeping only $1/\varepsilon$ divergences,
we find the renormalization constant
\begin{equation*}
Z = \exp\left[ i \frac{\alpha}{2u\varepsilon} \right]\,.
\end{equation*}
There are no self-energy corrections (because the interaction is instantaneous).
The anomalous dimension is
\begin{equation}
\Gamma = \frac{d\,\log Z}{d\,\log\mu} = - i \frac{\alpha}{u}\,.
\label{HH:u}
\end{equation}
So, the $1/u$ term in $\mathop{\mathrm{Im}} \Gamma(u-i\pi)$
is determined by the particle--antiparticle potential~\cite{KMO:93}.
Substituting $u=i\delta$ we reproduce the Euclidean result~(\ref{HH:delta}).

\subsection{Kinetic energy}
\label{S:Kin}

Now we shall discuss $1/M$ corrections to the HEET Lagrangian.
Unlike the leading term, these corrections depend on the electron spin.
In this Section, we'll discuss spin-0 electron case,
because it is simpler.
Then there is only one dimension-5 operator
which can be incorporated into the Lagrangian:
\begin{equation}
L = L_0 + \frac{1}{2M} C_k^0 O_k^0
= L_0 + \frac{1}{2M} C_k(\mu) O_k(\mu)\,,
\label{Kin:L}
\end{equation}
namely, the kinetic energy of the heavy electron
\begin{equation}
O_k^0 = \varphi_0^* \vec{D}\,^2 \varphi_0
= Z(\mu) O_k(\mu)\,.
\label{Kin:O}
\end{equation}
There is no need to include the operator $\varphi_0^* D_0^2 \varphi_0$:
it can be eliminated by a field redefinition
\begin{equation*}
\varphi_0 \to \varphi_0 + \frac{c}{2M} D_0 \varphi_0\,.
\end{equation*}
Generally speaking, any effective Lagrangian of a nonrelativistic field
can be written in a canonical form where the time derivative appears only in the leading term:
$L = \varphi_0^* D_0 \varphi_0 - V(\varphi_0)$;
the field energy $\int V(\varphi_0) d^3\vec{x}$ depends only on the field configuration
at a given moment, and hence $V(\varphi_0)$ contains only space derivatives.
All higher-dimensional terms containing $D_0$ can be eliminated by suitable field redefinitions.
The kinetic energy operator can be written in relativistic form:
\begin{equation}
O_k^0 = -  \varphi_0^* D_\bot^2 \varphi_0\,,
\label{Kin:0r}
\end{equation}
where $D_\bot^\mu=g_\bot^{\mu\nu}D_\nu$, $g_\bot^{\mu\nu}=g^{\mu\nu}-v^\mu v^\nu$.
Here $Z_k(\mu)$ is a minimal renormalization constant,
and $C_k(\mu)=Z(\mu)C_k^0$ is finite at $\varepsilon\to0$;
$M$ is the on-shell electron mass.
This Lagrangian gives the mass shell of the free electron $\varepsilon=C_k^0\vec{p}\,^2/(2M)$,
and hence $C_k^0=1$ at the tree level.

The kinetic-energy term gives the new vertices
\begin{equation}
\begin{split}
&\raisebox{-3.75mm}{\begin{picture}(22,6)
\put(11,5){\makebox(0,0){\includegraphics{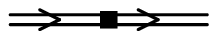}}}
\put(6,0){\makebox(0,0)[b]{$p$}}
\put(16,0){\makebox(0,0)[b]{$p$}}
\end{picture}} =
i \frac{C_k^0}{2M} p_\bot^2\,,\\
&\raisebox{-3.75mm}{\begin{picture}(22,16)
\put(11,9){\makebox(0,0){\includegraphics{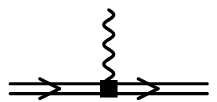}}}
\put(6,0){\makebox(0,0)[b]{$p$}}
\put(16,0){\makebox(0,0)[b]{$p'$}}
\put(11,16){\makebox(0,0)[t]{$\mu$}}
\end{picture}} =
i \frac{C_k^0}{2M} e'_0 (p+p')_\bot^\mu\,,\\
&\raisebox{-3.75mm}{\begin{picture}(22,16)
\put(11,9){\makebox(0,0){\includegraphics{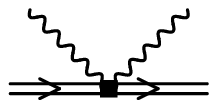}}}
\put(3,16){\makebox(0,0)[t]{$\mu$}}
\put(19,16){\makebox(0,0)[t]{$\nu$}}
\end{picture}} =
i \frac{C_k^0}{2M} e_0^{\prime2} g_\bot^{\mu\nu}\,.
\end{split}
\label{Kin:Feyn}
\end{equation}

Let's denote by $-i (C_k^0/(2M)) \Sigma_k(\omega,p_\bot^2)$
the sum of all bare one-particle-irreducible self-energy diagrams at the order $1/M$.
Each of these diagrams contains a single kinetic-energy vertex~(\ref{Kin:Feyn}).
The momentum $p_\bot$ flows through the static electron line.
No-photon kinetic vertices are quadratic in it;
one-photon vertices are linear;
2-photon vertices are independent of $p_\bot$.
The $p_\bot^2$ term comes from diagrams with a no-photon kinetic vertex.
Terms linear in $p_\bot$ vanish due to the rotational symmetry.
The coefficient of $p_\bot^2$ in a no-photon kinetic vertex is $i C_k^0/(2M)$.
Therefore, the coefficient of $p_\bot^2$ in the sum of all diagrams
is the sum of the leading-order HEET diagrams with a unit operator insertion
into each static electron propagator in turn.
This sum is just $-i d\Sigma/d\omega$, and hence
\begin{equation}
\Sigma_k(\omega,p_\bot^2) =
\frac{d\Sigma(\omega)}{d\omega} p_\bot^2 + \Sigma_{k0}(\omega)\,.
\label{Kin:Sigma}
\end{equation}

This result can also be derived in another way.
Let's consider the variation of $\Sigma$ for $v\to v+\delta v$
with an infinitesimal $\delta v$ ($v\cdot\delta v=0$).
There are two sources of this variation.
The expansion of the heavy-quark propagators $1/(p\cdot v+i0)$
produces insertions $i p_i\cdot\delta v$ into each propagator in turn.
Variations of the electron--photon vertices produce $i e'_0 \delta v^\mu$
for each vertex in turn.
Now let's consider the variation of $\Sigma_k$
for $p_\bot\to p_\bot+\delta p_\bot$ with an infinitesimal $\delta p_\bot$.
No-photon kinetic vertices produce $i (C_k^0/M) p_i\cdot\delta p_\bot$;
single-photon kinetic vertices produce $i (C_k^0/M) e'_0 \delta p_\bot^\mu$;
2-photon kinetic vertices do not change.
Therefore,
\begin{equation}
\frac{\partial\Sigma_k}{\partial p_\bot^\mu} =
2 \frac{\partial\Sigma}{\partial v^\mu}\,.
\label{Kin:Repar1}
\end{equation}
This is the Ward identity of reparametrization invariance.
Taking into account
$\partial\Sigma_k/\partial p_\bot^\mu = 2 (\partial\Sigma_k/\partial p_\bot^2) p_\bot^\mu$
and $\partial\Sigma/\partial v^\mu = (d\Sigma/d\omega) p_\bot^\mu$,
we obtain
\begin{equation}
\frac{\partial\Sigma_k}{\partial p_\bot^2} = \frac{d\Sigma}{d\omega}\,.
\label{Kin:Repar2}
\end{equation}
The right-hand side does not depend on $p_\bot^2$,
and hence we arrive at~(\ref{Kin:Sigma}).

The coefficients in the effective Lagrangian are obtained by equating
on-shell scattering amplitudes in full QED and in HEET
with the required accuracy in $1/M$.
A prerequisite for this matching is the requirement
that the mass shell itself is the same in both theories,
to the accuracy considered.
The mass shell is defined as the position of the pole of the full electron propagator.
In QED it is $p_0=\sqrt{M^2+\vec{p}\,^2}$,
where $M$ is the on-shell mass.
To the first order in $1/M$, this means $\omega=\vec{p}\,^2/(2M)$.
In HEET, the mass shell is the zero of the denominator
of the bare heavy-electron propagator:
\begin{equation}
\omega - \Sigma(\omega) - \frac{C_k^0}{2M}
\left[\vec{p}\,^2 - \frac{d\Sigma(\omega)}{d\omega} \vec{p}\,^2 + \Sigma_{k0}(\omega) \right]
= 0\,.
\label{Kin:mshell}
\end{equation}
We can expand this equation in $\omega$ up to linear terms;
$\Sigma(0)=0$, $(d\Sigma(\omega)/d\omega)_{\omega=0}=0$, $\Sigma_{k0}(0)=0$,
because these loop corrections are scale-free,
and the mass shell is
\begin{equation}
\omega = \frac{C_k^0}{2M} \vec{p}\,^2\,.
\label{Kin:mshell2}
\end{equation}
This is correct if $C_k^0 = Z_k^{-1}(\mu) C_k(\mu) = 1$.
The minimal renormalization constant $Z_k$ has to make $C_k(\mu)$ finite;
here this means
\begin{equation}
Z_k(\mu) = 1\,.
\label{Kin:Z}
\end{equation}
The kinetic-energy operator is not renormalized; its anomalous dimension is zero to all orders.
The coefficient of the kinetic-energy operator in the HEET Lagrangian is exactly unity,
\begin{equation}
C_k(\mu) = C_k^0 = 1\,,
\label{Kin:C}
\end{equation}
to all orders in perturbation theory, due to the reparametrization invariance!

On-shell scattering amplitudes in full QED expanded in $1/M$ to the first order
should be reproduced by the HEET Lagrangian.
Let's consider the simplest process ---
electron scattering in an external electromagnetic field.
In full QED the on-shell scattering amplitude of spin-0 electron
is determined by one form factor:
\begin{equation}
e_{\text{os}}\,\varphi^*(P') F(q^2) (P+P')^\mu \varphi(P)\,,
\label{Kin:F}
\end{equation}
where $P$ is the initial electron momentum,
$P'$ is the final one, $q=P'-P$,
and $e_{\text{os}}=e'_{\text{os}}=e'_0$ is the on-shell charge.
The current of a free spin-0 particle is $J^\mu = 2 P^\mu |\varphi(P)|^2$;
in the non-relativistic normalization $J^0=1$ this means $\varphi(P)=1/\sqrt{2E}$,
and this can be replaced by $1/\sqrt{2M}$ with the needed accuracy.
The form factor can be expanded as
\begin{equation}
F(q^2) = 1 + F'(0) \frac{q^2}{M^2} + \cdots
\label{Kin:F0}
\end{equation}
where $F(0)=1$ due to the Ward identity, and prime means the derivative in $q^2/M^2$
(${<}r^2{>} = 6 F'(0)/M^2$ is the charge radius squared).
Taking into account $P^{(\prime)} = Mv + p^{(\prime)}$
(to the first approximation $p\cdot v=p'\cdot v=0$),
we obtain the scattering amplitude up to $1/M$:
\begin{equation}
e_{\text{os}}\,\left[v^\mu + \frac{(p+p')_\bot^\mu}{2M}\right]\,.
\label{Kin:QED}
\end{equation}

In HEET all loop corrections to the scattering amplitude vanish (no scale).
At the tree level there is the leading vertex~(\ref{Feyn:vert})
and the one-photon kinetic-energy one~(\ref{Kin:Feyn}):
\begin{equation}
e'_0 \left[v^\mu + \frac{C_k^0}{2M} (p+p')_\bot^\mu\right]\,.
\label{Kin:HEET}
\end{equation}
We have again obtained~(\ref{Kin:C}).

We can also explicitly see reparametrization invariance of the Lagrangian.
Let
\begin{equation}
L_{v'} = \varphi_{v'}^* i v'\cdot D \varphi_{v'}
- \frac{C_k}{2M} \varphi_{v'}^* D_\bot^{\prime2} \varphi_{v'}
\label{Kin:Lv}
\end{equation}
be the HEET Lagrangian for $v'=v+\delta v$.
The field $\varphi_{v'}$ is related to $\varphi_v$ as
\begin{equation}
\varphi_{v'} = e^{i M\,\delta v\cdot x} \left(1 + \frac{i\,\delta v\cdot D}{2M}\right) \varphi_v\,.
\label{Kin:phi}
\end{equation}
Up to terms linear in $1/M$ and $\delta v$ we have
\begin{equation*}
L_{v'} = L_v - (C_k-1) \varphi_v^* i\,\delta v\cdot D \varphi_v\,;
\end{equation*}
so, the Lagrangian is invariant if $C_k=1$.

\subsection{Magnetic moment}
\label{S:Mag}

Now we return to the realistic case of spin-$\frac{1}{2}$ electron.
There are 2 dimension-5 operators which can appear in the Lagrangian at the $1/M$ level,
the kinetic energy and the magnetic moment interaction:
\begin{equation}
\begin{split}
&L = L_0 + \frac{1}{2M} C_k^0 O_k^0 + \frac{1}{2M} C_m^0 O_m^0
= L_0 + \frac{1}{2M} C_k(\mu) O_k(\mu) + \frac{1}{2M} C_m(\mu) O_m(\mu)\,,\\
&O_k^0 = h_0^+ \vec{D}\,^2 h_0 = - \bar{h}_{v0} D_\bot^2 h_{v0} = Z_k(\mu) O_k(\mu)\,,\\
&O_m^0 = - e_0 h_0^+ \vec{B}_0\cdot\vec{\sigma} h_0 = \frac{1}{2} e_0 \bar{h}_{v0} F^0_{\mu\nu} \sigma^{\mu\nu} h_{v0} = Z_m(\mu) O_m(\mu)\,,
\end{split}
\label{Mag:L}
\end{equation}
where $\sigma^{\mu\nu}=\frac{i}{2}[\gamma^\mu,\gamma^\nu]$
(in the $v$ rest frame, only spatial $\mu$, $\nu$ contribute).
As in the spin-0 case (Sect.~\ref{S:Kin}), $Z_k(\mu)=1$, $C_k(\mu)=C_k^0=1$.

The magnetic interaction term breaks the spin symmetry (Sect.~\ref{S:Photonia}).
It produces the vertex
\begin{equation}
\raisebox{0.25mm}{\begin{picture}(22,12)
\put(11,5){\makebox(0,0){\includegraphics{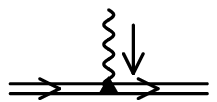}}}
\put(11,12){\makebox(0,0)[t]{$\mu$}}
\put(16,5){\makebox(0,0){$q$}}
\end{picture}} =
\frac{i e'_0 C_m^0}{2M} [\rlap/q,\gamma^\mu]\,.
\label{Mag:Feyn}
\end{equation}

The magnetic interaction coefficient $C_m$ is determined by matching the scattering amplitude
of an on-shell electron (with a physical polarization) in external magnetic field in full QED
(expanded up to the first order in $1/M$) and in HEET.
In full QED it is determined by 2 form factors:
\begin{equation}
\begin{split}
&e_{\text{os}}\,\bar{u}'(P') \left[ F_1(q^2) \gamma^\mu + F_2(q^2) \frac{[\rlap/q,\gamma^\mu]}{4M} \right] u(P)\\
&{} = e_{\text{os}}\,\bar{u}'(P') \left[ \left(F_1(q^2)+F_2(q^2)\right) \gamma^\mu - F_2(q^2) \frac{(P+P')^\mu}{2M} \right] u(P)\\
&{} = e_{\text{os}}\,\bar{u}'(P') \left[ F_1(q^2) \frac{(P+P')^\mu}{2M}
+ \left(F_1(q^2)+F_2(q^2)\right) \frac{[\rlap/q,\gamma^\mu]}{4M} \right] u(P)\,,
\end{split}
\label{Mag:FF}
\end{equation}
where
\begin{equation}
F_1(q^2) = 1 + F'_1(0) \frac{q^2}{M^2} + \cdots\,,\qquad
F_2(q^2) = F_2(0) + \cdots
\label{Mag:F12}
\end{equation}
($F_1(0)=1$ due to the Ward identity).
In QED the anomalous magnetic moment $F_2(0)$ is finite:
it contains no IR divergences at all orders in $\alpha$.
This is proved in Appendix~\ref{S:IR}.

The QED scattering amplitude~(\ref{Mag:FF})
expressed via the HEET spinors and expanded up to $1/M$ is
\begin{equation*}
e_{\text{os}}\,\bar{u}'_v(p') \left[ v^\mu + \frac{(p+p')_\bot^\mu}{2M}
+ \left(1+F_2(0)\right) \frac{i \sigma^{\mu\nu} q_\nu}{2M} \right] u_v(p)\,.
\end{equation*}
In HEET there are no loop corrections (no scale).
At the tree level there is the leading vertex~(\ref{Feyn:vert}),
the one-photon kinetic-energy one~(\ref{Kin:Feyn}),
and the magnetic-moment one~(\ref{Mag:Feyn}):
\begin{equation*}
e'_0\,\bar{u}'_v(p') \left[ v^\mu + \frac{C_k^0}{2M} (p+p')_\bot^\mu
+ \frac{C_m^0}{2M} i \sigma^{\mu\nu} q_\nu \right] u_v(p)\,.
\end{equation*}
Therefore
\begin{equation}
C_k^0 = 1\,,\qquad
C_m^0 = 1 + F_2(0)\,.
\label{Mag:Cm0}
\end{equation}
What's important here is the fact that $C_m^0$ is finite at $\varepsilon\to0$;
it needs no renormalization: $Z_m(\mu)=1$.
Therefore the magnetic interaction operator~(\ref{Mag:L}) does not renormalize;
its anomalous dimension is 0.
The magnetic interaction coefficient $C_m(\mu)=C_m^0=1+F_2(0)$ is the full electron magnetic moment
(in Bohr magnetons).
It is non-trivial: it contains all orders in $\alpha$.
It is not fixed by reparametrization invariance.

We can also explicitly see reparametrization invariance of the Lagrangian.
The field $h_{v'}$ is related to $h_v$ as
\begin{equation*}
h_{v'} = e^{i M\,\delta v\cdot x} \left(1 - \frac{\delta\rlap/v}{2} + \frac{i\,\delta v\cdot D}{2M}\right) h_v\,.
\end{equation*}
Up to terms linear in $1/M$ and $\delta v$ we have
\begin{equation*}
L_{v'} = L_v - (C_k-1) \bar{h}_v i\,\delta v\cdot D h_v\,.
\end{equation*}
The Lagrangian is invariant if $C_k=1$; $C_m$ is not constrained.

\subsection{$1/M^2$ corrections to the Lagrangian}
\label{S:L2}

Two dimension-6 operators appear in the HEET Lagrangian at the $1/M^2$ level,
the spin--orbit interaction $O_s$ and the Darwin interaction $O_d$:
\begin{equation}
\begin{split}
&L = L_0 + \frac{1}{2M} C_k^0 O_k^0 + \frac{1}{2M} C_m^0 O_m^0 + \frac{1}{4 M^2} C_s^0 O_s^0 + \frac{1}{4 M^2} C_d^0 O_d^0\,,\\
&O_s^0 = - \frac{i}{2} e'_0 h_0^+ \left( \vec{D}\times\vec{E}_0 - \vec{E}_0\times\vec{D} \right) \cdot \vec{\sigma} h_0
= - \frac{i}{2} e'_0 \bar{h}_{v0} \left[D_\bot^\mu,F_0^{\lambda\nu}\right]_+ v_\lambda \sigma_{\mu\nu} h_{v0}\,,\\
&O_d^0 = \frac{1}{2} e'_0 h_0^+ \left( \vec{D}\cdot\vec{E}_0 - \vec{E}_0\cdot\vec{D} \right) h_0
= \frac{1}{2} e'_0 \bar{h}_{v0} v^\mu \left[D_\bot^\nu,F_{0\mu\nu}\right] h_{v0}\,.
\end{split}
\label{L2:L}
\end{equation}
The scattering amplitude of an on-shell electron in an external electromagnetic field in HEET at the tree level is
\begin{equation}
e'_0\,\bar{u}'_v(p') \left[ v^\mu + C_k^0 \frac{(p+p')_\bot^\mu}{2M} + C_m^0 \frac{[\rlap/q,\gamma^\mu]}{4M}
+ C_d^0 \frac{q^2}{8M^2} v^\mu + C_s^0 \frac{[\rlap/p,\rlap/p']}{8M^2} v^\mu \right] u_v(p)\,.
\label{L2:HEET}
\end{equation}
All loop corrections vanish because are scale-free;
these corrections contain both UV and IR divergences, but they cancel each other.

In full QED the scattering amplitude is given by 2 form factors~(\ref{Mag:FF}).
A QED Dirac spinor $u(P)$ ($P=Mv+p$ is on shell, and hence $p\cdot v=-p^2/(2M)$)
is related to the corresponding HEET spinor $u_v(p)$
(which has only the upper 2 components in the $v$ rest frame, $\rlap/v u_v = u_v$)
by the Foldy--Wouthuysen transformation~\cite{BD:64}.
The 2-component spinor $u_v(p)$ is just the Dirac spinor in the $P$ rest frame,
and hence the Foldy--Wouthuysen transformation is simply the boost to the $v$ rest frame:
\begin{equation}
u(P) = c \left( 1 + \frac{\rlap/p}{2M} \right) u_v(p)\,.
\label{L2:FW}
\end{equation}
The normalization factor $c$ is determined by the requirement that
the particle density in the $v$ rest frame $\bar{u} \rlap/v u$
is given just by $\bar{u}_v u_v$:
\begin{equation}
c = \left[\left(1 - \frac{p^2}{2M^2}\right) \left(1 - \frac{p^2}{4M^2}\right)\right]^{-1/2}
= 1 + \frac{3}{8} \frac{p^2}{M^2} + \cdots
\label{L2:c}
\end{equation}
Then the amplitude~(\ref{Mag:FF}) can be rewritten with the $1/M^2$ accuracy as
\begin{equation}
\begin{split}
&e_{\text{os}}\,\bar{u}'_v(p') \biggl\{
F_1(q^2)
\left[v^\mu + \frac{(p+p')_\bot^\mu}{2M} - \frac{q^2 + [\rlap/p,\rlap/p']}{8M^2}\right]\\
&\qquad{} + \left(F_1(q^2) + F_2(q^2)\right)
\left[\frac{[\rlap/q,\gamma^\mu]}{4M} + \frac{q^2 + [\rlap/p,\rlap/p']}{4M^2}\right]
\biggr\} u_v(p)\,.
\end{split}
\label{L2:QED}
\end{equation}
Comparing it with the HEET amplitude~(\ref{L2:HEET}), we obtain
\begin{equation}
C_k^0 = 1\,,\qquad
C_m^0 = 1 + F_2(0)\,,\qquad
C_s^0 = 1 + 2 F_2(0)\,,\qquad
C_d^0 = 1 + 2 F_2(0) + 8 F_1'(0)\,.
\label{L2:C0}
\end{equation}

As discussed, in QED the anomalous magnetic moment $F_2(0)$ is IR finite to all orders.
Hence the spin--orbit interaction coefficient $C_s^0$ is finite at $\varepsilon\to0$
and needs no renormalization, just like $C_m^0$.
They are related by reparametrization invariance:
\begin{equation}
C_s = 2 C_m - 1\,.
\label{L2:repar}
\end{equation}
The electron magnetic moment $C_m = 1 + F_2(0)$ interacts with the magnetic field
in the electron rest frame;
this contribution to the spin--orbit interaction (the second line in~(\ref{L2:QED}))
is proportional to $C_m$.
But there is another contribution
due to the non-commutativity of the boosts to the $P$ and $P'$ rest frames
(the first line in~(\ref{L2:QED})).
This is a purely kinematic relativistic effect (Thomas precession);
it produces $-1$ in~(\ref{L2:repar}), and, naturally, there are no radiative corrections to it.
If we neglect radiative corrections ($F_2(0)$),
Thomas precession cancels $\frac{1}{2}$ of the magnetic-moment contribution.

The Darwin term is non-zero only inside the sources of the external electromagnetic field
(where $J^\nu = \partial_\mu F^{\mu\nu} \neq 0$).
Its coefficient is not fixed by reparametrization invariance:
it contains a new quantity $F_1'(0)$.
Unlike $F_2(0)$, $F_1'(0)$ is IR divergent starting from one loop.
Indeed, the full electron scattering cross section,
including both virtual corrections and real photon emission,
must be IR finite:
\begin{equation*}
1 + 2 F_1'(0) \frac{q^2}{M^2} + \int\limits_0^{\sim M} F(\omega)\,d\omega = \text{finite}\,,
\end{equation*}
where the soft photon emission probability $F(\omega)$ is given by~(\ref{HH:F})
with $\vartheta^2=-q^2/M^2$, and hence
\begin{equation}
F_1'(0) = - \frac{2}{3} \frac{\alpha}{4\pi\varepsilon} + \text{finite}\,.
\label{L2:IR}
\end{equation}
This divergence is IR;
as discussed above, loop corrections to the HEET scattering amplitude vanish
due to cancellation of UV and IR divergences.
IR behaviour of the full QED and HEET is the same;
therefore, $C_d^0$ contains the UV divergence,
which has to be removed by the UV renormalization constant $Z_d(\alpha(\mu))$.
This means that the Darwin interaction operator $O_d(\mu)$ and its coefficient $C_d(\mu)$
depend on $\mu$.

Let's briefly discuss an alternative method to derive the effective Lagrangian
which can be easily extended to higher orders in $1/M$ but only at the tree level
(see~\cite{N:94}).
We start from the QED Lagrangian
\begin{equation*}
L = \bar{\psi} (i\D-M) \psi\,.
\end{equation*}
The electron field can be written as
\begin{equation*}
\psi = e^{-iMv\cdot x} (h_v+H_v)\,,
\end{equation*}
where $\rlap/v h_v = h_v$, $\rlap/v H_v = -H_v$.
Then
\begin{equation*}
L = \bar{h}_v i v\cdot D h_v
+ \bar{H}_v i\D_\bot h_v
+ \bar{h}_v i\D_\bot H_v
- \bar{H}_v \left(2M+iv\cdot\D\right) H_v\,.
\end{equation*}
The equations of motion are
\begin{equation*}
\begin{split}
&i v\cdot D h_v = - i \D_\bot H_v\,,\\
&\left(2M + i v\cdot D\right) H_v = i \D_\bot h_v\,.
\end{split}
\end{equation*}
We can express the small field $H_v$ from the second equation~\cite{L:91}:
\begin{equation*}
H_v = \frac{i\D_\bot}{2M + i v\cdot D} h_v
= \frac{i\D_\bot}{2M} h_v + \mathcal{O}\left(\frac{1}{M^2}\right)
\end{equation*}
(this is equivalent to integrating $H_v$ out in the functional integral~\cite{MRR:92}).
Then
\begin{equation*}
L = \bar{h}_v i v\cdot D h_v - \frac{1}{2M} \bar{h}_v \D_\bot^2 h_v
= \bar{h}_v i v\cdot D h_v - \frac{1}{2M}  \bar{h}_v D_\bot^2 h_v
+ \frac{e}{4M} \bar{h}_v F_{\mu\nu} \sigma^{\mu\nu} h_v\,.
\end{equation*}
At higher orders in $1/M$, terms containing $v\cdot D$ appear;
they should be eliminated by appropriate field redefinitions.
One can also obtain the canonical Lagrangian containing $v\cdot D$ only in the leading term
by Foldy--Wouthuysen transformation of the field~\cite{KT:91}.
However, these algebraic methods cannot help us to derive loop corrections
in interaction coefficients in the Lagrangian.

\section{Muon magnetic moment}
\label{S:mu}

\begin{figure}[ht]
\begin{center}
\begin{picture}(74,22.232)
\put(16,13.732){\makebox(0,0){\includegraphics{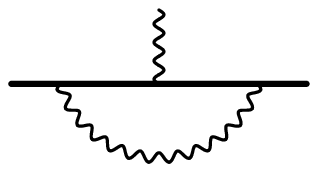}}}
\put(58,12.616){\makebox(0,0){\includegraphics{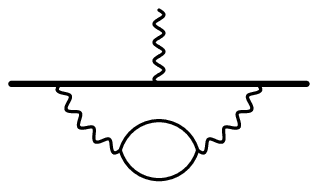}}}
\put(16,0){\makebox(0,0)[b]{a}}
\put(58,0){\makebox(0,0)[b]{b}}
\end{picture}
\end{center}
\caption{Muon anomalous magnetic moment:
(a) the leading diagram,
(b) the electron loop contribution}
\label{F:mu}
\end{figure}

Now we shall consider QED with muons having a large mass $M$
and electrons having a small mass $m$.
Namely, we'll discuss some contributions to the muon anomalous magnetic moment.
The leading one-loop term (Fig.~\ref{F:mu}a)
is given by the formula~(3.93) in~\cite{eft1}.
We shall concentrate on the electron-loop contribution (Fig.~\ref{F:mu}b).
The exact expression for it is~(3.94) in~\cite{eft1},
where we should substitute $m\to M$,
and $\Pi(k^2)$ is the electron-loop contribution to the photon self-energy.
According to the method of regions~\cite{S:02,J:11},
it is the sum of two contributions, hard and soft.

In the hard contribution $k\sim M$, $D_1\sim D_2\sim M^2$.
We can expand $\Pi(k^2)$ in Taylor series in $m^2$,
each term is a loop integral with a single scale $k^2\sim M^2$:
\begin{equation}
\Pi(k^2) = - 2 \frac{d-2}{d-1}
\frac{e_0^2 (-k^2)^{-\varepsilon}}{(4\pi)^{d/2}} G_1
\left[1 + \mathcal{O}\left(\frac{m^2}{k^2}\right) \right]\,,
\label{mu:Pi}
\end{equation}
where
\begin{equation*}
G_1 = \frac{\Gamma\left(2-\frac{d}{2}\right) \Gamma^2\left(\frac{d}{2}-1\right)}{\Gamma(d-2)}
= - \frac{2}{(d-3) (d-4)}
\frac{\Gamma(1+\varepsilon) \Gamma^2(1-\varepsilon)}{\Gamma(1-2\varepsilon)}\,.
\end{equation*}
Calculating the integral in $k$, we find the hard contribution
\begin{equation}
\frac{\mu_h}{\mu_0} =
32 \frac{(d-2) (d^2 - 7 d + 11)}{(d-1) (d-4) (d-5) (3d-8) (3d-10)}
\frac{e_0^2 M^{-2\varepsilon}}{(4\pi)^{d/2}} \Gamma(1+\varepsilon) R
\left[1 + \mathcal{O}\left(\frac{m^2}{M^2}\right) \right]\,,
\label{mu:h}
\end{equation}
where
\begin{equation*}
R = \frac{\Gamma(1+2\varepsilon) \Gamma^2(1-\varepsilon) \Gamma(1-4\varepsilon)}%
{\Gamma(1+\varepsilon) \Gamma(1-2\varepsilon) \Gamma(1-3\varepsilon)}
= 1 + \mathcal{O}(\varepsilon^2)\,.
\end{equation*}
Re-expressing via renormalized $\alpha(\mu)$ at $\mu=M$, we arrive at
\begin{equation}
\mu_0 + \mu_h = \frac{\alpha(M)}{2\pi}
\left[1 - \frac{25}{18} \frac{\alpha}{\pi} \right]\,.
\label{mu:hr}
\end{equation}
To this accuracy, $\alpha(M)=\alpha'(M)$,
where $\alpha'(M)$ is the \MS{} coupling in the effective theory without muons (see~\cite{eft1}):
\begin{equation*}
\alpha'(M) = \alpha'(m) \left(1 + \frac{2}{3} \frac{\alpha}{\pi} \log\frac{M}{m} \right)\,,
\end{equation*}
and $\alpha'(m)$ can be replaced by $\alpha_{\text{os}}$:
\begin{equation}
\mu_0 + \mu_h = \frac{\alpha_{\text{os}}}{2\pi}
\left[1 + \frac{2}{3} \frac{\alpha}{\pi}
\left( \log\frac{M}{m} - \frac{25}{12} \right) \right]\,.
\label{mu:hos}
\end{equation}
Of course, including other 2-loop diagrams will change the constant
added to the logarithm in~(\ref{mu:hos}).

Now we'll consider the soft region $k\sim m$, where
\begin{equation*}
D_1 = M D_s + D_2\,,\qquad
D_s = - 2 k\cdot v\,,\qquad
D_2 = - k^2
\end{equation*}
($D_s\sim m$, $D_2\sim m^2$).
We can expand the integrand in Taylor series in $1/M$;
each term is a loop integral with a single scale $m$:
\begin{equation}
\mu_s = \frac{- 2 i e_0^2}{(d-1) (d-2) M}
\int \frac{d^d k}{(2\pi)^d} \Pi(k^2)
\left[ \frac{8}{D_s^3} + \frac{d^2 - 4 d + 5}{D_s D_2}
+ \mathcal{O}\left(\frac{m}{M}\right) \right]\,.
\label{mu:s0}
\end{equation}
This is an HEET on-shell integral with a massive loop
($\Pi(k^2)$ where $k^2\sim m^2$).

\begin{figure}[ht]
\begin{center}
\begin{picture}(52,26)
\put(26,14.5){\makebox(0,0){\includegraphics{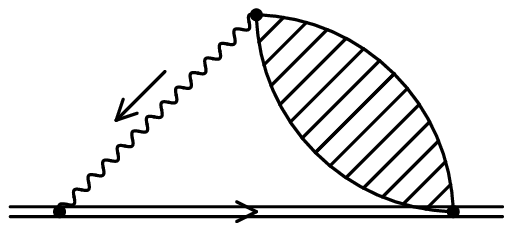}}}
\put(26,3){\makebox(0,0)[t]{$k$}}
\put(11,19){\makebox(0,0){$k$}}
\put(25,7){\makebox(0,0)[b]{1}}
\put(17.5,12.5){\makebox(0,0){2}}
\end{picture}
\end{center}
\caption{On-shell HEET diagrams}
\label{F:os1}
\end{figure}

Let's discuss the class of integrals (Fig.~\ref{F:os1})
\begin{equation}
F(n_1,n_2) = \frac{1}{i\pi^{d/2}} \int \frac{\Pi(k^2)\,d^d k}{D_1^{n_1} D_2^{n_2}}\,,\qquad
D_1 = -2k\cdot v-i0\,,\qquad
D_2 = -k^2-i0\,,
\label{mu:F}
\end{equation}
where $\Pi(k^2)$ is an arbitrary function~\cite{GSS:06}.
We can construct an identity in which $\Pi'(k^2)$ terms cancel:
\begin{equation*}
\frac{\partial}{\partial k} \cdot
\left( k - 2 \frac{D_2}{D_1} v \right)
\frac{\Pi(k^2)}{D_1^{n_1} D_2^{n_2}}
= \left[ d-n_1-2 - 4 (n_1+1) \frac{D_2}{D_1^2} \right]
\frac{\Pi(k^2)}{D_1^{n_1} D_2^{n_2}}\,.
\end{equation*}
Integrating it, we obtain an integration-by-parts relation
\begin{equation}
(d-n_1-2) F(n_1,n_2) =
4 (n_1+1) \mathbf{1}^{++} \mathbf{2}^- F(n_1,n_2)\,.
\label{mu:IBP}
\end{equation}

\begin{figure}[ht]
\begin{center}
\begin{picture}(92,52)
\put(46,26){\makebox(0,0){\includegraphics{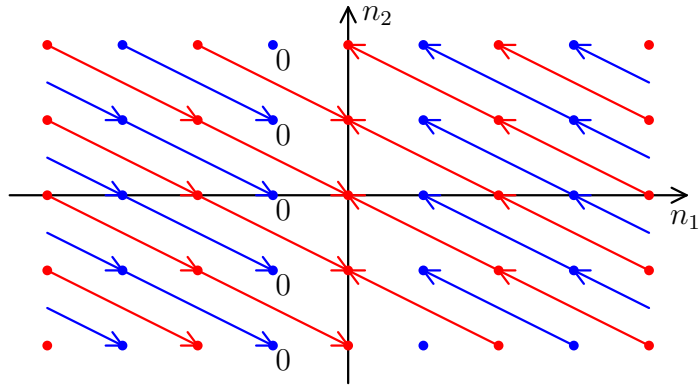}}}
\put(91,24){\makebox(0,0)[t]{$n_1$}}
\put(48,50){\makebox(0,0)[l]{$n_2$}}
\put(36.5,45.5){\makebox(0,0)[tl]{$0$}}
\put(36.5,35.5){\makebox(0,0)[tl]{$0$}}
\put(36.5,25.5){\makebox(0,0)[tl]{$0$}}
\put(36.5,15.5){\makebox(0,0)[tl]{$0$}}
\put(36.5,5.5){\makebox(0,0)[tl]{$0$}}
\end{picture}
\end{center}
\caption{Integration by parts}
\label{F:os2}
\end{figure}

Let's call integrals with even $n_1$ apparently even,
and with odd $n_1$ --- apparently odd
(they would be even and odd in $v$ if we neglected $i0$
in the denominator).
These two classes of integrals are not mixed
by the recurrence relation~(\ref{mu:IBP}).
We can use this relation to reduce all apparently even integrals
to vacuum integrals with $n_1=0$ (Fig.~\ref{F:os2}).
Apparently odd integrals with $n_1<0$ can be reduced to $n_1=-1$.
Substituting $n_1=-1$ to~(\ref{mu:IBP}),
we see that these integrals vanish,
and hence all integrals with odd $n_1<0$ vanish too.
Apparently odd integrals with $n_1>0$ can be reduced to $n_1=1$
(Fig.~\ref{F:os2});
however, they are not related to those with $n_1=-1$.

The solution of the recurrence relation can thus be written as
\begin{equation}
F(n_1,n_2) =
\left\{
\begin{array}{ll}
\displaystyle (-4)^{-n_1/2}
\frac{\Gamma\left(\frac{d}{2}\right)}{\Gamma\bigl(\frac{d-n_1}{2}\bigr)}
\frac{\Gamma\left(\frac{1-n_1}{2}\right)}{\Gamma\left(\frac{1}{2}\right)}
F\Bigl(0,n_2+\frac{n_1}{2}\Bigr)
& \mbox{even $n_1$}\,,\\
\displaystyle 2^{1-n_1}
\frac{\Gamma\left(\frac{d-1}{2}\right)}%
{\Gamma\left(\frac{n_1+1}{2}\right)\Gamma\bigl(\frac{d-n_1}{2}\bigr)}
F\Bigl(1,n_2+\frac{n_1-1}{2}\Bigr)
& \mbox{odd $n_1>0$}\,,\\
\displaystyle 0 & \mbox{odd $n_1<0$}\,.
\end{array}
\right.
\label{mu:sol}
\end{equation}
Some of these properties can be understood more directly.
If $n_1<0$, $i0$ in $D_1^{-n_1}$ can be safely neglected;
averaging this factor over $k$ directions, we obtain $0$ for odd $n_1$
and the upper formula in~(\ref{mu:sol}) for even $n_1$.
We see that this formula also holds for even $n_1>0$.

\begin{figure}[b]
\begin{center}
\begin{picture}(52,26)
\put(26,14.5){\makebox(0,0){\includegraphics{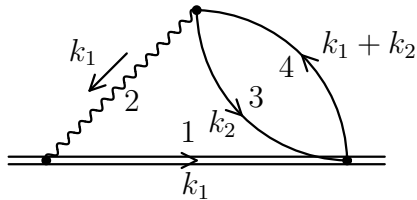}}}
\put(26,3){\makebox(0,0)[t]{$k_1$}}
\put(11,19){\makebox(0,0){$k_1$}}
\put(29.5,9.5){\makebox(0,0){$k_2$}}
\put(49,20){\makebox(0,0){$k_1+k_2$}}
\put(25,7){\makebox(0,0)[b]{1}}
\put(17.5,12.5){\makebox(0,0){2}}
\put(34,13){\makebox(0,0){3}}
\put(38,17){\makebox(0,0){4}}
\end{picture}
\end{center}
\caption{Two-loop on-shell HEET integrals with mass}
\label{F:os3}
\end{figure}

Now let's consider the 2-loop diagram (Fig.~\ref{F:os3})
\begin{equation}
F(n_1,n_2,n_3,n_4) = \frac{1}{(i\pi^{d/2})^2} \int
\frac{d^d k_1\,d^d k_2}{D_1^{n_1} D_2^{n_2} D_3^{n_3} D_4^{n_4}}\,,
\label{M2:def}
\end{equation}
where
\begin{align*}
&D_1 = -2 k_1\cdot v - i0\,,\quad
D_2 = -k_1^2 - i0\,,\\
&D_3 = 1 - k_2^2 - i0\,,\quad
D_4 = 1 - (k_1+k_2)^2 - i0\,.
\end{align*}
It is symmetric with respect to $3\leftrightarrow4$,
and vanishes if $n_3$ or $n_4$ is integer and non-positive.
It can be calculated using $\alpha$ parametrization~\cite{GSS:06}:
\begin{align}
&F(n_1,n_2,n_3,n_4) ={}
\label{mu:F2}\\
&\frac{\Gamma\bigl(\frac{n_1}{2}\bigr)
\Gamma\bigl(\frac{d-n_1}{2}-n_2\bigr)
\Gamma\bigl(\frac{n_1-d}{2}+n_2+n_3\bigr)
\Gamma\bigl(\frac{n_1-d}{2}+n_2+n_4\bigr)
\Gamma\bigl(\frac{n_1}{2}+n_2+n_3+n_4-d\bigr)}%
{2\Gamma(n_1) \Gamma(n_3) \Gamma(n_4)
\Gamma\bigl(\frac{d-n_1}{2}\bigr)
\Gamma(n_1+2n_2+n_3+n_4-d)}\,.
\nonumber
\end{align}
In full accordance with~(\ref{mu:sol}),
integrals $F(n_1,n_2,n_3,n_4)$ with even $n_1$
reduce to the well-known 2-loop vacuum integral $F(0,n_2+n_1/2,n_3,n_4)$,
and thus to the master integral
\begin{equation}
I_0^2 = \raisebox{-9.8mm}{\includegraphics{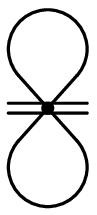}}\,;
\label{mu:even}
\end{equation}
those with odd $n_1<0$ vanish,
and with odd $n_1>0$ reduce to $F(1,n_2+(n_1-1)/2,n_3,n_4)$,
and thus to the master integral
\begin{equation}
J_0 = \raisebox{-5.8mm}{\includegraphics{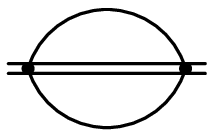}}
= 2^{4d-9} \pi^2 \frac{\Gamma(5-2d)}{\Gamma^2\left(2-\frac{d}{2}\right)}\,.
\label{mu:odd}
\end{equation}

Now we can easily calculate the leading soft contribution~(\ref{mu:s0}).
It is suppressed by the factor $m/M$, and is given by the apparently odd integral~(\ref{mu:odd}):
\begin{equation}
\mu_s = \frac{- 2 i e_0^2}{M}
\int \frac{d^d k}{(2\pi)^d} \frac{\Pi(k^2)}{D_s D_2}
= \frac{\alpha^2}{4} \frac{m}{M}\,.
\label{mu:s}
\end{equation}

\section{Heavy quark effective theory}
\label{S:QCD}

\subsection{HQET propagator}
\label{S:HQET}

QCD problems with a single heavy quark having small characteristic residual momentum
can be described by Heavy Quark Effective Theory (HQET)
which is the non-abelian version of the Bloch--Nordsieck effective theory.
At the leading order in $1/M$ its Lagrangian is~(\ref{Photonia:L});
we can rotate or switch off the heavy-quark spin without changing physics,
and use the simpler Lagrangian~(\ref{Photonia:spin0}).

Unlike the abelian case, the simple exponentiation~(\ref{HEET:S})
is no longer valid for the propagator in coordinate space.
A more complicated non-abelian exponentiation~\cite{GFT:84}
is valid instead:
\begin{align}
&S(t) = -i\theta(t) \exp \biggl[
C_F \frac{g_0^2}{(4\pi)^{d/2}} \left(\frac{it}{2}\right)^{2\varepsilon} S
+ C_F \frac{g_0^4}{(4\pi)^d} \left(\frac{it}{2}\right)^{4\varepsilon}
\left(C_A S_A + T_F n_l S_l\right)
\label{HQET:NAexp}\\
&{} + C_F \frac{g_0^6}{(4\pi)^{3d/2}} \left(\frac{it}{2}\right)^{6\varepsilon}
\Bigl(C_A^2 S_{AA}
+ C_F T_F n_l S_{Fl} + C_A T_F n_l S_{Al}
+ \left(T_F n_l\right)^2 S_{ll}\Bigr)
+ \cdots \biggr]\,.
\nonumber
\end{align}
Not all possible colour structures appear in the exponent,
but only maximally non-abelian (also called colour-connected) ones.

\begin{figure}[ht]
\begin{center}
\begin{picture}(107,37)
\put(53.5,18.5){\makebox(0,0){\includegraphics{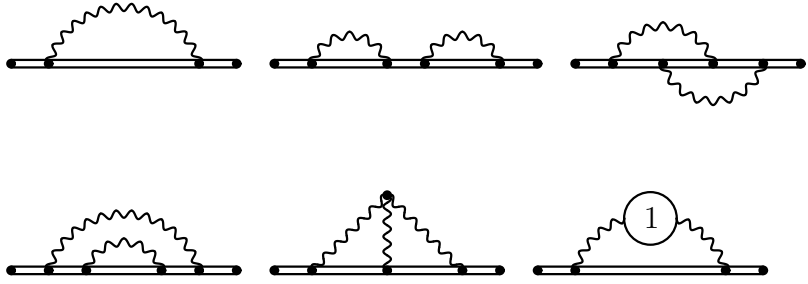}}}
\put(86,8){\makebox(0,0){1}}
\end{picture}
\end{center}
\caption{Heavy-quark propagator up to 2 loops}
\label{F:prop2}
\end{figure}

Diagrams for the HQET propagator up to 2 loops are shown in Fig.~\ref{F:prop2}.
If the colour factors of first 3 2-loop diagrams
were the same as that of the one-particle-reducible diagram,
i.\,e.\ equal to the square of the colour factor $C_F$
of the one-loop diagram (as in the abelian case),
then the sum of these diagrams would be equal to
$\frac{1}{2}$ of the square of the one-loop correction $S_F$.
However, the colour factor of the second 2-loop diagram
differs from $C_F^2$ by $-C_F C_A/2$,
which is the colour factor of the diagram with a 3-gluon vertex:
\begin{equation*}
\raisebox{-0.25mm}{\includegraphics{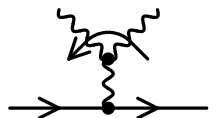}}
= \raisebox{-0.25mm}{\includegraphics{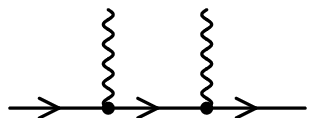}}
- \raisebox{-0.25mm}{\includegraphics{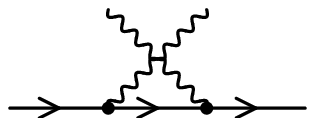}}\,,
\end{equation*}
or $[t^a,t^b] = i f^{abc} t^c$.
We should include this contribution
with $-C_F C_A/2$ instead of its full colour factor
into the term $S_{FA}$.
Of course, the diagram with 3-gluon vertex also contributes to $S_{FA}$.
The diagrams with the one-loop gluon self-energy
contribute to $S_{Fl}$ (quark loop) and $S_{FA}$ (gluon and ghost loops).

\begin{figure}[ht]
\begin{center}
\begin{picture}(62,30)
\put(31,16.5){\makebox(0,0){\includegraphics{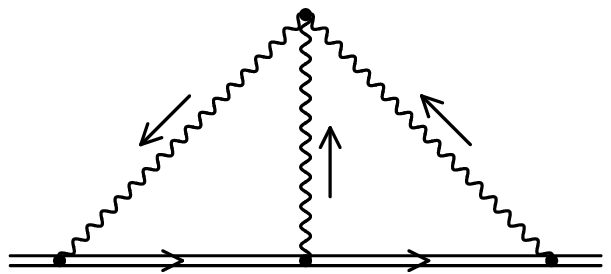}}}
\put(18.5,0){\makebox(0,0)[b]{{$k_{10}+\omega$}}}
\put(41,0){\makebox(0,0)[b]{{$k_{20}+\omega$}}}
\put(14.25,20.75){\makebox(0,0){{$k_1$}}}
\put(47.75,20.75){\makebox(0,0){{$k_2$}}}
\put(35,11.5){\makebox(0,0)[l]{{$k_1-k_2$}}}
\put(18.5,8){\makebox(0,0)[t]{{$n_1$}}}
\put(41,8){\makebox(0,0)[t]{{$n_2$}}}
\put(21,14){\makebox(0,0){{$n_3$}}}
\put(41,14){\makebox(0,0){{$n_4$}}}
\put(30,12){\makebox(0,0)[r]{{$n_5$}}}
\end{picture}
\end{center}
\caption{Two-loop diagram 1}
\label{F:d1}
\end{figure}

Now we shall discuss how to calculate 2-loop propagator diagrams in HQET~\cite{BG:91}.
There are 2 topologies of such diagrams.
The first one is (Fig.~\ref{F:d1})
\begin{align}
&\frac{1}{(i\pi^{d/2})^2} \int \frac{d^d k_1\,d^d k_2}%
{D_1^{n_1} D_2^{n_2} D_3^{n_3} D_4^{n_4} D_5^{n_5}} =
I(n_1,n_2,n_3,n_4,n_5) (-2\omega)^{2d-n_1-n_2-2(n_3+n_4+n_5)}\,,
\label{HQET:d1}\\
&D_1=-2(k_{10}+\omega)\,,\quad
D_2=-2(k_{20}+\omega)\,,\quad
D_3=-k_1^2\,,\quad
D_4=-k_2^2\,,\quad
D_5=-(k_1-k_2)^2\,.
\nonumber
\end{align}
It is symmetric with respect to $1\leftrightarrow2$ and $3\leftrightarrow4$,
and vanishes if two adjacent indices are $\le0$.

If $n_5=0$,
\begin{equation}
I(n_1,n_2,n_3,n_4,0) =
\raisebox{-7.25mm}{\begin{picture}(52,15.5)
\put(26,7.75){\makebox(0,0){\includegraphics{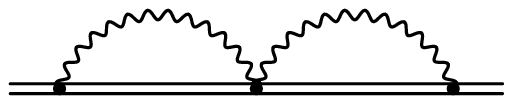}}}
\put(16,0){\makebox(0,0)[b]{{$n_1$}}}
\put(36,0){\makebox(0,0)[b]{{$n_2$}}}
\put(16,15.5){\makebox(0,0)[t]{{$n_3$}}}
\put(36,15.5){\makebox(0,0)[t]{{$n_4$}}}
\end{picture}}
= I(n_1,n_3) I(n_2,n_4)\,.
\label{HQET:d1n5}
\end{equation}
If $n_1=0$, the inner loop gives $G(n_3,n_5) (-p^2)^{d/2-n_3-n_5}$, and
\begin{equation}
\begin{split}
&I(0,n_2,n_3,n_4,n_5)
= \raisebox{-12mm}{\begin{picture}(52,25)
\put(26,14){\makebox(0,0){\includegraphics{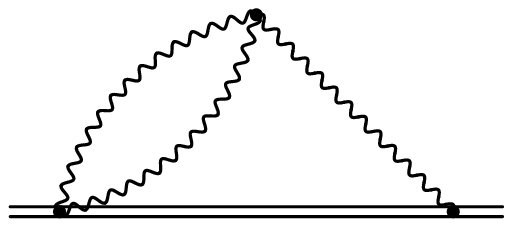}}}
\put(26,0){\makebox(0,0)[b]{$n_2$}}
\put(38.5,15.5){\makebox(0,0){$n_4$}}
\put(10.5,18.5){\makebox(0,0){$n_3$}}
\put(21.5,9){\makebox(0,0){$n_5$}}
\end{picture}}\\
&{}= \raisebox{-11mm}{\begin{picture}(32,23)
\put(16,11.5){\makebox(0,0){\includegraphics{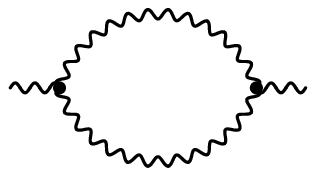}}}
\put(16,0){\makebox(0,0)[b]{$n_5$}}
\put(16,23){\makebox(0,0)[t]{$n_3$}}
\end{picture}} \times
\raisebox{-7.25mm}{\begin{picture}(32,16)
\put(16,7.75){\makebox(0,0){\includegraphics{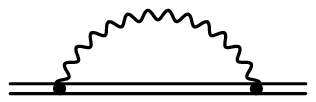}}}
\put(16,0){\makebox(0,0)[b]{$n_2$}}
\put(16,16){\makebox(0,0)[t]{$n_4+n_3+n_5-d/2$}}
\end{picture}}\\
&{}= G(n_3,n_5) I(n_2,n_4+n_3+n_5-d/2)
\end{split}
\label{HQET:d1n1}
\end{equation}
(the case $n_2=0$ is symmetric).
If $n_3=0$, the inner loop gives $I(n_1,n_5) (-2\omega)^{d-n_1-2n_5}$, and
\begin{equation}
\begin{split}
&I(n_1,n_2,0,n_4,n_5)
= \raisebox{-11mm}{\begin{picture}(52,23)
\put(26,11.5){\makebox(0,0){\includegraphics{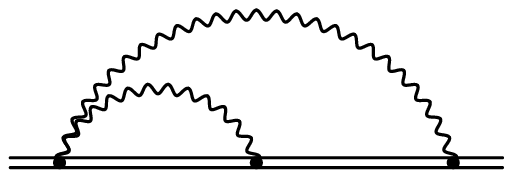}}}
\put(16,0){\makebox(0,0)[b]{$n_1$}}
\put(36,0){\makebox(0,0)[b]{$n_2$}}
\put(26,23){\makebox(0,0)[t]{$n_4$}}
\put(25.5,10.5){\makebox(0,0){$n_5$}}
\end{picture}}\\
&{}= \raisebox{-7.25mm}{\begin{picture}(32,16)
\put(16,7.75){\makebox(0,0){\includegraphics{fig33.eps}}}
\put(16,0){\makebox(0,0)[b]{$n_1$}}
\put(16,16){\makebox(0,0)[t]{$n_5$}}
\end{picture}} \times
\raisebox{-7.25mm}{\begin{picture}(32,16)
\put(16,7.75){\makebox(0,0){\includegraphics{fig33.eps}}}
\put(16,0){\makebox(0,0)[b]{$n_2+n_1+2n_5-d$}}
\put(16,16){\makebox(0,0)[t]{$n_3$}}
\end{picture}}\\
&{}= I(n_1,n_5) I(n_2+n_1+2n_5-d,n_4)
\end{split}
\label{HQET:d1n3}
\end{equation}
(the case $n_4=0$ is symmetric).

But what can we do when all the indices are positive?
We use integration by parts~\cite{BG:91}.
Applying $(\partial/\partial k_2)\cdot k_2$ or $(\partial/\partial k_2)\cdot(k_2-k_1)$
to the integrand of~(\ref{HQET:d1}) we obtain vanishing integrals.
On the other hand, we can calculate the derivatives explicitly;
using $2 k_2\cdot v=-D_2-2\omega$ and $2(k_2-k_1)\cdot k_2=D_3-D_4-D_5$,
we get~(\ref{HQET:d1}) with
\begin{align*}
&d-n_2-n_5-2n_4 - 2 \omega \frac{n_2}{D_2} + \frac{n_5}{D_5}(D_3-D_4)\,,\\
&d-n_2-n_4-2n_5 + \frac{n_2}{D_2}D_1 + \frac{n_4}{D_4}(D_3-D_5)
\end{align*}
inserted under the integral sign.
This means that $I$~(\ref{HQET:d1}) satisfies the recurrence relations
\begin{align}
&\left[ d-n_2-n_5-2n_4 + n_2\2+ + n_5\5+(\3--\4-) \right] I = 0\,,
\label{HQET:IBP1}\\
&\left[ d-n_2-n_4-2n_5 + n_2\2+\1- + n_4\4+(\3--\5-) \right] I = 0\,.
\label{HQET:IBP2}
\end{align}
Applying $(\partial/\partial k_2)\cdot v$ we derive
\begin{equation}
\left[ -2n_2\2+ + n_4\4+(\2--1) + n_5\5+(\2--\1-) \right] I = 0\,.
\label{HQET:IBP3}
\end{equation}
There is also the homogeneity relation.
Let's apply $\omega(d/d\omega)$ to~(\ref{HQET:d1}):
\begin{equation}
\left[ 2(d-n_3-n_4-n_5)-n_1-n_2 + n_1\1+ + n_2\2+ \right] I = 0\,.
\label{HQET:Hom}
\end{equation}
This is the sum of the $(\partial/\partial k_2)\cdot k_2$ relation~(\ref{HQET:IBP1})
and the symmetric $(\partial/\partial k_1)\cdot k_1$ one.

\begin{figure}[ht]
\begin{center}
\begin{picture}(150,50)
\put(25,25){\makebox(0,0){\includegraphics[width=50mm]{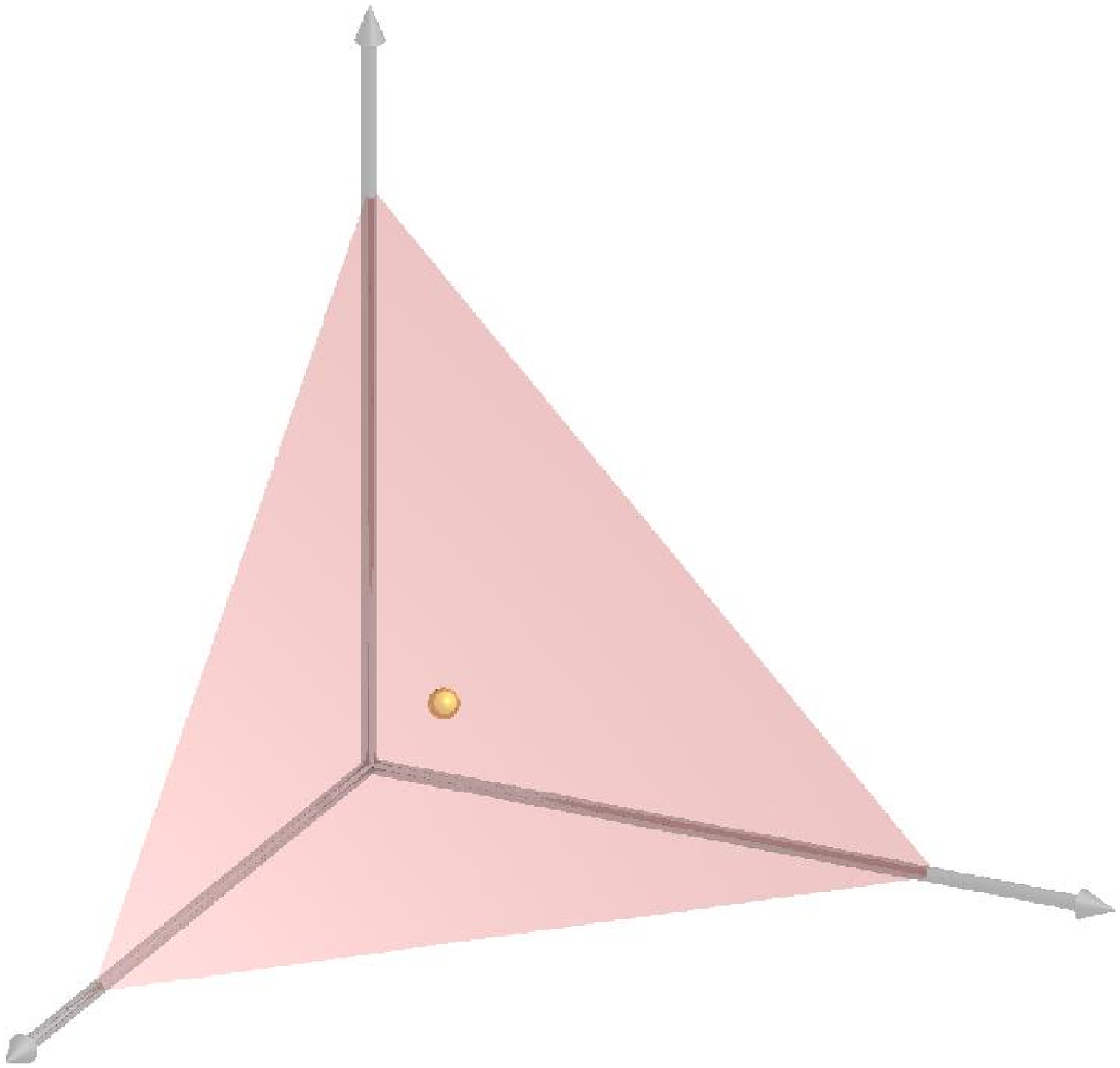}}}
\put(75,25){\makebox(0,0){\includegraphics[width=50mm]{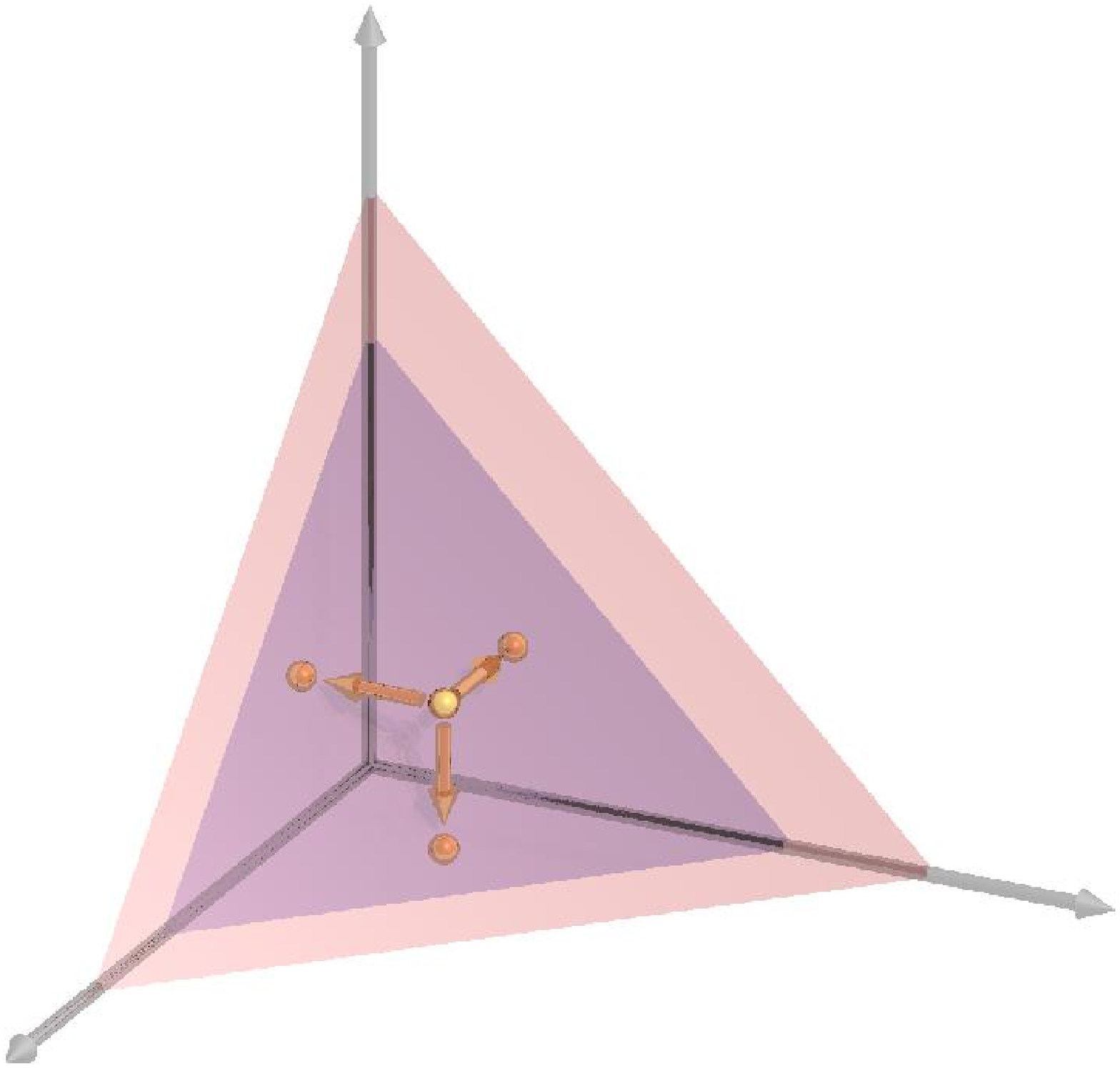}}}
\put(125,25){\makebox(0,0){\includegraphics[width=50mm]{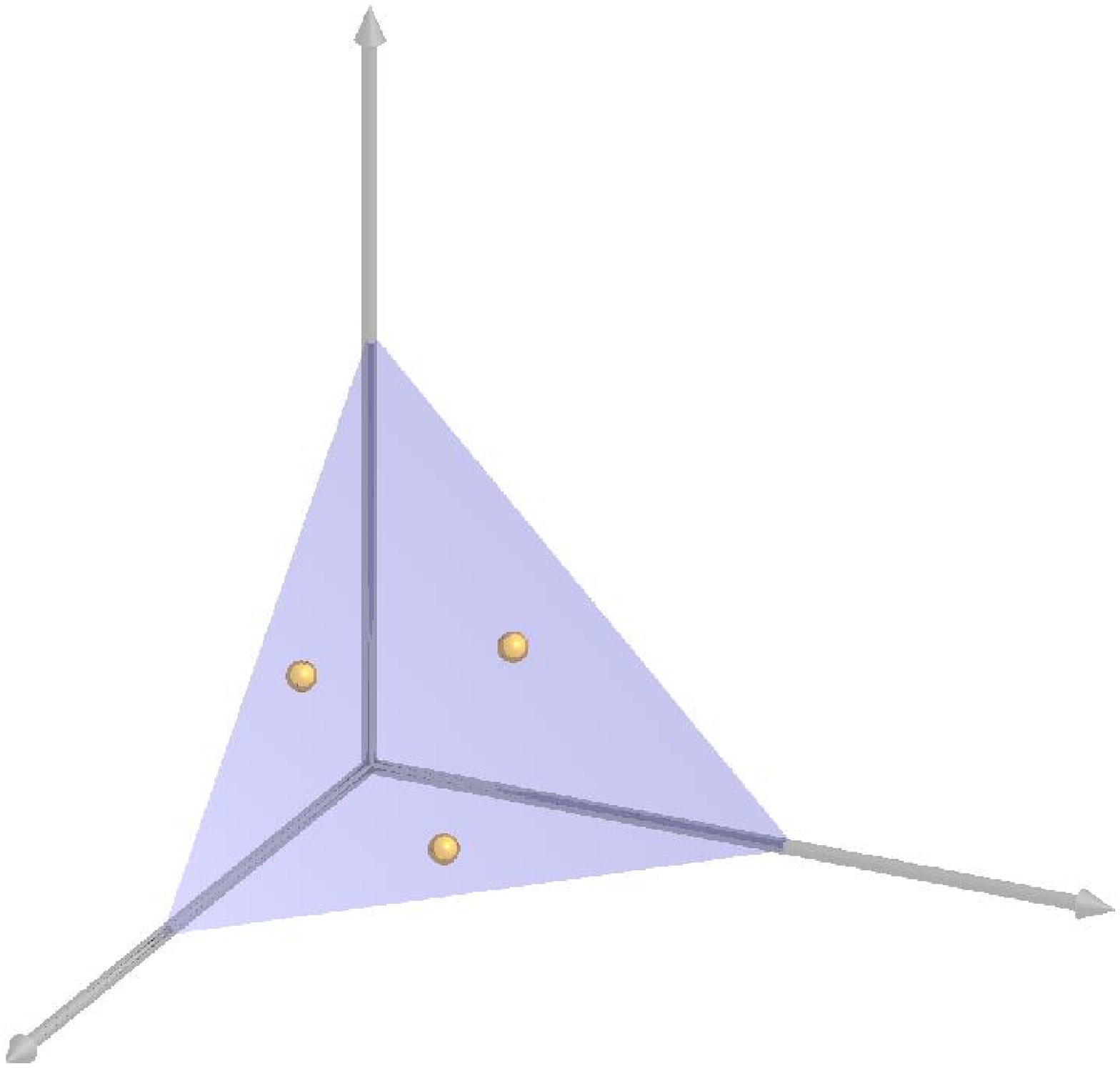}}}
\end{picture}
\end{center}
\caption{Integration by parts}
\label{F:IBP}
\end{figure}

A useful recurrence relation can be obtained by subtracting the $\1-$ shifted homogeneity relation~(\ref{HQET:Hom})
from the $(\partial/\partial k_2)\cdot(k_2-k_1)$ relation~(\ref{HQET:IBP2}):
\begin{equation}
\bigl[d-n_1-n_2-n_4-2n_5+1
- \bigl(2(d-n_3-n_4-n_5)-n_1-n_2+1\bigr)\1-
+ n_4\4+(\3--\5-) \bigr] I = 0\,.
\label{HQET:David}
\end{equation}
Let's solve it for $I(n_1,n_2,n_3,n_4,n_5)$:
\begin{equation*}
I = \frac{(2(d-n_3-n_4-n_5)-n_1-n_2+1)\1- + n_4\4+(\5--\3-)}%
{d-n_1-n_2-n_4-2n_5+1} I\,.
\end{equation*}
Each application of this relation reduces $n_1+n_3+n_5$ by 1 (Fig.~\ref{F:IBP}).
Therefore, after a finite number of steps, any integral $I$
will reduce to the boundary cases~(\ref{HQET:d1n5}), (\ref{HQET:d1n1}), (\ref{HQET:d1n3}).

\begin{figure}[ht]
\begin{center}
\begin{picture}(84,50)
\put(42,25){\makebox(0,0){\includegraphics{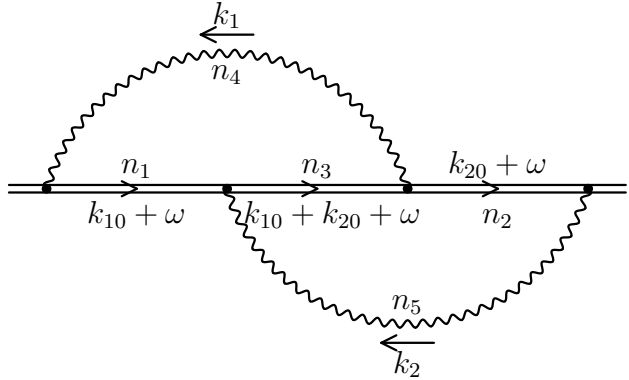}}}
\put(30,50){\makebox(0,0)[t]{{$k_1$}}}
\put(54,0){\makebox(0,0)[b]{{$k_2$}}}
\put(18,23.5){\makebox(0,0)[t]{{$k_{10}+\omega$}}}
\put(44,23.5){\makebox(0,0)[t]{{$k_{10}+k_{20}+\omega$}}}
\put(66,26.5){\makebox(0,0)[b]{{$k_{20}+\omega$}}}
\put(18,26.5){\makebox(0,0)[b]{{$n_1$}}}
\put(42,26.5){\makebox(0,0)[b]{{$n_3$}}}
\put(66,23.5){\makebox(0,0)[t]{{$\vphantom{k}n_2$}}}
\put(30,41.5){\makebox(0,0)[t]{{$n_4$}}}
\put(54,8.5){\makebox(0,0)[b]{{$n_5$}}}
\end{picture}
\end{center}
\caption{Two-loop diagram 2}
\label{F:d2}
\end{figure}

The second topology is (Fig.~\ref{F:d2})
\begin{equation}
\begin{split}
&\frac{1}{(i\pi^{d/2})^2} \int \frac{d^d k_1\,d^d k_2}%
{D_1^{n_1} D_2^{n_2} D_3^{n_3} D_4^{n_4} D_5^{n_5}} =
J(n_1,n_2,n_3,n_4,n_5) (-2\omega)^{2d-n_1-n_2-n_3-2(n_4+n_5)}\,,\\
&D_1=-2(k_{10}+\omega)\,,\quad
D_2=-2(k_{20}+\omega)\,,\quad
D_3=-2(k_{10}+k_{20}+\omega)\,,\\
&D_4=-k_1^2\,,\quad
D_5=-k_2^2\,.
\end{split}
\label{HQET:d2}
\end{equation}
It is symmetric with respect to $1\leftrightarrow2$, $4\leftrightarrow5$;
it vanishes if $n_4\le0$ or $n_5\le0$ or two adjacent heavy indices ($1\ldots3$) are $\le0$.

If $n_3=0$, it is given by~(\ref{HQET:d1n5});
if $n_1=0$ --- by~(\ref{HQET:d1n3}) (the case $n_2=0$ is symmetric).
In general, we use the fact that the denominators are linearly dependent:
$D_1 + D_2 - D_3 = -2\omega$.
Therefore
\begin{equation}
J = (\1-+\2--\3-) J\,.
\label{HQET:parfrac}
\end{equation}
This relation reduces $n_1+n_2+n_3$ by 1 (Fig.~\ref{F:IBP});
therefore, after a finite number of steps, any integral $J$
will reduce to the boundary cases.
In principle, the integral~(\ref{HQET:d2}) can contain a numerator $(k_1\cdot k_2)^n$
which does not reduce to the denominators.
This wider class of integrals can also be easily calculated~\cite{G:00}.

\begin{figure}[ht]
\begin{center}
\begin{picture}(84,17)
\put(16,8.5){\makebox(0,0){\includegraphics{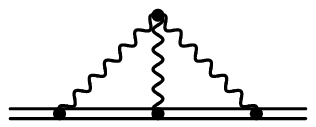}}}
\put(63,8.5){\makebox(0,0){\includegraphics{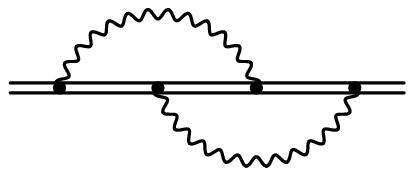}}}
\end{picture}
\end{center}
\caption{Generic topologies of 2-loop propagator diagrams in HQET}
\label{F:Top2}
\end{figure}

Let's summarize.
There are 2 generic topologies of 2-loop propagator integrals in HQET (Fig.~\ref{F:Top2}).
For all integer indices $n_i$ they can be reduced to 2 master integrals
\begin{equation}
\raisebox{-3.75mm}{\includegraphics{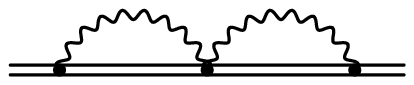}} = I_1^2\,,\qquad
\raisebox{-4.5mm}{\includegraphics{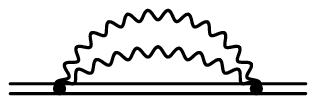}} = I_2
\end{equation}
(all $n_i=1$ here).
Here the $L$-loop HQET sunset is
\begin{equation}
\raisebox{-5.5mm}{\begin{picture}(32,12)
\put(16,6){\makebox(0,0){\includegraphics{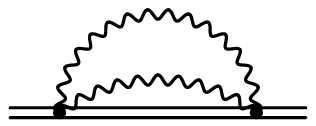}}}
\put(16,7.6666667){\makebox(0,0){{$\cdots$}}}
\end{picture}}
= I_L
= \Gamma(2L+1-Ld) \Gamma^L\left(\tfrac{d}{2}-1\right)
= \frac{\Gamma(1+2L\varepsilon)\Gamma^L(1-\varepsilon)}{(1-L(d-2))_{2L}}\,.
\label{HQET:sunset}
\end{equation}

Now it is easy to calculate the heavy-quark self-energy up to 2 loops~\cite{BG:91}:
\begin{equation}
\begin{split}
&\Sigma(\omega) = - C_F \frac{g_0^2 (-2\omega)^{1-2\varepsilon}}{(4\pi)^{d/2}} (d-3) I_1 A\\
&{} + C_F \frac{g_0^4 (-2\omega)^{1-4\varepsilon}}{(4\pi)^d} \biggl\{
- 16 \frac{(d-2) (2d-5)}{(d-4) (d-6)} I_2 P
- C_F \frac{4 (d-3)^2 (2d-5)}{d-4} I_2 A^2\\
&\qquad{} + \left( C_F - \frac{C_A}{2} \right) 2 (d-3)
\left[ (d-3) I_1^2 - 2 (2d-5) I_2 \right] A^2\\
&\qquad{} - C_A (d-3) \left[ (d-3) I_1^2
+ 2 \frac{2d-5}{d-4} I_2 \right] A (1-a_0) \biggr\}\,,
\end{split}
\label{HQET:Sigma2}
\end{equation}
where
\begin{equation*}
A = a_0 - 1 - \frac{2}{d-3}\,,
\end{equation*}
and the one-loop self-energy insertion into the gluon propagator is proportional to
\begin{equation*}
P = T_F n_l - \frac{3d-2 + (d-1) (2d-7) \xi - \frac{1}{4} (d-1) (d-4) \xi^2}{4 (d-2)} C_A
\end{equation*}
($\xi = 1-a_0$).

The heavy-quark propagator is
\begin{equation}
\begin{split}
&\omega S(\omega) = 1
+ C_F \frac{g_0^2 (-2\omega)^{-2\varepsilon}}{(4\pi)^{d/2}} 2 (d-3) I_1 A\\
&{} + C_F \frac{g_0^4 (-2\omega)^{-4\varepsilon}}{(4\pi)^d} \biggl\{
32 \frac{(d-2) (2d-5)}{(d-4) (d-6)} T_F n_l I_2\\
&\qquad{} + 8 \frac{(d-3) (2d-5) (2d-7)}{d-4} A^2 C_F I_2
- 4 (d-3) A C_A I_1^2\\
&\qquad{} + 8 \frac{(2d-5) (2d-7)}{(d-3) (d-4) (d-6)}\\
&\qquad\biggl[ \frac{(d-2)^2 (d-5)}{(d-3) (2d-7)}
+ (d^2 - 4d + 5) A
- \frac{1}{4} (d-3) (d^2 - 9d + 16) A^2 \biggr] C_A I_2 \biggr\}\,.
\end{split}
\label{HQET:Sw}
\end{equation}
Fourier-transforming it to coordinate space we obtain
\begin{align}
&S(t) = - i \theta(t) \exp \biggl\{
C_F \frac{g_0^2}{(4\pi)^{d/2}} \left(\frac{it}{2}\right)^{2\varepsilon} \Gamma(-\varepsilon) A
\nonumber\\
&{} + C_F \frac{g_0^4}{(4\pi)^d} \left(\frac{it}{2}\right)^{4\varepsilon} \Gamma^2(-\varepsilon)
\biggl[ 2 \frac{d-2}{(d-3) (d-6) (2d-7)} T_F n_l
\nonumber\\
&\quad{} + \frac{1}{2 (d-3)^2 (d-6)} \biggl( \frac{(d-2)^2 (d-5)}{(d-3) (2d-7)}
+ (d^2 - 4d + 5) A
- \frac{1}{4} (d-3) (d^2 - 9d + 16) A^2 \biggr) C_A
\nonumber\\
&\quad{} - \frac{A}{d-3} \frac{\Gamma^2(1+2\varepsilon)}{\Gamma(1+4\varepsilon)} C_A
\biggr] \biggr\}\,.
\label{HQET:St}
\end{align}
This result has the structure~(\ref{HQET:NAexp})
thus providing a strong check of~(\ref{HQET:Sigma2}).
After re-expressing the propagator~(\ref{HQET:St}) via the renormalized quantities $\alpha_s(\mu)$, $a(\mu)$
it still has the exponential form with the same colour structures.
This means that the wave-function renormalization constant has such a form, too,
and in the anomalous dimensions only maximally non-abelian (colour-connected) structures appear:
\begin{equation}
\gamma_h = 2 C_F (a-3) \frac{\alpha_s}{4\pi}
+ C_F \left[C_A \left(\frac{a^2}{2}+4a-\frac{179}{6}\right) + \frac{32}{3} T_F n_l \right]
\left(\frac{\alpha_s}{4\pi}\right)^2
+ \cdots
\label{HQET:gamma}
\end{equation}

The heavy--quark propagator and $\gamma_h$ are calculated up to 3 loops in HQET~\cite{CG:03}
(there you can find a detailed discussion of the colour structures in the exponent~(\ref{HQET:NAexp})
at 3 loops);
$\gamma_h$ was first found earlier~\cite{MR:00} from an on-shell massive QCD calculation
and the requirement that the matching coefficient $z(\mu)$ (Sect.~\ref{S:El}) is finite.

\subsection{Heavy--light currents}
\label{S:HL}

Now we shall consider the QCD operators
\begin{equation}
j_0 = \bar{q}_0 \Gamma Q_0 = Z_j(\mu) j(\mu)\,,
\label{HL:j}
\end{equation}
where $\Gamma$ is a Dirac matrix.
They can be expressed via HQET operators:
\begin{equation}
j(\mu) = e^{-iMv\cdot x} \left[ C_\Gamma(\mu) \tilde{\jmath}(\mu)
+ \frac{1}{2M} \sum_i B_i(\mu) O_i(\mu) + \cdots \right]\,,
\label{HL:exp}
\end{equation}
where
\begin{equation}
\tilde{\jmath}_0 = \bar{q}_0 \Gamma h_{v0} = \tilde{Z}_j(\mu) \tilde{\jmath}(\mu)
\label{HL:jtilde}
\end{equation}
is the HQET heavy--light current,
and $O_i$ are dimension-4 operators with appropriate quantum numbers.
We shall not discuss $1/M$ corrections here.

First we discuss the HQET currents~(\ref{HL:jtilde}).
The Dirac matrix $\Gamma$ can be moved outside any diagram with such a current.
Therefore, we can consider the current $\tilde{\jmath}_0=\bar{q}_0\varphi_{v0}$
with the spin-0 heavy quark instead.
The vertex function
\begin{equation}
\tilde{\Gamma}(\omega,p) = 1 + \tilde{\Lambda}(\omega,p)
= \raisebox{-4.75mm}{\begin{picture}(32,7)
\put(16,3.5){\makebox(0,0){\includegraphics{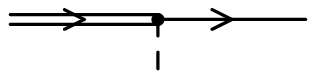}}}
\put(8.5,4){\makebox(0,0)[t]{$\omega$}}
\put(23.5,4){\makebox(0,0)[t]{$p$}}
\end{picture}}
+ \raisebox{-4.75mm}{\begin{picture}(32,14.5)
\put(16,7.25){\makebox(0,0){\includegraphics{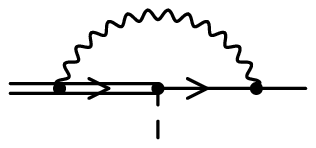}}}
\end{picture}}
+ \cdots
\label{HL:Gamma}
\end{equation}
should be $\tilde{Z}_\Gamma \tilde{\Gamma}_r$,
where $\tilde{Z}_\Gamma$ is a minimal renormalization constant,
and $\tilde{\Gamma}_r$ is finite at $\varepsilon\to0$.
The UV divergence of $\tilde{\Lambda}(\omega,p)$ does not depend on its external momenta,
and we may set them to 0:
\begin{equation*}
\begin{split}
\tilde{\Lambda}(0,0) &{}= - i C_F g_0^2 \int \frac{d^d k}{(2\pi)^d}
\frac{\gamma^\mu \rlap/k v^\nu}{k_0 (k^2)^2} \left[g_{\mu\nu} - (1-a_0) \frac{k_\mu k_\nu}{k^2}\right]
= - i C_F g_0^2 a_0 \int \frac{d^d k}{(2\pi)^d} \frac{1}{(k^2)^2}\\
&{}= C_F a \frac{\alpha_s}{4\pi\varepsilon}\,,
\end{split}
\end{equation*}
where we took into account $\rlap/k=k_0\gamma^0-\vec{k}\cdot\vec{\gamma}$,
and the integral of $\vec{k}$ vanishes
(of course, some IR regulator is implied here).
Then
\begin{equation*}
\tilde{Z}_j = Z_q^{1/2} Z_h^{1/2} \tilde{Z}_\Gamma = 1 + \frac{3}{2} C_F \frac{\alpha_s}{4\pi\varepsilon}
\end{equation*}
is gauge independent, and we obtain the 1-loop anomalous dimension.

The 2-loop vertex $\Lambda(\omega,0)$ can be calculated using the methods described in Sect.~\ref{S:HQET};
the anomalous dimension is~\cite{BG:91,JM:91}
\begin{equation}
\tilde{\gamma}_j = - 3 C_F \frac{\alpha_s}{4\pi}
+ C_F \left[ \left(-\frac{8}{3}\pi^2+\frac{5}{2}\right) C_F
+ \left(\frac{2}{3}\pi^2-\frac{49}{6}\right) C_A + \frac{10}{3} T_F n_l
\right] \left(\frac{\alpha_s}{4\pi}\right)^2\,.
\label{HL:gamma}
\end{equation}
The 3-loop term has been calculated in~\cite{CG:03}.

Now we shall discuss the QCD/HQET matching.
There are 8 Dirac structures giving non-vanishing quark currents in 4 dimensions:
\begin{equation}
\begin{split}
&\Gamma = 1\,,\quad
\gamma^0\,,\quad
\gamma^i\,,\quad
\gamma^i \gamma^0\,,\\
&\gamma^{[i} \gamma^{j]}\,,\quad
\gamma^{[i} \gamma^{j]} \gamma^0\,,\quad
\gamma^{[i} \gamma^{j} \gamma^{k]}\,,\quad
\gamma^{[i} \gamma^{j} \gamma^{k]} \gamma^0\,,
\end{split}
\label{HL:Dirac}
\end{equation}
Those in the second row can be obtained from the first row
by multiplying by the 't~Hooft--Veltman $\gamma_5^{\text{HV}}$.
We are concerned with flavour non-singlet currents only,
therefore, we may also use the anticommuting $\gamma_5^{\text{AC}}$
(there is no anomaly).
The renormalized QCD currents with different prescriptions for $\gamma_5$ 
are related by~(A.18), (A.19) in~\cite{eft1}.
The anomalous dimension of the HQET current~(\ref{HL:jtilde})
does not depend on the Dirac structure $\Gamma$.
Therefore, there are no factors similar to $Z_{P,A}$ in HQET.
Multiplying $\Gamma$ by $\gamma_5^{\text{AC}}$
does not change the matching coefficient.
Therefore, the matching coefficients for the currents
in the second row of~(\ref{HL:Dirac})
are not independent;
they can be obtained from those for the first row:
\begin{equation}
\begin{split}
Z_P(\mu) ={}&
\frac{C_{\gamma_5^{\text{AC}}}(\mu)}{C_{\gamma_5^{\text{HV}}}(\mu)} =
\frac{C_1(\mu)}{C_{\gamma^0 \gamma^1 \gamma^2 \gamma^3}(\mu)}\,,\\
Z_A(\mu) ={}&
\frac{C_{\gamma_5^{\text{AC}} \gamma^0}(\mu)}{C_{\gamma_5^{\text{HV}} \gamma^0}(\mu)} =
\frac{C_{\gamma^0}(\mu)}{C_{\gamma^1 \gamma^2 \gamma^3}(\mu)} =
\frac{C_{\gamma_5^{\text{AC}} \gamma^3}(\mu)}{C_{\gamma_5^{\text{HV}} \gamma^3}(\mu)} =
\frac{C_{\gamma^3}(\mu)}{C_{\gamma^0 \gamma^1 \gamma^2}(\mu)}\,,\\
Z_T(\mu) ={}&
\frac{C_{\gamma_5^{\text{AC}} \gamma^0 \gamma^1}(\mu)}{C_{\gamma_5^{\text{HV}} \gamma^0 \gamma^1}(\mu)} =
\frac{C_{\gamma^0 \gamma^1}(\mu)}{C_{\gamma^2 \gamma^3}(\mu)} =
\frac{C_{\gamma_5^{\text{AC}} \gamma^2 \gamma^3}(\mu)}{C_{\gamma_5^{\text{HV}} \gamma^2 \gamma^3}(\mu)} =
\frac{C_{\gamma^2 \gamma^3}(\mu)}{C_{\gamma^0 \gamma^1}(\mu)} = 1\,.
\end{split}
\label{HL:Ratios}
\end{equation}
In particular, two matching coefficients are equal:
\begin{equation}
C_{\gamma^i \gamma^0}(\mu) = C_{\gamma^{[j} \gamma^{k]}}(\mu)\,.
\label{HL:CT}
\end{equation}

In order to find the matching coefficients $C_\Gamma(\mu)$,
we equate on-shell matrix elements of
the left- and right-hand side of~(\ref{HL:exp}).
They are obtained by considering transitions
of the heavy quark with momentum $P=Mv+p$
to the light quark with momentum $k$:
\begin{equation}
{<}q(k)|j(\mu)|Q(Mv+p){>} =
C_\Gamma(\mu) {<}q(k)|\tilde{\jmath}(\mu)|Q_v(p){>}
+ \mathcal{O}\left(\frac{p,k}{M}\right)\,.
\label{HL:match}
\end{equation}
Both on-shell matrix elements of the renormalized currents are UV-finite;
both contain IR divergences, which are the same on the left- and right-hand sides.
The on-shell matrix elements are
\begin{equation}
\begin{split}
{<}q(k)|j(\mu)|Q(P){>} &{}=
\bar{u}(k) \Gamma(P,k) u(P)\,
Z_j^{-1}(\mu) Z_Q^{1/2} Z_q^{1/2}\,,\\
{<}q(k)|\tilde{\jmath}(\mu)|h_v(p){>} &{}=
\bar{u}(k) \tilde{\Gamma}(p_0,k) u_v(p)\,
\tilde{Z}_j^{-1}(\mu) Z_h^{1/2} \tilde{Z}_q^{1/2}\,,
\end{split}
\label{HL:onshell}
\end{equation}
where $\Gamma(P,k)$ and $\tilde{\Gamma}(p_0,k)$ are the bare vertex functions,
$Z_Q$ and $Z_q$ are the on-shell wave-function renormalization constants
of the heavy and the light quark in QCD,
$Z_h$ is the on-shell wave-function renormalization constant
of the HQET quark field $h_v$,
and $\tilde{Z}_q$ differs from $Z_q$ because there are no $Q$ loops in HQET.
The difference between $u(Mv+p)$ and $u_v(p)$ is of order $p/M$,
and can be neglected.
It is most convenient to use $p=k=0$,
then the $\mathcal{O}(1/M)$ term is absent.
The QCD vertex
\begin{equation}
\Gamma(Mv,0) = 1 + \Lambda(Mv,0)
= \raisebox{-4.75mm}{\begin{picture}(32,7)
\put(16,3.5){\makebox(0,0){\includegraphics{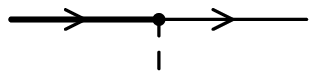}}}
\put(8.5,4){\makebox(0,0)[t]{$Mv$}}
\put(23.5,4){\makebox(0,0)[t]{$0$}}
\end{picture}}
+ \raisebox{-4.75mm}{\begin{picture}(32,14.5)
\put(16,7.25){\makebox(0,0){\includegraphics{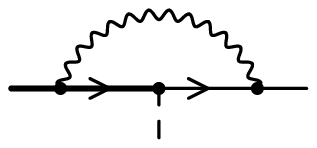}}}
\end{picture}}
+ \cdots
\label{HL:GammaQCD}
\end{equation}
has 2 Dirac structures:
\begin{equation*}
\Gamma(Mv,0) = \Gamma \cdot (A + B \gamma^0)\,.
\end{equation*}
This leads to
\begin{equation*}
\bar{u}(0) \Gamma(Mv,0) u(Mv) = \bar{\Gamma}\,\bar{u}(0) \Gamma u(Mv)
\quad\text{with}\quad
\bar{\Gamma} = A + B\,.
\end{equation*}
The HQET vertex has just one Dirac structure.
Therefore,
\begin{equation}
C_\Gamma(\mu) =
\frac{\bar{\Gamma} Z_j^{-1}(\mu) Z_Q^{1/2} Z_q^{1/2}}%
{\tilde{\Gamma}(0,0) \tilde{Z}_j^{-1}(\mu) Z_h^{1/2} \tilde{Z}_q^{1/2}}\,.
\label{HL:main}
\end{equation}
Here $M$ is the on-shell mass of the heavy quark
(because the external heavy quark with $p^2=M^2$
should be on its mass shell).
Therefore, mass-counterterm vertices have to be taken into account
on all $Q$ lines.
If all flavours except $Q$ are massless,
all loop corrections to $\tilde{\Gamma}(0,0)$,
$Z_h$, and $\tilde{Z}_q$ contain no scale
and hence vanish: $\tilde{\Gamma}(0,0)=1$,
$\tilde{Z}_Q=1$, $\tilde{Z}_q=1$.
The QCD quantities $\Gamma(Mv,0)$, $Z_Q$, and $Z_q$ contain a single scale $M$.
The ratio $Z_q/\tilde{Z}_q$ is the light-quark field decoupling coefficient
(Sect.~3.5 in~\cite{eft1}).

There exists an exact relation~\cite{BG:95}
between the matching coefficients $C_1(\mu)$ and $C_{\gamma^0}(\mu)$.
Namely, the renormalized vector and scalar currents are related by
\begin{equation}
i \partial_\alpha j^\alpha = M(\mu) j(\mu)\,,
\label{HL:div}
\end{equation}
where $M(\mu)$ is the $\overline{\text{MS}}$ mass
of the heavy quark $Q$.
Taking the on-shell matrix element of this equality
between the heavy quark with $P=Mv$ and the light quark with $k=0$
and re-expressing both QCD matrix elements via the matrix element
of the HQET current with $\Gamma=1$, we obtain
\begin{equation}
M C_{\gamma^0}(\mu) = M(\mu) C_1(\mu)\,.
\label{HL:mm}
\end{equation}

Now we are going to calculate the QCD vertex $\bar{\Gamma}$
with the 1-loop accuracy.
Initially, we make no assumptions about properties of the Dirac matrix $\Gamma$.
The one-loop diagram~(\ref{HL:GammaQCD}) can be written
as a sum of terms of the form
\begin{equation}
\bar{u}(0) \gamma_{\mu_1}\dots\gamma_{\mu_l} \Gamma
\gamma_{\nu_1}\dots\gamma_{\nu_r} u(Mv) \cdot
I^{\mu_1\dots\mu_l;\nu_1\dots\nu_r}\,,
\label{HL:form}
\end{equation}
where $I$ is some integral over the loop momentum,
$l$ is even, and $l+r\le4$.
After the integration, $I^{\mu_1\dots\mu_l;\nu_1\dots\nu_r}$
can contain only $g^{\mu\nu}$ and $v^\alpha$.
The resulting contractions of pairs of $\gamma$-matrices on the left,
and of pairs on the right, merely produce additional terms
of the form~(\ref{HL:form}), with smaller values of $l+r$.
Before performing the remaining contractions,
one may anticommute $\gamma$-matrices, so as to arrange things
such that $\rlap/v$ occurs only on the extreme left
or on the extreme right, with the contracted indices in between
occurring in opposite orders on the left and right of $\Gamma$.
The additional terms arising from the anticommutations
have fewer $\gamma$-matrices, with $l$ remaining even.
Repeating this procedure for all values of $l+r$,
from 4 down to 0, we may cast the 1-loop diagram in the form
\begin{equation*}
\begin{split}
&\bar{u}(0) \left[ \Gamma(x'_1+x'_2\rlap/v)
+ \rlap/v\gamma_\mu\Gamma\gamma^\mu(x'_3+x'_4\rlap/v)
+ \gamma_\mu\gamma_\nu\Gamma\gamma^\nu\gamma^\mu x'_5
\right] u(Mv)\,;\\
&\bar{\Lambda} = \bar{u}(0) \left( \sum_i x_i L_i \Gamma R_i \right) u(Mv)\,,
\end{split}
\end{equation*}
where
\begin{gather*}
x_1 = x'_1+x'_2\,,\quad
x_2 = x'_3+x'_4\,,\quad
x_3 = x'_5\,,\\
L_i\times R_i = 1\times1\,,\;
\rlap/v\gamma^\mu\times\gamma_\mu\,,\;
\gamma^\mu\gamma^\nu\times\gamma_\nu\gamma_\mu\,.
\end{gather*}
The coefficients $x_i$ can be found by calculating
the double traces
\begin{equation*}
\begin{split}
&y_i = \left<\bar{L}_i\times\bar{R}_i,\Lambda\right>\,,\qquad
\left<\bar{L}_i\times\bar{R}_i,L\Gamma R\right> \equiv
\frac{1}{4} \Tr \bar{L} L \cdot \frac{1}{4} \Tr \bar{R} R\,,\\
&\bar{L}_i\times\bar{R}_i =
1\times(1+\rlap/v)\,,\;
\gamma^\alpha\rlap/v\times(1+\rlap/v)\gamma_\alpha\,,\;
\gamma^\alpha\gamma^\beta\times(1+\rlap/v)\gamma_\beta\gamma_\alpha\,.
\end{split}
\end{equation*}
Solving the linear system, we obtain
\begin{equation}
\left(\begin{array}{c}x_1\\x_2\\x_3\end{array}\right) =
\frac{1}{2(d-1)(d-2)}
\left(\begin{array}{ccc}
(d-2)(3d-2) & 0  & -(d-2) \\
0           & 2d & -2     \\
-(d-2)      & -2 & 1
\end{array}\right)
\left(\begin{array}{c}y_1\\y_2\\y_3\end{array}\right)\,.
\label{HL:sol}
\end{equation}

Now we assume
\begin{equation}
\rlap/v\Gamma = \sigma \Gamma\rlap/v\,,\quad
\sigma = \pm1\,,\quad
\gamma^\mu\Gamma\gamma_\mu = 2 \sigma h(d) \Gamma\,,
\label{HL:h}
\end{equation}
where
\begin{equation}
h(d) = \eta \left(n-\frac{d}{2}\right)\,,\quad
\eta = (-1)^{n+1} \sigma
\label{HL:hd}
\end{equation}
for the antisymmetrized product of $n$ $\gamma$-matrices.
The effect of each contraction is then to produce a factor $2\sigma h$.
Terms with an odd number of contractions necessarily contain $\rlap/v$
on the left, which yields an extra $\sigma$ when moved to the right,
where it merely gives $\rlap/v u(Mv)=u(Mv)$.
Thus the result involves only powers of $h$:
\begin{equation*}
\bar{\Lambda} = x_1 + x_2\cdot2h + x_3(2h)^2\,.
\end{equation*}
Substituting the solution~(\ref{HL:sol}) for $x_i$, we obtain $\bar{\Lambda} = \left<P,\Lambda\right>$,
\begin{align*}
&P = \frac{1}{2(d-1)(d-2)} \Bigl[
(d-2)(3d^2-2-4h^2) 1\times(1+\rlap/v)\\
&{} + 4h(d-2h) \gamma^\alpha\rlap/v\times(1+\rlap/v)\gamma_\alpha
- \bigl(d-2-4h(1-h)\bigr)
\gamma^\alpha\gamma^\beta\times(1+\rlap/v)\gamma_\beta\gamma_\alpha
\Bigr]\,.
\end{align*}
Now we can apply the projector $P$ to the integrand
of the one-loop diagram~(\ref{HL:GammaQCD})
and reduce it to a scalar expression quadratic in $h$:
\begin{equation}
\begin{split}
&\bar{\Lambda} = \frac{i C_F g_0^2}{2(d-1)} \int \frac{d^d k}{(2\pi)^d}
\frac{2(d-1)+(d D_2/M^2+4)h-2(D_2/M^2+4)h^2}{D_1 D_2}\,,\\
&D_1 = M^2 - (k+Mv)^2\,,\quad
D_2 = - k^2\,,
\end{split}
\label{HL:La1I}
\end{equation}
where terms with $D_1$ in the numerator have been omitted as they yield 0.
The integral can be easily calculated once, and for all currents.
We obtain the one-loop result
\begin{equation}
\bar{\Lambda} = - C_F \frac{g_0^2 M^{-2\varepsilon}}{(4\pi)^{d/2}} \Gamma(\varepsilon)
\frac{(1-h)(d-2+2h)}{(d-2)(d-3)}\,.
\label{HL:La1}
\end{equation}
This on-shell vertex is gauge-invariant.

At last, we can combine all pieces of~(\ref{HL:main}).
With 1-loop accuracy~\cite{EH:90}
\begin{equation}
C_\Gamma(M) = 1 + C_F \frac{\alpha_s(M)}{4\pi}
\left[ 3(n-2)^2 + (2-\eta)(n-2) - 4 \right]
+ \cdots
\label{HL:C1}
\end{equation}
They satisfy the relations~(\ref{HL:Ratios}) and~(\ref{HL:mm}).
The 2-loop corrections have been calculated in~\cite{BG:95,G:98},
and the 3-loop ones in~\cite{BGMPSS:10}.

We are now in the position to apply our results to the matrix elements
between a $B$ or $B^*$ meson with momentum $p$ and the vacuum.
They are defined through
\begin{equation}
\begin{split}
{<}0| \left(\bar{q} \gamma_5^{\text{AC}} Q\right)_\mu |B{>} &{}=
- i M_B f_B^P(\mu)\,,\\
{<}0| \bar{q} \gamma^\alpha \gamma_5^{\text{AC}} Q |B{>} &{}=
i f_B P^\alpha\,,\\
{<}0| \bar{q} \gamma^\alpha Q |B^*{>} &{}=
i M_{B^*} f_{B^*} e^\alpha\,,\\
{<}0| \left(\bar{q} \sigma^{\alpha\beta} Q\right)_\mu |B^*{>} &{}=
f_{B^*}^{T}(\mu) (P^\alpha e^\beta - P^\beta e^\alpha)\,,
\end{split}
\label{HL:matel}
\end{equation}
where $e^\alpha$ is the $B^*$ polarization vector.
The corresponding HQET matrix elements are
\begin{equation}
\begin{split}
{<}0| \left(\bar{q}\gamma_5^{\text{AC}} h_v\right)_\mu |B(\vec{p}\,){>}\strut_{\text{nr}} &{}=
- i F(\mu)\,,
\nonumber\\
{<}0| \left(\bar{q} \vec{\gamma} h_v\right)_\mu |B^*(\vec{p}\,){>}\strut_{\text{nr}} &{}=
i F(\mu) \vec{e}\,,
\end{split}
\label{HL:HQET}
\end{equation}
where the single-meson states are normalized by the non-relativistic condition
\begin{equation*}
\strut_{\text{nr}}{<}B(\vec{p}\,')|B(\vec{p}\,){>}\strut_{\text{nr}}
= (2\pi)^3 \delta(\vec{p}\,'-\vec{p}\,)\,,
\end{equation*}
in contrast to the usual relativistic one
\begin{equation*}
{<}B(\vec{p}\,')|B(\vec{p}\,){>} = (2\pi)^3 2 P_0 \delta(\vec{p}\,'-\vec{p}\,)\,.
\end{equation*}
We also remind the reader that $\bar{q} \Gamma \gamma^0 h_v = \bar{q} \Gamma h_v$,
so that there are only two currents.
These two matrix elements are characterized by a single hadronic parameter $F(\mu)$
due to the heavy-quark spin symmetry.
As a result, we have
\begin{equation}
\begin{split}
f_B &{}= \sqrt{\frac{2}{M_B}} C_{\gamma^0}(\mu) F(\mu)
\left[1 + \mathcal{O}\left(\frac{\Lambda_{\text{QCD}}}{M}\right)\right]\,,\\
f_{B^*} &{}= \sqrt{\frac{2}{M_{B^*}}} C_{\gamma^i}(\mu) F(\mu)
\left[1 + \mathcal{O}\left(\frac{\Lambda_{\text{QCD}}}{M}\right)\right]\,,
\end{split}
\label{HL:fB}
\end{equation}
and similar formulas for $f^P_B(\mu)$, $f^T_{B^*}(\mu)$.
These matrix elements are $\sim1/\sqrt{M}$,
up to effects of anomalous dimensions and power corrections.

Taking the matrix element of~(\ref{HL:div}) we obtain~\cite{BG:95}
\begin{equation}
\frac{f_B^P(\mu)}{f_B} = \frac{M_B}{M(\mu)}\,.
\label{HL:mmB}
\end{equation}
Here $M_B=M+\bar{\Lambda} + \mathcal{O}(\Lambda_{\text{QCD}}^2/M)$
where $\bar{\Lambda}$ is the residual energy of the ground-state $B$ meson
in the limit $M\to\infty$.
Neglecting $1/M$ corrections, we see that this equation coincides with~(\ref{HL:mm}).

The ratio
\begin{equation}
\frac{f_{B^*}}{f_B}
= \frac{C_{\gamma^i}(\mu)}{C_{\gamma^0}(\mu)} + \mathcal{O}\left(\frac{\Lambda_{\text{QCD}}}{M}\right)
= 1 - 2 C_F \frac{\alpha_s(M)}{4\pi}
+ \mathcal{O}\left(\alpha_s^2,\frac{\Lambda_{\text{QCD}}}{M}\right)
\label{HL:fBstar}
\end{equation}
does not depend on $\mu$.
The 2-loop correction has been calculated in~\cite{BG:95},
and the 3-loop one in~\cite{BGMPSS:10}.

Let's consider the ratio $f_B/f_D$.
It is convenient to use $\mu=M_b$ in~(\ref{HL:fB})
because then there are no large logarithms.
The hadronic matrix element $F(\mu)$ here is $F^{(4)}(\mu)$,
the matrix element in HQET with $n_l=4$;
its evolution is determined by the anomalous dimension~(\ref{HL:gamma}).
A similar formula can be written for $f_D$;
$F^{(4)}(M_c)$ is related to $F^{(3)}(M_c)$ by a decoupling relation~\cite{G:98,GSS:06},
their difference is $\mathcal{O}(\alpha_s^2)$.
Therefore,
\begin{equation}
\frac{f_B}{f_D} = \sqrt{\frac{M_c}{M_b}}
\left(\frac{\alpha_s^{(4)}(M_c)}{\alpha_s^{(4)}(M_b)}\right)^{6/25}
\left[1 + \mathcal{O}\left(\alpha_s,\frac{\Lambda_{\text{QCD}}}{M_{b,c}}\right)\right]\,.
\label{HL:fBD}
\end{equation}
Perturbative corrections up to $\alpha_s^2$ have been obtained in~\cite{CG:03}.

We can also look at re-expressing QCD operators via HQET ones
from the point of view of the method of regions.
Let's consider the decay $Q\to ql\bar{\nu}$
with the $\alpha_s$ accuracy.
Its matrix element is
\begin{equation}
\left(Z_Q^{\text{os}}\right)^{1/2}
\raisebox{-4.75mm}{\begin{picture}(32,7)
\put(16,3.5){\makebox(0,0){\includegraphics{qhl0.eps}}}
\put(8.5,4){\makebox(0,0)[t]{$Mv$}}
\put(23.5,4){\makebox(0,0)[t]{$\vphantom{M}p$}}
\end{picture}}
+ \raisebox{-4.75mm}{\begin{picture}(32,14.5)
\put(16,7.25){\makebox(0,0){\includegraphics{qhl1.eps}}}
\put(3.5,4){\makebox(0,0)[t]{$Mv$}}
\put(28.5,4){\makebox(0,0)[t]{$\vphantom{M}p$}}
\end{picture}}
\label{HL:QqQCD}
\end{equation}
($Z_q^{\text{os}}=1$ at 1 loop because it has no scale, $p^2=0$).
Let's consider the case of a small light-quark energy $p_0\ll M$
(nearly all energy goes to the virtual $W$ and then to the lepton pair).
The 1-loop QCD diagram here contains 2 scales, $M$ and $p_0$.
It is given by the sum of 2 contributions.
In the hard region $k\sim M$,
we expand the integrand in the Taylor series in $p$
(we are going to keep only the leading term, the value at $p=0$).
In the soft region $k\sim p$,
we expand the integrand in the Taylor series in $1/M$
(again we keep only the leading term).
The result is
\begin{equation}
\raisebox{-4.75mm}{\begin{picture}(32,14.5)
\put(16,7.25){\makebox(0,0){\includegraphics{qhl1.eps}}}
\put(3.5,4){\makebox(0,0)[t]{$Mv$}}
\put(28.5,4){\makebox(0,0)[t]{$\vphantom{M}p$}}
\end{picture}}
= \raisebox{-4.75mm}{\begin{picture}(32,14.5)
\put(16,7.25){\makebox(0,0){\includegraphics{qhl1.eps}}}
\put(3.5,4){\makebox(0,0)[t]{$Mv$}}
\put(28.5,4){\makebox(0,0)[t]{$\vphantom{M}0$}}
\end{picture}}
+ \raisebox{-4.75mm}{\begin{picture}(32,14.5)
\put(16,7.25){\makebox(0,0){\includegraphics{hl1.eps}}}
\put(3.5,4){\makebox(0,0)[t]{$0$}}
\put(28.5,4){\makebox(0,0)[t]{$\vphantom{0}p$}}
\end{picture}}\,.
\label{HL:regions}
\end{equation}
The hard contribution
(together with the 1-loop term in $Z_Q^{\text{os}}$
which also contains only the hard scale $M$)
produces the QCD/HQET matching coefficient;
the soft one is the HQET diagram.
The decay matrix element becomes
\begin{equation}
\label{HL:QqHQET}
\raisebox{-4.75mm}{\begin{picture}(32,7)
\put(16,3.5){\makebox(0,0){\includegraphics{hl0.eps}}}
\put(8.5,4){\makebox(0,0)[t]{$0$}}
\put(23.5,4){\makebox(0,0)[t]{$\vphantom{0}p$}}
\put(16,7.5){\makebox(0,0)[b]{$C$}}
\end{picture}}
+ \raisebox{-4.75mm}{\begin{picture}(32,14.5)
\put(16,7.25){\makebox(0,0){\includegraphics{hl1.eps}}}
\put(3.5,4){\makebox(0,0)[t]{$0$}}
\put(28.5,4){\makebox(0,0)[t]{$\vphantom{0}p$}}
\end{picture}}\,.
\end{equation}
Higher corrections in the hard contribution correspond to
higher-dimensional HQET operators (with their matching coefficients)
in the expansion of the QCD current.
Higher corrections in the soft contribution correspond to
HQET diagrams with insertions of $1/M^n$ suppressed terms in the HQET Lagrangian.

\subsection{Heavy--heavy currents}
\label{S:HQETHH}

\begin{figure}[ht]
\begin{center}
\begin{picture}(90,57)
\put(21,46){\makebox(0,0){\includegraphics{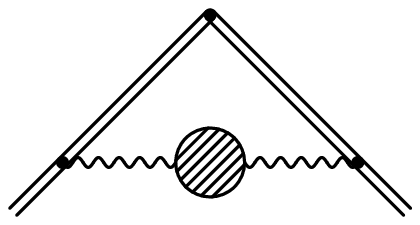}}}
\put(69,46){\makebox(0,0){\includegraphics{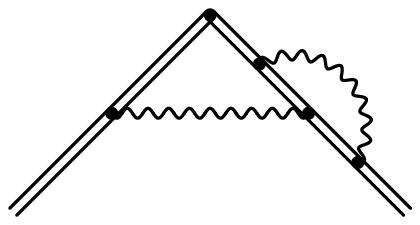}}}
\put(21,14){\makebox(0,0){\includegraphics{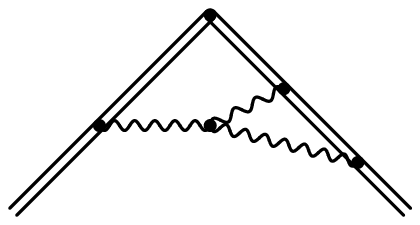}}}
\put(69,14){\makebox(0,0){\includegraphics{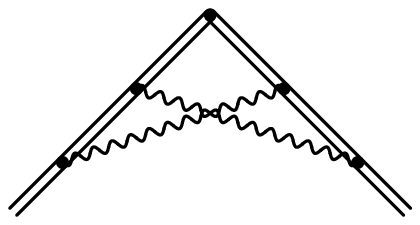}}}
\put(21,32){\makebox(0,0)[b]{a}}
\put(69,32){\makebox(0,0)[b]{b}}
\put(21,0){\makebox(0,0)[b]{c}}
\put(69,0){\makebox(0,0)[b]{d}}
\end{picture}
\end{center}
\caption{Diagrams contributing to the exponent}
\label{F:HH2}
\end{figure}

Not only the static-quark propagator~(\ref{HQET:NAexp}) (straight Wilson line)
but also the Green function with a heavy--heavy current insertion (Wilson line with an angle)
is given by an exponent of a series containing only
maximally non-abelian (colour-connected) structures~\cite{GFT:84}:
\begin{equation}
G(t,t';\vartheta) =
\exp \left[ C_F \frac{g_0^2}{(4\pi)^{d/2}} F
+ C_F \frac{g_0^4}{(4\pi)^d} \left( C_A F_A + T_F n_l F_l \right)
+ \cdots \right]\,.
\label{HQETHH:NAExp}
\end{equation}
If the colour factors of all 2-loop diagrams with two gluons attached
to the heavy-quark line were equal to $C_F^2$ (as in the abelian case),
they would produce the $F^2$ term in the expansion of the exponential.
In the non-abelian case, the colour factors of some diagrams
also contain a non-abelian part $C_F C_A$,
which should be taken into account separately (these parts contribute to $F_A$).
The ratio~(\ref{HH:Ratio}) can be written similarly.
Only the diagrams with the $J_0$ vertex inside the correction (Fig.~\ref{F:HH2})
contribute to $\mathcal{F}_A$ and $\mathcal{F}_l$
(the diagrams of Figs.~\ref{F:HH2}b,d should be taken with
the non-abelian part of their colour factors, $C_F C_A$).
Therefore, the renormalization constant has a similar structure:
\begin{equation*}
\begin{split}
Z_J = \exp \biggl[&
C_F \frac{\alpha_s}{4\pi\varepsilon} \left( f(\vartheta) - f(0) \right)\\
&{} + C_F \left(\frac{\alpha_s}{4\pi\varepsilon}\right)^2
\Bigl( C_A \left( f_A(\vartheta) - f_A(0) \right)
+ T_F n_l \left( f_l(\vartheta) - f_l(0) \right)
\Bigr) + \cdots \biggr]\,.
\end{split}
\end{equation*}
In particular, this means that the 2-loop anomalous dimension
contains no $C_F^2$ term.

These diagrams have been calculated in~\cite{KR:87}
(except the easiest one, Fig.~\ref{F:HH2}a with the quark-loop correction,
which gives $f_l(\vartheta)$ and was found in~\cite{K:89};
see Sect.~8.4 in~\cite{G:04} for a simple derivation).
The result obtained in~\cite{KR:87} contained two single integrals
which were not expressed via known functions.
One of them has been calculated in~\cite{G:04} (see~(7.15));
the last one has been calculated in~\cite{K:09},
and the result has been written in terms of logarithms, $\Li2$ and $\Li3$ only:
\begin{equation}
\begin{split}
&\Gamma(\vartheta) = 4 C_F (\vartheta\coth\vartheta-1) \frac{\alpha_s}{4\pi}\\
&{} + 8 C_F \Biggl\{
\left[ C_A \left( \frac{67}{18} - \frac{\pi^2}{6} \right) - \frac{10}{9} T_F n_l \right] (\vartheta\coth\vartheta-1)\\
&\hphantom{{}+8C_F\Biggl\{\Biggr.}
{} + C_A \Biggl[
\coth^2\vartheta \left( \Li3\left(e^{-2\vartheta}\right) + \vartheta \Li2\left(e^{-2\vartheta}\right)
+ \frac{\vartheta^3}{3} + \frac{\pi^2}{6} \vartheta^2 - \zeta_3 \right)\\
&\hphantom{{}+8C_F\Biggl\{{}+C_A\Biggl[\Biggr.\Biggr.}
{} + \coth\vartheta \left( \Li2\left(e^{-2\vartheta}\right) - 2 \vartheta \log\left(1 - e^{-2\vartheta}\right)
- \frac{\vartheta^3}{3} - \vartheta^2 - \frac{\pi^2}{6} \vartheta - \frac{\pi^2}{6} \right)\\
&\hphantom{{}+8C_F\Biggl\{{}+C_A\Biggl[\Biggr.\Biggr.}
+ \vartheta^2 + \frac{\pi^2}{6} + 1 \Biggr]\Biggr\}
\left(\frac{\alpha_s}{4\pi}\right)^2 + \cdots
\end{split}
\label{HQETHH:gamma2}
\end{equation}
It is, of course, even with respect to $\vartheta\to-\vartheta$,
though some polylogarithmic identities are needed to prove this.

At small angles~(\ref{HH:G0})
\begin{equation}
\Gamma_0 = \frac{4}{3} C_F \frac{\alpha_s}{4\pi}
+ C_F \left[ C_A \left(\frac{376}{27}-\frac{8}{9}\pi^2\right)
- \frac{80}{27} T_F n_l \right]
\left(\frac{\alpha_s}{4\pi}\right)^2 + \cdots
\label{HQETHH:gamma0}
\end{equation}
At large angles, $\Gamma(\vartheta)$ is linear in $\vartheta$~(\ref{HH:Ginf})
to all orders in $\alpha_s$~\cite{KR:87}, and
\begin{equation}
\Gamma_\infty = 4 C_F \frac{\alpha_s}{4\pi}
+ C_F \left[ C_A \left(\frac{268}{9}-\frac{4}{3}\pi^2\right)
- \frac{80}{9} T_F n_l \right]
\left(\frac{\alpha_s}{4\pi}\right)^2 + \cdots
\label{HQETHH:gammainf2}
\end{equation}
This quantity is related~\cite{K:89} to the asymptotics of the evolution kernel $P_{qq}(z)$ at $z\to1$:
\begin{equation}
P_{qq}(z) = \Gamma_\infty \left(\frac{1}{1-z}\right)_+ + C \delta(1-z) + \mathcal{O}((1-z)^0)\,;
\label{HQETHH:Pqq}
\end{equation}
$P_{qq}(z)$ is currently known up to 3 loops~\cite{MVV:04},
and hence $\Gamma_\infty$ is also known with the same accuracy~\cite{N:05}.
The asymptotics of the Brodsky--Lepage evolution kernel $V_{qq}(x,y)$ at $x-y\to0$
is also governed by $\Gamma_\infty$.

The imaginary part of $\Gamma(\delta-i\pi)$ at $\delta\to0$ is determined by the quark--antiquark potential~\cite{KMO:93};
it is currently known up to 3 loops~\cite{SSS:10}.

It is remarkable that one of the finest perturbative-HQET papers~\cite{KR:87}
was written several years before the HQET gold rush of 1990--91.

We don't consider matching full QCD currents and the effective-theory operator $J$ here,
see, e.\,g., \cite{N:94} and Chapter~7 in~\cite{G:04}.

\subsection{Chromomagnetic interaction}
\label{S:CM}

\begin{figure}[ht]
\begin{center}
\begin{picture}(62,31)
% a
\put(13,18.5){\makebox(0,0){\includegraphics{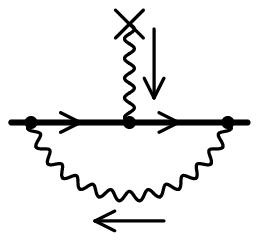}}}
\put(13,0){\makebox(0,0)[b]{a}}
\put(13,5){\makebox(0,0){$k$}}
\put(7,21){\makebox(0,0){$k+P$}}
\put(24,21){\makebox(0,0){$k+P+q$}}
\put(17.5,26.5){\makebox(0,0){$q$}}
% b
\put(49,18.25){\makebox(0,0){\includegraphics{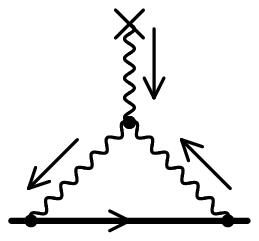}}}
\put(49,0){\makebox(0,0)[b]{b}}
\put(49,5){\makebox(0,0){$k+P$}}
\put(39,16){\makebox(0,0){$k$}}
\put(62,16){\makebox(0,0){$k-q$}}
\put(53.5,26.5){\makebox(0,0){$q$}}
\end{picture}
\end{center}
\caption{One-loop vertex}
\label{F:Vert1Loop}
\end{figure}

In order to find the chromomagnetic interaction coefficient $C_m$ in the HQET Lagrangian,
we match the amplitudes of scattering of an on-shell heavy quark in an external chromomagnetic field
in QCD and HQET.
It is convenient to use the background field method~\cite{A:81}.
At one loop~\cite{EH:90b} (Fig.~\ref{F:Vert1Loop})
\begin{equation}
F_2(0) = \frac{g_0^2 M^{-2\varepsilon}}{(4\pi)^{d/2}} \frac{\Gamma(\varepsilon)}{2(d-3)}
\left[ 2 (d-4) (d-5) C_F - (d^2-8d+14) C_A \right]\,.
\label{CM:F2}
\end{equation}
The diagram Fig.~\ref{F:Vert1Loop}b is IR divergent,
unlike the abelian case (Sect.~\ref{S:Mag}).

If all flavours except $Q$ are massless, all loop corrections in HQET vanish.
Both QCD and HQET scattering amplitudes are renormalized and hence UV-finite;
both have IR divergences.
These divergences are the same,
because HQET has been constructed to reproduce the IR behaviour of QCD.
Vanishing loop corrections in HQET have UV and IR divergences which cancel each other.
The UV divergences of $C_m^0$ are removed by $Z_m^{-1}(\mu)$;
the IR ones match those of $1+F_2(0)$.

The $1/\varepsilon$ term in~(\ref{CM:F2}) gives the 1-loop anomalous dimension~\cite{EH:90b}
of the chromomagnetic operator $O_m$.
The 2-loop contribution has been calculated in~\cite{ABN:97,CG:97}:
\begin{equation}
\gamma_m = 2 C_A \frac{\alpha_s}{4\pi}
+ \frac{4}{9} C_A (17 C_A - 13 T_F n_l) \left(\frac{\alpha_s}{4\pi}\right)^2
+ \cdots
\label{CM:gamma}
\end{equation}
The 3-loop term has been derived in~\cite{GMPS:07}.
Of course, this anomalous dimension vanishes in the abelian case (Sect.~\ref{S:Mag}):
the HEET operator $O_m$ does not renormalize.

The renormalized chromomagnetic coefficient with the 1-loop accuracy is~\cite{EH:90b}
\begin{equation}
C_m(\mu) = 1 + 2 \left( - C_A \log\frac{M}{\mu} + C_F + C_A \right) \frac{\alpha_s(M)}{4\pi} + \cdots
\label{CM:C}
\end{equation}
The 2-loop correction has been found in~\cite{CG:97}, and the 3-loop one in~\cite{GMPS:07}.
It is convenient to use $C_m(M)$ (containing no large logarithms)
as the initial condition for the renormalization group equation.
In the abelian case $C_m(\mu)$ does not depend on $\mu$,
and is just the electron magnetic moment.

The most prominent physical effect caused by the chromomagnetic interaction
is the mass splittings of hadronic doublets which are degenerate at $M=\infty$
due to the heavy-quark spin symmetry.
The mass splitting $M_{B^*}-M_B$ is $\sim1/M_b$;
therefore, $M_{B^*}^2-M_B^2$ is constant (up to power corrections):
\begin{equation}
M_{B^*}^2 - M_B^2 = \frac{4}{3} C_m^{(4)}(\mu) \mu_{G(4)}^2(\mu)
+ \mathcal{O}\left(\frac{\Lambda_{\text{QCD}}}{M_b}\right)\,,
\label{CM:MB}
\end{equation}
where the index ``(4)'' means that we are considering the $n_l=4$ flavour HQET,
and $\mu_{G(4)}^2(\mu)$ is the matrix element of $O_m(\mu)$ over the ground-state meson.
It is most natural to choose $\mu=M_b$ in~(\ref{CM:MB})
because then $C_m$ contains no large logarithms.
A similar formula can be written for $D$ mesons.
The running of $\mu_{G(n_l)}^2(\mu)$ is governed by the anomalous dimension~(\ref{CM:gamma}).
The matrix elements $\mu_{G(4)}^2(M_c)$ and $\mu_{G(3)}^2(M_c)$ are related by decoupling, see~\cite{G:04};
their difference is $\mathcal{O}(\alpha_s^2)$.
In the leading logarithmic approximation we obtain from the 1-loop anomalous dimension~(\ref{CM:gamma})
\begin{equation}
\frac{M_{B^*}^2 - M_B^2}{M_{D^*}^2 - M_D^2} =
\left(\frac{\alpha_s^{(4)}(M_c)}{\alpha_s^{(4)}(M_b)}\right)^{-9/25}
\left[1 + \mathcal{O}\left(\alpha_s,\frac{\Lambda_{\text{QCD}}}{M_{b,c}}\right)\right]\,.
\label{CM:R}
\end{equation}
This agrees well with the experimental value 0.88.
Unfortunately, higher perturbative corrections~\cite{GMPS:07} are large and negative;
no convergence is seen, and the agreement with the experiment becomes much worse.
Also the $\Lambda_{\text{QCD}}/M_c$ power correction is expected to be quite large.

\textbf{Acknowledgements}.
I am grateful to
S.~Bekavac,
D.\,J.~Broadhurst,
K.\,G.~Chetyrkin,
A.~Czarnecki,
A.\,I.~Davydychev,
T.~Huber,
A.\,V.~Kotikov,
R.\,N.~Lee,
D.~Ma\^{\i}tre,
P.~Marquard,
M.~Neubert,
J.\,H.~Piclum,
D.~Seidel,
A.\,V.~Smirnov,
V.\,A.~Smirnov,
M.~Steinhauser,
O.\,I.~Yakovlev
for collaboration on various HQET related projects, and to
K.~Melnikov,
A.~Penin
for a useful discussion of the proof that the electron anomalous magnetic moment in QED
is IR finite to all orders.
The work was partially supported by RFBR (grant 12-02-00106-a)
and by Russian Ministry of Education and Science.

\appendix
\section{One-loop self-energy diagram}
\label{S:App1}

\begin{figure}[b]
\begin{center}
\begin{picture}(54,21)
\put(27,11){\makebox(0,0){\includegraphics{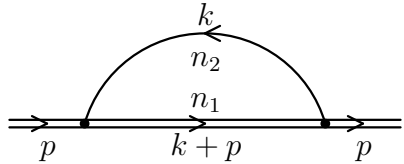}}}
\put(27,0){\makebox(0,0)[b]{$k+p$}}
\put(27,21){\makebox(0,0)[t]{$k$}}
\put(6,0){\makebox(0,0)[b]{$p$}}
\put(48,0){\makebox(0,0)[b]{$p$}}
\put(27,9){\makebox(0,0)[t]{$n_1$}}
\put(27,12){\makebox(0,0)[b]{$n_2$}}
\end{picture}
\end{center}
\caption{One-loop propagator diagram}
\label{F:hm1}
\end{figure}

Here we shall discuss the massive diagram (Fig.~\ref{F:hm1}):
\begin{equation}
\begin{split}
&I_{n_1 n_2}(m,p_0) =
\frac{1}{i\pi^{d/2}} \int \frac{d^d k}{D_1^{n_1} D_2^{n_2}}\,,\\
&D_1 = -2(k+p)_0 - i0\,,\quad
D_2 = m^2 - k^2 - i0\,.
\end{split}
\label{App1:def}
\end{equation}
It has a cut from the threshold $\omega=m$ to $+\infty$.
It vanishes at integer $n_2\le0$ because of the HQET loop.
At integer $n_1\le0$ it is the vacuum diagram~(2.10) in~\cite{eft1}
with a numerator; in particular,
\begin{equation*}
I_{0 n}(m,\omega) = m^{d-2n} V(n)\,.
\end{equation*}
The limit $m\to0$ is smooth if it does not produce IR divergence:
\begin{equation}
\lim_{m\to0} I_{n_1 n_2}(m,\omega) = (-2\omega)^{d-2n_1-n_2} I(n_1,n_2)
\quad\text{if}\quad
n_2 < \frac{d}{2}\,,
\label{App1:m0}
\end{equation}
see~(\ref{L1:I}).

Using the HQET Feynman parametrization~(\ref{L1:Feyn}), we obtain
\begin{equation}
I_{n_1 n_2}(m,\omega) =
\frac{\Gamma\left(n_1+n_2-\frac{d}{2}\right)}{\Gamma(n_1) \Gamma(n_2)}
\int_0^\infty y^{n_1-1} (y^2 - 2 \omega y + m^2)^{d/2-n_1-n_2}\,dy\,.
\label{App1:Feyn}
\end{equation}
It is easy to calculate this integral at $\omega=0$:
\begin{equation}
\begin{split}
&I_{n_1 n_2}(m,0) = I_0(n_1,n_2) m^{d-n_1-2n_2}\,,\\
&I_0(n_1,n_2) =
\frac{\Gamma\left(\frac{n_1}{2}\right)
\Gamma\left(\frac{n_1-d}{2}+n_2\right)}%
{2 \Gamma(n_1) \Gamma(n_2)} =
\frac{\pi^{1/2}}{2^{n_1}}
\frac{\Gamma\left(\frac{n_1-d}{2}+n_2\right)}%
{\Gamma\left(\frac{n_1+1}{2}\right) \Gamma(n_2)}
\end{split}
\label{App1:I0}
\end{equation}
(it vanishes at odd negative integer $n_1$
because the integrand is odd in $k$).
It is also easy to find this diagram at the threshold.
At $\omega<0$ the result is~\cite{GHM:07}
\begin{equation}
\begin{split}
I_{n_1 n_2}(m,\omega) =& m^{d-n_1-2n_2}
\frac{\Gamma\left(n_1+n_2-\frac{d}{2}\right) \Gamma(n_1+2n_2-d)}%
{\Gamma(n_2) \Gamma(2(n_1+n_2)-d)}\\
&{}\times{}_2 F_1 \left( \left.
\begin{array}{c}
\frac{n_1}{2}, \frac{n_1-d}{2} + n_2\\
n_1 + n_2 - \frac{d-1}{2}
\end{array}
\right| 1 - \frac{\omega^2}{m^2} \right)\,.
\end{split}
\label{App1:I}
\end{equation}
The point $\omega=0$ is regular;
when we go from a small $\omega<0$ to $\omega>0$
along some path in the complex plane,
we make a full cycle around the branch point of the hypergeometric function,
and arrive at another Riemann sheet.
A similar hypergeometric representation has been derived in~\cite{Z:02}.

\begin{figure}[ht]
\begin{center}
\begin{picture}(42,42)
\put(21,21){\makebox(0,0){\includegraphics{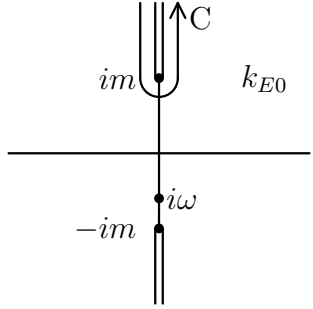}}}
\put(22,15){\makebox(0,0)[l]{$i\omega$}}
\put(18,31){\makebox(0,0)[r]{$im$}}
\put(18,11){\makebox(0,0)[r]{$-im$}}
\put(25,39){\makebox(0,0)[l]{C}}
\put(35,31){\makebox(0,0){$k_{E0}$}}
\end{picture}
\end{center}
\caption{Integration contour}
\label{F:Contour}
\end{figure}

This diagram can also be calculated by first taking the integral
in $d^{d-1}\vec{k}$ (after the Wick rotation)~\cite{G:08} (cf.~\cite{K:92}):
\begin{equation*}
I_{n_1 n_2}(m,\omega) =
\frac{\Gamma\left(n_2-\frac{d-1}{2}\right)}{\pi^{1/2} \Gamma(n_2)}
\int_{-\infty}^{+\infty} d k_{E0}
\frac{(k_{E0}^2+m^2)^{(d-1)/2-n_2}}{(- 2 \omega - 2 i k_{E0})^{n_1}}\,.
\end{equation*}
If $\omega<0$, we can deform the integration contour (Fig.~\ref{F:Contour}):
\begin{equation}
I_{n_1 n_2}(m,\omega) = 2 \frac{\Gamma\left(n_2-\frac{d-1}{2}\right)}{\pi^{1/2} \Gamma(n_2)}
\cos\left[\pi \left(\frac{d}{2} - n_2\right)\right]
\int_m^\infty dk
\frac{(k^2-m^2)^{(d-1)/2-n_2}}{(2k - 2\omega)^{n_1}}\,.
\label{App1:K}
\end{equation}
This integral is
\begin{equation}
\begin{split}
I_{n_1 n_2}(m,\omega) =& m^{d-n_1-2n_2}
\frac{\Gamma\left(n_1+n_2-\frac{d}{2}\right) \Gamma(n_1+2n_2-d)}%
{\Gamma(n_2) \Gamma(2(n_1+n_2)-d)}\\
&{}\times{}_2 F_1 \left( \left.
\begin{array}{c}
n_1,n_1+2n_2-d\\
n_1+n_2-\frac{d-1}{2}
\end{array}
\right| \frac{1}{2} \left(1 + \frac{\omega}{m}\right) \right)\,.
\end{split}
\label{App1:I2}
\end{equation}
Using the quadratic transformation~(15.8.18)~\cite{AS},
we again arrive at~(\ref{App1:I});
note, however, that~(\ref{App1:I2}) is valid not only
for $\omega<0$, but also for $\omega\in(0,m)$.

At $m=0$ (\ref{App1:K}) gives the massless result
\begin{equation}
I(n_1,n_2) = \frac{2^{2n_2-d+1}}{\pi^{1/2}}
\cos\left[\pi\left(\frac{d}{2}-n_2\right)\right]
\frac{\Gamma(d-2n_2) \Gamma(n_1+2n_2-d)
\Gamma\bigl(n_2-\frac{d-1}{2}\bigr)}%
{\Gamma(n_1) \Gamma(n_2)}\,.
\label{App1:Im0}
\end{equation}
Using the well-known properties of the $\Gamma$ function
\begin{equation*}
\Gamma(2x) = \pi^{-1/2} 2^{2x-1} \Gamma(x) \Gamma\bigl(x+\tfrac{1}{2}\bigr)\,,\qquad
\Gamma(x) \Gamma(1-x) = \frac{\pi}{\sin\pi x}\,,
\end{equation*}
we can simplify this result to~(\ref{L1:I}).

We can investigate asymptotics of our diagram
using the method of regions~\cite{S:02,J:11}.
If $\omega\ll m$, there is one integration region $k\sim m$.
We may expand $D_1^{-n_1}$ in $\omega$ and obtain
\begin{equation}
\begin{split}
&I_{n_1 n_2}(m,\omega) = m^{d-n_1-2n_2}
\sum_{n=0}^\infty I_0(n_1+n,n_2) \frac{(n_1)_n}{n!}
\left(\frac{2\omega}{m}\right)^n\\
&{} = m^{d-n_1-2n_2} I_0(n_1,n_2) \Biggl[
{}_2 F_1 \left( \left.
\begin{array}{c}
\frac{n_1}{2},\frac{n_1-d}{2}+n_2\\\frac{1}{2}
\end{array}
\right| \frac{\omega^2}{m^2} \right)\\
&{} + \frac{\Gamma\left(\frac{n_1+1}{2}\right)
\Gamma\left(\frac{n_1-d+1}{2}+n_2\right)}%
{\Gamma\left(\frac{n_1}{2}\right)
\Gamma\left(\frac{n_1-d}{2}+n_2\right)}\,
\frac{2\omega}{m}\,
{}_2 F_1 \left( \left.
\begin{array}{c}
\frac{n_1+1}{2},\frac{n_1-d+1}{2}+n_2\\\frac{3}{2}
\end{array}
\right| \frac{\omega^2}{m^2} \right)
\Biggr]\,.
\end{split}
\label{App1:w0}
\end{equation}
This is a regular Taylor series in $\omega$;
the fractional power of $m$ follows from dimension counting.
At $\omega<0$ it is equivalent to~(\ref{App1:I}).

Now let's consider $-\omega\gg m$.
There are two regions: hard $k\sim\omega$ and soft $k\sim m$,
$I_{n_1 n_2}(m,\omega)=I_h+I_s$.
This is called OPE, see~\cite{S:02} for more detail.
In the hard region, we may expand $D_2^{-n_2}$ in $m^2$:
\begin{equation}
\begin{split}
I_h &= (-2\omega)^{d-n_1-2n_2}
\sum_{n=0}^\infty I(n_1,n_2+n) \frac{(n_2)_n}{n!}
\left(-\frac{m^2}{4\omega^2}\right)^n\\
&= (-2\omega)^{d-n_1-2n_2} I(n_1,n_2)\,
{}_2 F_1 \left( \left.
\begin{array}{c}
\frac{n_1-d}{2}+n_2,\frac{n_1-d+1}{2}+n_2\\n_2+1-\frac{d}{2}
\end{array}
\right| \frac{m^2}{\omega^2} \right)\,.
\end{split}
\label{App1:hard}
\end{equation}
This is a regular Taylor series in $m^2$;
the fractional power of $-2\omega$ follows from dimension counting.
In the OPE terms, this is the (one-loop) coefficient function
of the unit operator.
In the soft region, we may expand $D_1^{-n_1}$ in $k$
(all odd terms vanish):
\begin{equation}
\begin{split}
I_s &= m^{d-2n_2} (-2\omega)^{-n_1}
\sum_{n=0}^\infty I_0(-2n,n_2) \frac{(n_1)_{2n}}{(2n)!}
\left(\frac{m^2}{4\omega^2}\right)^n\\
&= m^{d-2n_2} (-2\omega)^{-n_1}
V(n_2)\,
{}_2 F_1 \left( \left.
\begin{array}{c}
\frac{n_1}{2},\frac{n_1+1}{2}\\\frac{d}{2}-n_2+1
\end{array}
\right| \frac{m^2}{\omega^2} \right)\,.
\end{split}
\label{App1:soft}
\end{equation}
This is a regular Taylor series in $\omega$
(after extraction of the leading $(-2\omega)^{-n_1}$);
the fractional power of $m$ follows from dimension counting.
In the OPE terms,
this is the series of perturbative (one-loop) vacuum averages
of local operators (with $2n$ derivatives)
accompanied by their tree-level coefficient functions.
Now we see that the leading term in $I_h$~(\ref{App1:hard})
dominates over the leading term in $I_s$~(\ref{App1:soft})
at $m\to0$ if $n_2<d/2$, cf.~(\ref{App1:m0}).

Finally, we shall calculate this diagram using the Mellin--Barnes method.
It is easy to check the identity
\begin{equation}
\frac{1}{(a+b)^n} = \frac{a^{-n}}{\Gamma(n)} \frac{1}{2\pi i}
\int_{-i\infty}^{+i\infty} dz\,\Gamma(-z) \Gamma(n+z)
\left(\frac{b}{a}\right)^z\,.
\label{App1:MB}
\end{equation}
Here the integration contour is chosen in such a way
that all poles of $\Gamma(\cdots+z)$
(they are called \emph{left poles})
are to the left of the contour,
and all poles of $\Gamma(\cdots-z)$
(they are called \emph{right poles})
are to the right of it.
Indeed, closing the contour to the right
we get the expansion of the left-hand side in $b/a$;
closing it to the left --- the expansion in $a/b$.
In particular, it is often convenient to write
the massive propagator in the form~\cite{BD:91}
\begin{equation}
\begin{split}
\frac{1}{(m^2-p^2)^n} &= \frac{m^{-2n}}{\Gamma(n)} \frac{1}{2\pi i}
\int_{-i\infty}^{+i\infty} dz\,\Gamma(-z) \Gamma(n+z)
\left(\frac{-p^2}{m^2}\right)^z\,,\\
\raisebox{-3mm}{\begin{picture}(22,8)
\put(11,4){\makebox(0,0){\includegraphics{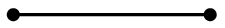}}}
\put(11,6){\makebox(0,0){$n$}}
\end{picture}} &=
\frac{m^{-2n}}{\Gamma(n)} \frac{1}{2\pi i}
\int_{-i\infty}^{+i\infty} dz\,\Gamma(-z) \Gamma(n+z)
m^{-2z}
\raisebox{-3mm}{\begin{picture}(22,8)
\put(11,4){\makebox(0,0){\includegraphics{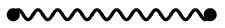}}}
\put(11,6){\makebox(0,0){$-z$}}
\end{picture}}\,.
\end{split}
\label{App1:BD}
\end{equation}
A massive line can be replaced by a massless one
(raised to the power $-z$)
at the price of one extra integration in $z$.

Now it is easy to calculate our massive diagram
using the massless result~(\ref{L1:I}):
\begin{equation}
I_{n_1 n_2}(m,\omega) =
\frac{m^{-2n_2} (-2\omega)^{d-n_1}}{\Gamma(n_1) \Gamma(n_2)}
\frac{1}{2\pi i} \int_{-i\infty}^{+i\infty} dz\,
\Gamma(n_1-d-2z) {\textstyle\Gamma\left(\frac{d}{2}+z\right)} \Gamma(n_2+z)
\left(\frac{-2\omega}{m}\right)^{2z}\,.
\label{App1:IMB}
\end{equation}
If we close the integration contour to the right,
then the sum over residues of the right poles
gives us the expansion in $\omega/m$.
In our case, their is one series of right poles
$z_n=(n+n_1-d)/2$ ($n=0$, 1, 2\dots),
and we obtain~(\ref{App1:w0}).
If we close the integration contour to the left,
then the sum over residues of the left poles
gives us the expansion in $m/\omega$,
thus providing the analytic continuation of~(\ref{App1:w0}).
In our case, there are two series of left poles,
$z^h_n=-n-n_2$ and $z^s_n=-n-\frac{d}{2}$;
they produce $I_h$~(\ref{App1:hard}) and $I_s$~(\ref{App1:soft}),
correspondingly.

\section{Electron field renormalization in QED}
\label{S:Appe}

Suppose we consider a gauge in which the free photon propagator
has a longitudinal part $\Delta(k) k_\mu k_\nu$:
\begin{equation*}
D^0_{\mu\nu}(k) =
\frac{1}{k^2} \left(g_{\mu\nu} - \frac{k_\mu k_\nu}{k^2}\right)
+ \Delta(k) k_\mu k_\nu
\end{equation*}
(then the full photon propagator $D_{\mu\nu}(k)$
has the same longitudinal part).
Then the full electron propagator has the form~\cite{LK:56}
\begin{equation}
S(x) = S_L(x) e^{- i e_0^2 (\tilde{\Delta}(x) - \tilde{\Delta}(0))}\,,\quad
\tilde{\Delta}(x) = \int \Delta(k) e^{-ikx} \frac{d^d k}{(2\pi)^d}\,,
\label{Appe:LK}
\end{equation}
where $S_L(x)$ is the Landau-gauge propagator.
This property follows from the simple gauge transformation
of the electron field in QED,
and does not generalize to non-abelian theories.
Various derivations of this formula are considered, e.\,g.,
in~\cite{F:56,JZ:59,BS:80,MK:80,MR:00}.
In the usual covariant gauge $\Delta(k)=a_0/(k^2)^2$;
therefore, $\tilde{\Delta}(0)=0$ in dimensional regularization.

The electron field renormalization does not depend on its mass.
For simplicity, we shall consider the massless electron,
whose propagator has a single Dirac structure:
\begin{equation*}
S(x) = S_0(x) e^{\sigma(x)}\,,
\end{equation*}
where
\begin{equation*}
S_0(x) = \frac{\Gamma(d/2)}{2 \pi^{d/2}}
\frac{\rlap/x}{(-x^2+i0)^{d/2}}
\end{equation*}
is the $d$-dimensional free massless electron propagator.
Then
\begin{equation*}
\sigma(x) = \sigma_L(x) + a_0 \frac{e_0^2}{(4\pi)^{d/2}}
\left(-\frac{x^2}{4}\right)^\varepsilon \Gamma(-\varepsilon)\,.
\end{equation*}
Re-expressing via renormalized quantities, we have
\begin{equation*}
\sigma(x) = \sigma_L(x) + a(\mu) \frac{\alpha(\mu)}{4\pi}
\left(-\frac{\mu^2 x^2}{4}\right)^\varepsilon
e^{\gamma_E\varepsilon} \Gamma(-\varepsilon)\,.
\end{equation*}
This should be equal to $\log Z_\psi + \sigma_r(x)$,
where $\log Z_\psi$ contains all negative powers of $\varepsilon$,
and $\sigma_r(x)$ --- all non-negative ones.
Therefore,
\begin{equation}
\log Z_\psi(\alpha,a) =
\log Z_L(\alpha) - a \frac{\alpha}{4\pi\varepsilon}\,,
\label{Appe:Z}
\end{equation}
where the Landau-gauge $Z_L(\alpha)$ starts from $\alpha^2$.
In QED
\begin{equation*}
\frac{d\log(a(\mu) \alpha(\mu))}{d\log\mu} = - 2 \varepsilon
\end{equation*}
exactly, because $Z_A Z_\alpha=1$.
Hence the anomalous dimension
\begin{equation}
\gamma_\psi(\alpha,a) =
2 a \frac{\alpha}{4\pi} + \gamma_L(\alpha)
\label{Appe:gamma}
\end{equation}
contains $a$ only in the one-loop term.

Let's see how this works up to 2 loops.
In momentum space, for the massless electron we have
\begin{equation*}
S(p) = \frac{1}{\rlap/p-\Sigma(p)}\,,\quad
\Sigma(p) = \rlap/p \Sigma_V(p^2)\,,\quad
\Sigma_V(p^2) = s_1 \frac{e_0^2 (-p^2)^{-\varepsilon}}{(4\pi)^{d/2}}
+ s_2 \frac{e_0^4 (-p^2)^{-2\varepsilon}}{(4\pi)^d} + \cdots
\end{equation*}
In coordinate space we obtain, using~(\ref{L1:Fl1}),
\begin{equation*}
\begin{split}
S(x) = S_0(x) \biggl[& 1
+ s_1
\frac{\Gamma(2-2\varepsilon)}{\Gamma(2-\varepsilon) \Gamma(1+\varepsilon)}
\frac{e_0^2}{(4\pi)^{d/2}}
\left(-\frac{x^2}{4}\right)^\varepsilon\\
&{} + (s_1^2 + s_2)
\frac{\Gamma(2-3\varepsilon)}{\Gamma(2-\varepsilon) \Gamma(1+2\varepsilon)}
\frac{e_0^4}{(4\pi)^d}
\left(-\frac{x^2}{4}\right)^{2\varepsilon}
+ \cdots \biggr]\,,
\end{split}
\end{equation*}
or
\begin{equation*}
\begin{split}
&\sigma(x) = s_1
\frac{\Gamma(2-2\varepsilon)}{\Gamma(2-\varepsilon) \Gamma(1+\varepsilon)}
\frac{e_0^2}{(4\pi)^{d/2}}
\left(-\frac{x^2}{4}\right)^\varepsilon\\
&{} + \left[ (s_1^2 + s_2)
\frac{\Gamma(2-3\varepsilon)}{\Gamma(2-\varepsilon) \Gamma(1+2\varepsilon)}
- \frac{1}{2}
\left( s_1
\frac{\Gamma(2-2\varepsilon)}{\Gamma(2-\varepsilon) \Gamma(1+\varepsilon)}
\right)^2
\right]
\frac{e_0^4}{(4\pi)^d}
\left(-\frac{x^2}{4}\right)^{2\varepsilon}
+ \cdots
\end{split}
\end{equation*}
Calculation of $s_1$ and $s_2$ is discussed in~\cite{G:07} in detail.
Substituting them, we arrive at
\begin{equation*}
\begin{split}
&\sigma(x) = a_0 \Gamma(-\varepsilon)
\frac{e_0^2}{(4\pi)^{d/2}}
\left(-\frac{x^2}{4}\right)^\varepsilon\\
&{} + \left[ 2 \frac{(d-2) (d-4)}{(d-3) (d-6) (3d-8)} n_f
- \frac{1}{2} \frac{d+4}{3d-8}
- \frac{(d-6) (3d-10)}{4 (d-3)^2} R \right]
\Gamma^2(-\varepsilon)
\frac{e_0^4}{(4\pi)^d}
\left(-\frac{x^2}{4}\right)^{2\varepsilon}\\
&{} + \cdots
\end{split}
\end{equation*}
where $n_f$ is the number of lepton flavours
($n_f=1$ in the usual QED), and
\begin{equation*}
R =
\frac{\Gamma(1-3\varepsilon) \Gamma(1-\varepsilon) \Gamma^2(1+\varepsilon)}%
{\Gamma^2(1-2\varepsilon) \Gamma(1+2\varepsilon)}
= 1 + 6 \zeta_3 \varepsilon^3 + \cdots
\end{equation*}
The one-loop term is linear in $a_0$;
the coefficient of $a_0$ agrees with our derivation
based on~(\ref{Appe:LK}).
The one-loop correction to the propagator
vanishes in the Landau gauge $a_0=0$;
this means that $\sigma_L(x)$ starts from $\alpha^2$.
All $a_0$-dependent terms have cancelled in the 2-loop term.
Thus we have completely checked $a_0$-dependent terms
in the 2-loop self-energy $s_2$.
Re-expressing $\sigma(x)$ via renormalized quantities,
we obtain
\begin{equation*}
\log Z_\psi(\alpha,a) =
- a \frac{\alpha}{4\pi\varepsilon}
+ \left(n_f + \frac{3}{4}\right) \frac{1}{\varepsilon}
\left(\frac{\alpha}{4\pi}\right)^2
+ \cdots
\end{equation*}
and hence
\begin{equation}
\gamma_\psi(\alpha,a) = 2 a \frac{\alpha}{4\pi}
- (4 n_f + 3) \left(\frac{\alpha}{4\pi}\right)^2
+ \cdots
\end{equation}
in full agreement with the general arguments.
This anomalous dimension has been calculated
up to 4 loops~\cite{CR:00}.

\section{Electron magnetic moment is IR finite}
\label{S:IR}

Let $\Gamma(\lambda)$ be the vertex
(expanded in $q$ up to the linear term and projected onto the magnetic-moment structure)
with an IR cutoff $\lambda\ll M$.
Then the magnetic moment is
\begin{equation}
\mu(\lambda) = Z_\psi^{\text{os}}(\lambda) \Gamma(\lambda)\,.
\label{IR:mu}
\end{equation}
We want to calculate $\mu(\lambda')$ for $\lambda'\ll\lambda$.
Essential contributions to $\Gamma(\lambda')$ are given by the skeleton diagrams
\begin{align}
&\Gamma(\lambda')
= \raisebox{-0.25mm}{\includegraphics{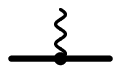}}
+ \raisebox{-4mm}{\includegraphics{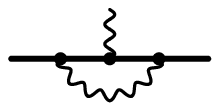}}
\label{IR:Gamma}\\
&{} + \raisebox{-11.5mm}{\includegraphics{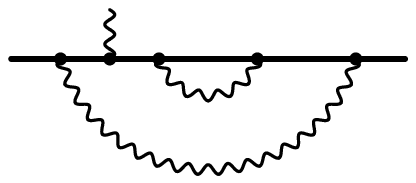}}
+ \raisebox{-11.5mm}{\includegraphics{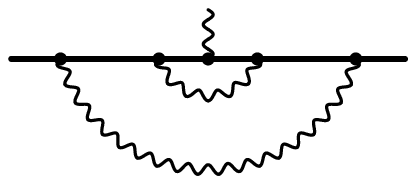}}
+ \raisebox{-11.5mm}{\includegraphics{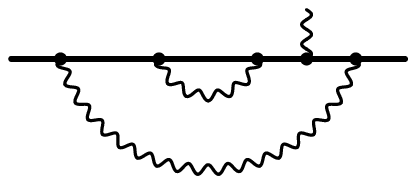}}
\nonumber\\
&{} + \raisebox{-7.75mm}{\includegraphics{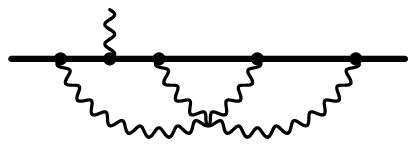}}
+ \raisebox{-7.75mm}{\includegraphics{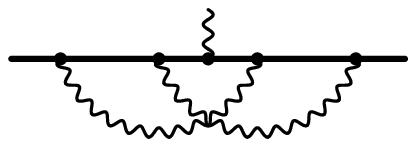}}
+ \raisebox{-7.75mm}{\includegraphics{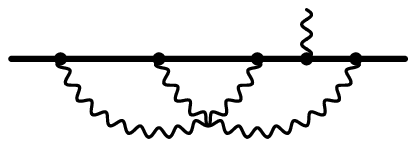}}
+ \cdots
\nonumber
\end{align}
where both ends of each soft photon line are attached to the external electron line.
When the momenta $k_i$ of all $L$ soft photon lines $\to0$,
the residual momenta $p_i$ of all $2L$ soft electron lines also $\to0$,
thus producing an IR divergence.
Each photon propagator in~(\ref{IR:Gamma})
is the full photon propagator with a small momentum $k_i<\lambda$,
and is equal to $Z_A^{\text{os}}(\lambda) D_0(k_i)$.
Each electron propagator in~(\ref{IR:Gamma})
is the full electron propagator with a small residual momentum $p_i<\lambda$,
and is equal to $Z_\psi^{\text{os}}(\lambda) S_0(Mv+p_i)$.
Each virtual-photon vertex in~(\ref{IR:Gamma})
is the full vertex with nearly on-shell momenta,
and is equal to $Z_\Gamma^{\text{os}}(\lambda) e_0 \gamma^\mu$.
Finally, each external-photon vertex in~(\ref{IR:Gamma})
is the full vertex $\Gamma(\lambda)$.

Let's multiply $\Gamma(\lambda')$~(\ref{IR:Gamma}) by $Z_\psi^{\text{os}}(\lambda)$.
Then each virtual-photon vertex will contain
$Z_\psi^{\text{os}}(\lambda) \bigl[Z_A^{\text{os}}(\lambda)\bigr]^{1/2} Z_\Gamma^{\text{os}}(\lambda) = e_{\text{os}}(\lambda)$,
and the external-photon vertex will be
$Z_\psi^{\text{os}}(\lambda) \Gamma(\lambda) = \mu(\lambda)$.
IR divergences in~(\ref{IR:Gamma}) can be reproduced in the effective theory:
\begin{align}
&Z_\psi^{\text{os}}(\lambda) \Gamma(\lambda')
= \raisebox{-0.25mm}{\includegraphics{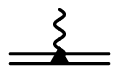}}
+ \raisebox{-4mm}{\includegraphics{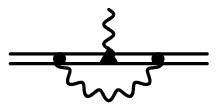}}
\label{IR:GammaHEET}\\
&{} + \raisebox{-11.5mm}{\includegraphics{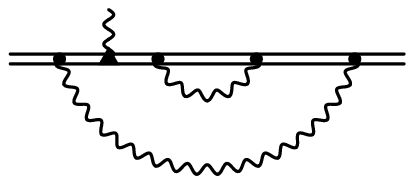}}
+ \raisebox{-11.5mm}{\includegraphics{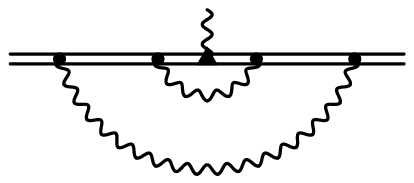}}
+ \raisebox{-11.5mm}{\includegraphics{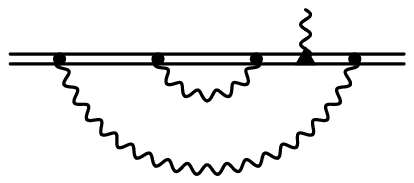}}
\nonumber\\
&{} + \raisebox{-7.75mm}{\includegraphics{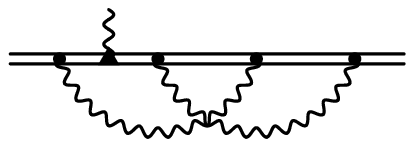}}
+ \raisebox{-7.75mm}{\includegraphics{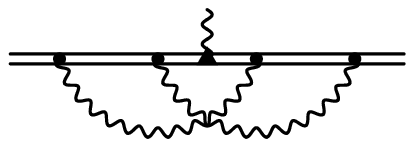}}
+ \raisebox{-7.75mm}{\includegraphics{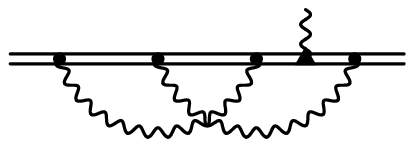}}
+ \cdots
\nonumber
\end{align}
(propagators are free,
virtual-photon vertices contain $e_{\text{os}}(\lambda)$,
and external-photon vertices contain $\mu(\lambda)$).
The electron lines attached to the external-photon vertex may be considered on-shell and the polarizations physical
(the electron-propagator numerators $(1+\gamma^0)/2$ just project onto the upper spinor components),
so that this vertex can be written in terms of 2 form factors~(\ref{Mag:FF});
we select the magnetic-moment structure $i\sigma^{\mu\nu}q_\nu/(2M)$.
It contains $q$; hence $q$ may be neglected everywhere else.
We arrive at
\begin{align}
&Z_\psi^{\text{os}}(\lambda) \Gamma(\lambda') = \mu(\lambda) \Biggl[
\raisebox{-0.25mm}{\includegraphics{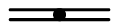}}
+ \raisebox{-4mm}{\includegraphics{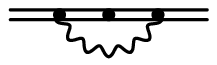}}
\label{IR:GammaHEET2}\\
&{} + \raisebox{-11.5mm}{\includegraphics{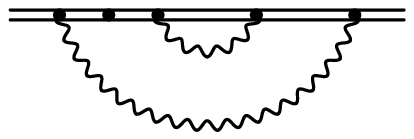}}
+ \raisebox{-11.5mm}{\includegraphics{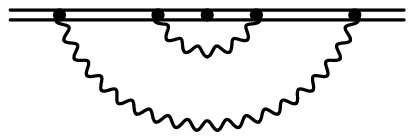}}
+ \raisebox{-11.5mm}{\includegraphics{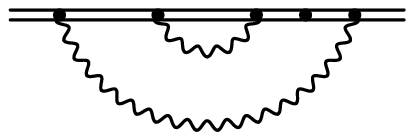}}
\nonumber\\
&{} + \raisebox{-7.75mm}{\includegraphics{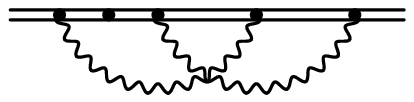}}
+ \raisebox{-7.75mm}{\includegraphics{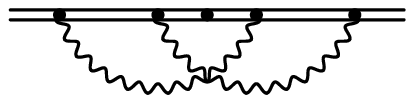}}
+ \raisebox{-7.75mm}{\includegraphics{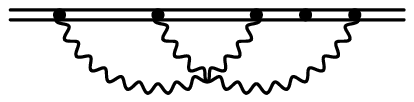}}
+ \cdots
\Biggr]\,.
\nonumber
\end{align}

We also need $Z_\psi^{\text{os}}(\lambda')$.
The on-shell electron wave-function renormalization is (see~(\ref{El:Zos}))
\begin{equation}
Z_\psi^{\text{os}} = \frac{1}{1 - \Sigma_0'(0)}\,,\qquad
\Sigma_0(\omega) = \frac{1}{4} \Tr (1+\rlap/v) \Sigma((M+\omega)v)\,.
\label{IR:Zos}
\end{equation}
Essential contributions to $\bigl[Z_\psi^{\text{os}}(\lambda')\bigr]^{-1}$ are given by the skeleton diagrams
\begin{align}
&\bigl[Z_\psi^{\text{os}}(\lambda')\bigr]^{-1}
= \raisebox{-0.25mm}{\includegraphics{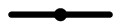}}
+ \raisebox{-4mm}{\includegraphics{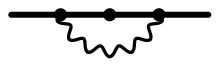}}
\label{IR:Z}\\
&{} + \raisebox{-11.5mm}{\includegraphics{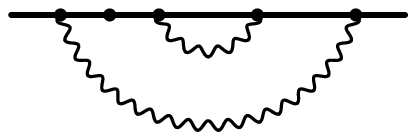}}
+ \raisebox{-11.5mm}{\includegraphics{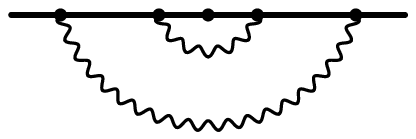}}
+ \raisebox{-11.5mm}{\includegraphics{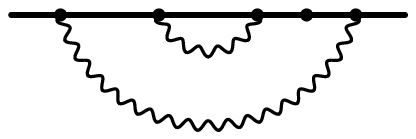}}
\nonumber\\
&{} + \raisebox{-7.75mm}{\includegraphics{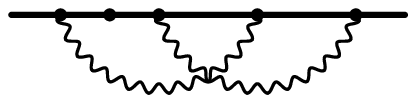}}
+ \raisebox{-7.75mm}{\includegraphics{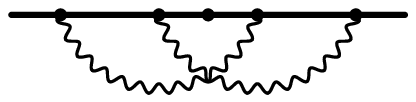}}
+ \raisebox{-7.75mm}{\includegraphics{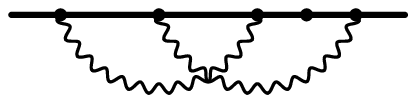}}
+ \cdots
\nonumber
\end{align}
Let's multiply it by $Z_\psi^{\text{os}}(\lambda)$:
\begin{align}
&Z_\psi^{\text{os}}(\lambda) \bigl[Z_\psi^{\text{os}}(\lambda')\bigr]^{-1}
= \raisebox{-0.25mm}{\includegraphics{ird0.eps}}
+ \raisebox{-4mm}{\includegraphics{ird1.eps}}
\label{IR:ZHEET}\\
&{} + \raisebox{-11.5mm}{\includegraphics{ird21.eps}}
+ \raisebox{-11.5mm}{\includegraphics{ird22.eps}}
+ \raisebox{-11.5mm}{\includegraphics{ird23.eps}}
\nonumber\\
&{} + \raisebox{-7.75mm}{\includegraphics{ird31.eps}}
+ \raisebox{-7.75mm}{\includegraphics{ird32.eps}}
+ \raisebox{-7.75mm}{\includegraphics{ird33.eps}}
+ \cdots
\nonumber
\end{align}
(propagators are free, vertices contain $e_{\text{os}}(\lambda)$).
At last, dividing~(\ref{IR:GammaHEET2}) by~(\ref{IR:ZHEET}), we obtain
\begin{equation}
\mu(\lambda')
= \frac{Z_\psi^{\text{os}}(\lambda) \Gamma(\lambda')}{Z_\psi^{\text{os}}(\lambda) \bigl[Z_\psi^{\text{os}}(\lambda')\bigr]^{-1}}
= \mu(\lambda)\,.
\label{IR:result}
\end{equation}
Thus we have proved that the electron magnetic moment in QED is not sensitive to an IR cutoff.

\end{document}